\documentclass[11pt,a4paper]{article}                           

\usepackage{a4}          

\usepackage[UKenglish]{babel}
\usepackage{amsmath}
\usepackage{amssymb}
\usepackage[usenames]{color} 
\usepackage{multirow}
\usepackage{graphicx}
\usepackage{epstopdf}
\usepackage{array}
\usepackage{hhline}
\usepackage{microtype}
\usepackage[bf]{caption}
\usepackage{cite}
\usepackage{color}
\usepackage[usenames,dvipsnames]{xcolor}
 
\usepackage{ifpdf}
\ifpdf
  \usepackage[pdftex,
    pdftitle={SU(5) Tops with (Multiple) U(1)'s},
    pdfauthor={},
    pdfsubject={F-theory compactifications},
    bookmarksopen, bookmarksnumbered, bookmarksopenlevel=2]{hyperref}
\fi
\def\hybrid{\topmargin -20pt    \oddsidemargin 0pt
        \headheight 0pt \headsep 0pt
        \textwidth 6.25in       
        \textheight 9 in       
        \marginparwidth .875in
        \parskip 5pt plus 1pt 
          \jot = 1.5ex
   }
\hybrid
\numberwithin{equation}{section}
\numberwithin{table}{section}

\newcommand{\beq}{\begin{equation}}  \newcommand{\eeq}{\end{equation}}
\newcommand{\bal}{\begin{aligned}}   \newcommand{\eal}{\end{aligned}}
\newcommand{\bea}{\begin{eqnarray}}  \newcommand{\eea}{\end{eqnarray}}

\newcommand{\GUT}{\mathrm{GUT}}

\newcommand{\nn}{\nonumber}

\newcommand{\cK}{\mathcal{K}}

\newcommand{\T}[1]{\textmd{#1}}
\newcommand{\C}{\text{C}}

\newcommand{\B}{\text{B}}

\newcommand{\tw}{\text{w}}
\newcommand{\tu}{\text{u}}
\newcommand{\tv}{\text{v}}
\newcommand{\tU}{\text{U}}

\newcommand{\un}{{\bf 1}}

\newcommand{\f}{{\bf 5}}
\newcommand{\fb}{{\bf \bar{5}}}
\newcommand{\te}{{\bf 10}}
\newcommand{\teb}{{\bf \bar{10}}}
\newcommand{\op}{\oplus}

\usepackage{cancel}

\newcommand{\be}{\begin{equation}}
\newcommand{\ee}{\end{equation}}

\newcommand{\executeiffilenewer}[3]{%
 \ifnum\pdfstrcmp{\pdffilemoddate{#1}}%
 {\pdffilemoddate{#2}}>0%
 {\immediate\write18{#3}}\fi%
}

\begin{document}

\baselineskip=14pt
\parskip 5pt plus 1pt

\vspace*{-1.5cm}
\begin{flushright}    
  {\small

  }
\end{flushright}

\vspace{2cm}
\begin{center}        
  {\LARGE $SU(5)$ Tops with Multiple U(1)s in F-theory}
\end{center}

\vspace{0.75cm}
\begin{center}        
Jan Borchmann$^{1}$, Christoph Mayrhofer$^{1}$, Eran Palti$^{1,2}$ and Timo Weigand$^{1}$
\end{center}

\vspace{0.15cm}
\begin{center}        
  \emph{$^{1}$ Institut f\"ur Theoretische Physik, Ruprecht-Karls-Universit\"at, \\
             Heidelberg, Germany}
             \\[0.15cm]
   \emph{$^{2}$ Centre de Physique Theorique, Ecole Polytechnique,\\ 
                CNRS, 
                Palaiseau, France
   }
\end{center}

\vspace{2cm}


\begin{abstract}

We study F-theory compactifications with up to two Abelian gauge group factors that are based on elliptically fibered Calabi-Yau 4-folds describable as generic hypersurfaces.  
Special emphasis is put on elliptic fibrations based on generic ${\rm Bl}_2 \mathbb P^2[3]$-fibrations. These exhibit a Mordell-Weil group of rank two corresponding to two extra rational sections which give rise to two Abelian gauge group factors. We show that an alternative description of the same geometry as a complete intersection makes the existence of a holomorphic zero-section manifest, on the basis of which we compute the $U(1)$ generators and a class of gauge fluxes. We analyse the fibre degenerations responsible for the appearance of localised charged matter states, whose charges, interactions and chiral index we compute geometrically.
We implement an additional $SU(5)$ gauge group by constructing the four inequivalent toric tops giving rise to $SU(5) \times U(1) \times U(1)$ gauge symmetry and analyse the matter content. We demonstrate that notorious non-flat points can be avoided in well-defined Calabi-Yau 4-folds. These methods are applied to the remaining possible hypersurface fibrations with one generic Abelian gauge factor. 
We analyse the local limit of our  $SU(5) \times U(1) \times U(1)$ models and
show that one of our models is not embeddable into $E_8$ due to recombination of matter curves that cannot be described as a Higgsing of $E_8$. We argue that such recombination forms a general mechanism that opens up new model building possibilities in F-theory.

\end{abstract}

\begin{center}
\emph{We dedicate this work to Matthis Florian.}
\end{center}

\thispagestyle{empty}
\clearpage
\setcounter{page}{1}


\newpage

\tableofcontents
\section{Introduction}

This article is devoted to a systematic construction of singular elliptically fibred Calabi-Yau 4-folds $ Y_4: T^2 \rightarrow B_3$ with several independent sections and their resolutions  $\hat Y_4$. Such geometries are the basis for the analysis of 4-dimensional F-theory \cite{Vafa:1996xn,Morrison:1996na} compactifications with both non-Abelian and Abelian gauge group factors. A considerable amount of recent work has been devoted to studying singular elliptically fibred 4-folds and their explicit resolutions realising 4-dimensional $SU(5)$ GUT models \cite{Donagi:2008ca,Beasley:2008dc,Beasley:2008kw,Donagi:2008kj} in globally defined frameworks \cite{Blumenhagen:2009yv,Esole:2011sm,Marsano:2011hv,Krause:2011xj,Grimm:2011fx} (see e.g. \cite{Chen:2010ts,Collinucci:2010gz,Knapp:2011wk,Morrison:2011mb,Collinucci:2012as,Kuntzler:2012bu,Tatar:2012tm,Lawrie:2012gg,Braun:2013cb,Hayashi:2013lra,Grassi:2013kha} for recent work analysing other gauge groups and related aspects of the geometry of elliptic fibrations).
 In addition to the non-Abelian sector, $U(1)$ symmetry groups are known to play a major role in phenomenological string model building, and F-theory is no exception. 
In particular, the selection rules associated with $U(1)$ symmetries have featured prominently in the phenomenology of F-theory GUT models, see             ~\cite{Marsano:2009gv,Marsano:2009wr,Dudas:2009hu,King:2010mq,Dudas:2010zb,Marsano:2010sq,Ludeling:2011en,Dolan:2011iu,Marsano:2011nn,Callaghan:2011jj,Dolan:2011aq,Palti:2012aa,Kerstan:2012cy,Palti:2012dd,Mayrhofer:2013ara} and the reviews \cite{Weigand:2010wm,Maharana:2012tu} for further references. Abelian gauge groups depend in an intricate way on the details of the global geometry \cite{Hayashi:2010zp,Grimm:2010ez}. This has lead to an exciting and fruitful interplay between more formal and applied aspects of string model building.

The appearance of Abelian gauge group factors in F-theory is tied to the existence of extra sections of the elliptic fibration in addition to the universal or zero-section.
A non-degenerate section generally maps every point in the base $B_3$ to a point in the fibre. If an elliptic fibration possesses a holomorphic section, i.e. a section which is non-degenerate over the entire $B_3$ and varies holomorphically, this provides an embedding of the base $B_3$ as a holomorphic divisor into the elliptic fibration $Y_4$. 
In the context of F-theory compactifications, this therefore defines the physical compactification space $B_3$. 
In addition, an elliptic fibration may exhibit extra sections. The relation between such extra sections and $U(1)$ symmetries is roughly this:
As we will see in detail, the presence of extra sections, beyond the universal section defining the base, renders the 4-fold singular in codimension two by inducing an $SU(2)$ singularity in the fibre over certain curves. This is related to the fact that the extra sections degenerate in codimension. The resolution of these singularities gives rise to 
elements $ \tw \in H^{1,1}(\hat Y_4)$ which are not in the pullback of $H^{1,1}(B_3)$ nor are equivalent to the universal section. 
Expansion of the M-theory 3-form $C_3$ in terms of such 2-forms as $C_3 = A \wedge \tw + \ldots$ is well-known to 
yield a $U(1)$ gauge potential in the low-energy effective theory by F/M-theory duality \cite{Morrison:1996na}. After the resolution, the extra sections wrap entire fibre components over certain curves in $B_3$, and therefore form rational (as opposed to holomorphic) sections.
Therefore studying elliptic 4-folds with several sections is the basis for studying Abelian gauge symmetry in F-theory. 
Note that the group law on the elliptic curve endows the set of sections with a group structure, the so-called Mordell-Weil group. We are henceforth interested in fibrations with Mordell-Weil rank greater than or equal to one. 

Elliptic fibrations with a holomorphic section can be written in Weierstra\ss{}  form  
\begin{eqnarray} \label{Weier}
y^2 = x^3 + f x z^4 + g z^6.
\end{eqnarray}
The fibre coordinates $[x : y : z]$ are homogeneous coordinates on the fibre ambient space $\mathbb P_{2,3,1}$, and $f$ and $g$ are defined on the base as sections of $\bar {\cal K}^4$ and $\bar {\cal K}^6$, respectively (with $\bar {\cal K}$ the anti-canonical bundle of the base $B_3$). 
As it stands, if $f$ and $g$ are generic there are no non-Abelian singularities in the fibration \eqref{Weier}, and since our Abelian gauge groups also come from resolving $SU(2)$ loci, neither do we have any massless Abelian gauge fields. Inducing non-Abelian singularities by making $f$ and $g$ non-generic is in principle a well understood procedure. It is possible to read off the singularity type from the vanishing order of $f$ and $g$ over a divisor. 
From the above the construction of Abelian gauge groups amounts to finding the possible restrictions on $f$ and $g$ which lead to extra (rational) sections. 
This is a less well understood procedure and is the topic of much current investigation \cite{Grimm:2010ez,GrassiPerduca,Choi:2012pr,Morrison:2012ei,Mayrhofer:2012zy,Braun:2013yti,Borchmann:2013jwa,Cvetic:2013nia,Grimm:2013oga,Braun:2013nqa,Cvetic:2013uta}.\footnote{Early work on Abelian gauge groups and/or multi-section fibrations in the context of 6-dimensional F-theory compactifications includes \cite{Klemm:1996hh,Aldazabal:1996du,Candelas:1997pv,Berglund:1998va}.} As yet no classification of the Abelian sector is known, and current work essentially relies on particular methods of finding forms of $f$ and $g$ that allow for additional sections.

Often these methods write the elliptic fibration not in Weierstra\ss{}  form, but choose a different representation of an elliptic curve 
where the additional sections can be more easily identified, and eventually transform it to Weierstra\ss{}  form through rational maps. 
In fact, there are many more representations of an elliptic curve either as a hypersurface or as a complete intersection of some ambient space.
A representation, e.g. as some other hypersurface than the Weierstra\ss{}  model with relatively mild restrictions on the coefficients can map to a Weierstra\ss{}  model with highly non-generic $f$ and $g$. In this sense finding restrictions on the elliptic curve such as to create extra sections may be easier if one starts with a different representation of the elliptic curve.

The initial such systematic studies of Abelian symmetries in the presence of non-Abelian ones were done by writing the fibration in Tate form \cite{Grimm:2010ez,Braun:2011zm,Mayrhofer:2012zy,Grimm:2011fx}. A Tate model is defined as the hypersurface in $\mathbb P_{2,3,1}[6]$ given by
\begin{eqnarray}
P_T: y^2 = x^3 + a_1 x y z + a_2 x^2 z^2 + a_3 y z^3 + a_4 x z^4 + a_6 z^6.
\end{eqnarray}
Tate's algorithm \cite{Bershadsky:1996nh} gives a prescription for engineering extra singularities in the fibre over a divisor $w=0$ on $B_3$ corresponding to non-Abelian gauge groups by restricting $a_i = a_{i,n_i} w^{n_i}$ in a well-defined way. Note that while every Tate model can be brought into Weierstra\ss{}  form, the converse is generally not true \cite{Katz:2011qp}. 

Possibly the simplest example of elliptic 4-folds with one extra section is obtained by setting $a_6 \equiv 0$ \cite{Grimm:2010ez}. This gives rise to an independent extra section at 
\begin{eqnarray}
{\rm Sec}_1: \qquad [x : y : z] = [0 : 0 : z].
\end{eqnarray}
The fibration becomes singular at this point in the fibre over the curve $a_3 = a_4 =0$ in $B_3$. The singularity can be resolved by a blow-up in the ambient space, $x \rightarrow x \, s, \quad y \rightarrow y \, s$ \cite{Grimm:2010ez,Krause:2011xj,Grimm:2011fx}, which leads to a smooth space given as a hypersurface in a ${\rm Bl}_1 {\mathbb P}_{2,3,1}$-fibration over $B_3$.
Alternatively, the resolution can be performed 
in a way similar to the conifold resolution \cite{Braun:2011zm}, in which case the smooth space is described as a complete intersection within a six-dimensional complex space.
Both ways amount to a small resolution on $\hat Y_4$. In the resolved space the extra section gives rise to a new divisor responsible for the presence of an Abelian gauge potential. E.g.\ in the blow-up procedure of \cite{Grimm:2010ez,Krause:2011xj,Grimm:2011fx}  the resolved section corresponds to the divisor $s=0$.

This construction of extra sections based on the Tate model was systematically generalised in  \cite{Mayrhofer:2012zy}.
Inspired by the forms of coefficients that appear in local models of F-theory, a systematic factorisation of  $P_T$ was described which guarantees the existence of (possibly multiple independent) extra sections. Combined with a non-Abelian gauge group of e.g. $SU(5)$ type along a divisor such factorised Tate models lead  to $SU(5)$ GUT models with up to four generic $U(1)$ factors. This algorithm was worked out explicitly for the two possible inequivalent models of $SU(5) \times U(1)$ symmetry that result in this fashion. One of these is a generalisation the $SU(5) \times U(1)$ model of \cite{Grimm:2010ez,Krause:2011xj,Grimm:2011fx}, while the other one realises a so-called Peccei-Quinn symmetry and is the first example of an $SU(5) \times U(1)$-fibration with a split ${\bf 10}$-curve.

The Tate model is still a $\mathbb P_{[2,3,1]}[6]$-fibration. An alternative starting point for multi-section fibrations is to use different representations of the elliptic fibre either as a different hypersurface or, more generally, as a complete intersection. 
In \cite{Morrison:2012ei} it was shown that a large class of elliptic fibrations with a Mordell-Weil group of rank 1 (corresponding to one extra section and thus one extra generic $U(1)$), written as a hypersurface with generic coefficients, can be brought into the form of a special $\textmd{Bl}_{1}\mathbb P_{[1,1,2]}[4]$-fibration
\begin{equation}\label{eq:211-hse-ng}
  \B\,\textmd v^2\, \textmd w +  s\, \textmd w^2  =\C_3\,\textmd v^3\,\textmd  u +  \C_2\, s \, \textmd v^2 \,\textmd  u^2 +  \C_1 \, s^2 \,\textmd v\,\textmd  u^3 +   \C_0\, s^3\,\textmd  u^4\,.
\end{equation}
Here $[\textmd u : \textmd v : \textmd w]$ represent homogeneous coordinates of the fibre ambient space $\mathbb P_{[1,1,2]}$ and $\B, \C_i$ are suitable sections of $\bar {\cal K}$. Let us set $s=1$ for a moment. Then \eqref{eq:211-hse-ng} represents, up to coordinate redefinitions, the most generic quartic polynomial in $\mathbb P_{[1,1,2]}$ leading to an elliptic curve, except that the term $\textmd v^4$ is missing.\footnote{Note, however, that the coefficient of $\tw^2$ has not been allowed to vary.} This non-genericity is responsible for the presence of the two sections  \cite{Morrison:2012ei}
\begin{eqnarray}
&& {\rm Sec}_0: \qquad  [\textmd u : \textmd v : \textmd w] = [0 : \textmd v : 0], \\
&& {\rm Sec}_1: \qquad  [\textmd u : \textmd v : \textmd w] = [0 : \textmd v : -\B \textmd v^2],
\end{eqnarray}
where ${\rm Sec}_0$ represents the universal holomorphic section. 
As it turns out the fibration with $s=1$ exhibits singularities in the fibre over  codimension-two loci in the base $B_3$. This is remedied by introducing the blow-up coordinate $s$ via $\textmd  u \rightarrow \textmd u \, s, \quad \textmd  w \rightarrow \textmd w \, s$. 
In the resolved space the divisor $s=0$ represents an extra rational section. In fact the 2-section models proposed in \cite{Grimm:2010ez} are a special case of fibrations of type \eqref{eq:211-hse-ng}.
Furthermore one can find the explicit birational map transforming \eqref{eq:211-hse-ng} into Weierstra\ss{}  form or Tate form and map the universal and the additional section to sections of $P_T$ - see \cite{Morrison:2012ei} for details.

The elliptic fibration \eqref{eq:211-hse-ng} is, in fact, based on one out of the 16 possible ways to write a torus as a hypersurface (as opposed to a complete intersection) of a toric ambient space. These have been analysed in detail in \cite{Bouchard:2003bu}. 
With otherwise generic $\B, \C_i$ the fibration \eqref{eq:211-hse-ng} gives rise to gauge group $U(1)$, and in order to describe extra non-Abelian gauge groups $\B, \C_i$ must be restricted further. A prescription to achieve this in the language of toric geometry is given by the construction of tops introduced in ~\cite{Candelas:1996su} and classified in \cite{Bouchard:2003bu}. 
In \cite{Mayrhofer:2012zy}  the factorised $SU(5) \times U(1)$ Tate models have also been brought into the general form \eqref{eq:211-hse-ng} and non-Abelian singularities have been analysed with the help of such tops. In particular the resulting non-generic form of the coefficients $\B, \C_i$ is such as to allow for multiple $\te$ matter curves after combining the toric blow-up with a small resolution into a complete intersection.

A systematic starting point for the construction of elliptic fibrations with several sections is to focus first on these possible hypersurface descriptions of the elliptic fibre.
To stay within the general logic behind the constructions of \cite{Grimm:2010ez,Braun:2011zm,Mayrhofer:2012zy} we are particularly interested in that part of the structure of 
Abelian gauge groups in F-theory which can be analysed in a manner independent of the explicit choice for the base space $B_3$.
This will lead to generic statements of the elliptic fibre and result in a number of conditions which a base $B_3$ has to satisfy in order to give rise to a well-defined F-theory compactification with the properties under consideration.
By genericity of the construction we  mean that all of the results we will derive are guaranteed to hold for suitable choices of $B_3$ which satisfy the preconditions that we will specify.
Further restrictions on the input parameters, which may require special choices of $B_3$, can lead to extra, non-generic structure. Indeed, in  \cite{Braun:2013nqa} a survey of possible constructions of toric elliptic fibrations based on the polygons of \cite{Bouchard:2003bu} has been performed (see \cite{Braun:2013yti} for a specific example thereof) and a classification of the maximal and minimal number of independent sections (including non-generic situations in the above sense) is given. Previously \cite{GrassiPerduca} had classified possible toric and non-toric sections in elliptic fibrations given as hypersurfaces.

In this article we consider fibrations with two extra independent sections based on the  $\textmd{Bl}_{2}\mathbb P^2[3]$ representation of the elliptic fibre. These describe the most generic fibrations with at least 2 additional $U(1)$ symmetries that can be written as a hypersurface with generic coefficients. In the letter \cite{Borchmann:2013jwa} we have already presented some of the main results on these $\textmd{Bl}_{2}\mathbb P^2[3]$ fibrations including a construction of all $SU(5)\times U(1)\times U(1)$ models based on this fibration which can be achieved via the classification of tops. This is due to the fact that $\textmd{Bl}_{2}\mathbb P^2[3]$ corresponds to one of the polygons of  \cite{Bouchard:2003bu}. In this work, apart from spelling out the most important  derivations behind these results, we substantially add to the analysis of  \cite{Borchmann:2013jwa}. 
Some parts of sections 2 and 3 of this article have some overlap with the work presented in \cite{Cvetic:2013nia,Cvetic:2013uta}, which also studies $\textmd{Bl}_{2}\mathbb P^2[3]$ fibrations (and exemplifies the incorporation of $SU(5)\times U(1)\times U(1)$ via one of the tops analysed in \cite{Borchmann:2013jwa}), but the methods of \cite{Cvetic:2013nia,Cvetic:2013uta} and of our work oftentimes differ.

In Section~\ref{sec:fthe2u1s} we begin with an exposition of $\textmd{Bl}_{2}\mathbb P^2[3]$-fibrations without extra non-Abelian gauge groups. We describe the logic that leads to this presentation of the elliptic fibre and analyse the properties of the extra two rational sections.
A drawback of the simple description of the elliptic fibration as a hypersurface is that that no holomorphic zero section exists which can be described as the vanishing locus of a divisor pulled back from the ambient space. 
In Section~\ref{sec:holzerosec} we show that this is merely an artefact of the specific presentation of the elliptic fibre as a hypersurface and work out an alternative description as a complete intersection in which a holomorphic zero section becomes manifest. Thus, a holomorphic embedding of the base is possible. We use this embedding to define the two      $U(1)$ generators presented already in \cite{Borchmann:2013jwa}.
In Section~\ref{eq:U1singlets} we work out the birational map from the $\textmd{Bl}_{2}\mathbb P^2[3]$-fibration to the Weierstra\ss{}  model, also presented already in \cite{Borchmann:2013jwa}, and use this to analyse the spectrum of localised matter states charged under the two $U(1)$ gauge groups as well as their Yukawa interactions.
In Section~\ref{sec:U(1)Fluxes} we use our holomorphic embedding of the base to construct a simple class of chirality-inducing  $G_4$-fluxes. Apart from the two $U(1)$-fluxes we find another $G_4$-flux that can be understood geometrically as the Hodge dual to one of the matter surfaces hosting the charged singlets. 
We compute the chiral index of the singlet states with respect to these fluxes by exploiting the geometric properties of these fluxes in a manner completely independent of the base space $B_3$.

In Section~\ref{sec:Tops-all} we detail the implementation of an extra $SU(5)$-singularity in the fibre over a base divisor  which is conventionally called $S_{}$. There are 5 inequivalent tops which lead to different $SU(5) \times U(1) \times U(1)$ models. One of them is pathological in that it leads to a non-flat fibre in codimension-two, and is henceforth discarded. Section \ref{sec:topsintro} describes the general construction of tops \cite{Candelas:1996su,Bouchard:2003bu}, which is then applied in Section~\ref{sec_Top1} to one out of the remaining 4 inequivalent ways of realising an $SU(5) \times U(1) \times U(1)$ in our framework. This allows us to analyse the $SU(5)$ matter spectrum, the $U(1)$ charges and the Yukawa interactions of the states. 
An extra complication arises because the restrictions on the fibration necessary to induce an $SU(5)$ singularity lead to points on $B_3$ over which the complex fibre dimension jumps from one to two. To avoid tensionless strings such non-flat points are to be avoided. In Section~\ref{sec:Flatness} we exemplify that this is indeed possible for the specific fibration of Section~\ref{sec_Top1} by restricting the fibration in a suitable manner. We prove, for the example of $B_3= {\mathbb P}^3$, that this procedure gives rise to a well-defined elliptically fibred 4-fold. 

In Appendix~\ref{app-SU5} we work out all the $SU(5)$ tops  for those of the 16 polygons of \cite{Bouchard:2003bu} which give rise to generic elliptic fibrations with one or two extra sections, provide the matter spectrum, the Abelian generators and list the $U(1)$ charges. In Appendix~\ref{app-SU4} we extend this analysis to the corresponding $SU(4)$ tops. The motivation for this is that in \cite{Mayrhofer:2012zy} the first examples of $SU(5)$ elliptic fibrations with two ${\bf 10}$-curves have been constructed as a complete intersection, starting from an $SU(4)$ top but with non-generic coefficients such as to enhance to $SU(5)$.

In Section~\ref{sec:embe8} we explore whether the matter content, gauge symmetries, and Yukawa couplings of our global models can be embedded into a Higgsed $E_8$ gauge theory. 
Local model building in F-theory has been based on a classification scheme of models which arises from possible Higgsings of an $E_8$ theory to $SU(5)_{GUT}$, which is usefully written through an intermediate breaking $E_8 \rightarrow SU(5)_{GUT} \times SU(5)_{\perp}$. This is the underlying structure behind the spectral cover constructions which were imported from the Heterotic string and introduced to F-theory in \cite{Donagi:2009ra,Marsano:2009gv}, and which were heavily used in the literature subsequently. Having constructed global models it is therefore natural to ask whether they fit into this local model framework. 

There are two aspects of our models which must be recreated in a valid embedding into $E_8$. We specify an embedding through the embedding of the two global $U(1)$ symmetries of our models into the Cartan of $SU(5)_{\perp}$. Then the charges of all the matter curves in the theory should appear in a decomposition of the adjoint of $E_8$ into $SU(5)_{\GUT} \times U(1)^4$. This is the requirement that the charges are embeddable in $E_8$ and we find that this is possible for all of our models (and all the other global models constructed in the literature to date). The second requirement is that the Yukawa couplings of our global models can arise as gauge invariant couplings of the states coming from the decomposition of the adjoint of $E_8$. We find that one of our models, the one based on top 4, fails to meet this requirement. In this sense we are presenting the first example of an $SU(5)$ F-theory model which cannot be embedded into $E_8$ even in this group theory sense. More precisely we find that recreating 
the Yukawa couplings requires that two $\f$-matter curves are recombined but that there is no GUT singlet in the adjoint of $E_8$ which has the correct charge to perform this recombination. 

We go on to argue that the required recombination singlet which goes beyond $E_8$ is in fact present in F-theory models. It is part of a class of singlets which includes an example that was identified already in the global models of \cite{Mayrhofer:2012zy}. Such singlets have the combined charges of two singlets that come from the adjoint of $E_8$. We argue that they are present and can recombine any two $\f$-matter curves in F-theory models, something which goes beyond a Higgsed $E_8$ theory where generically there are pairs of $\f$-matter curves that cannot form a gauge invariant cubic coupling with GUT singlets. We show that the presence of such singlets which go beyond $E_8$ is tied to the fact that the presence of a coupling of type $\un \, \f \, \fb$, which appears at points on $S_{}$, can not be determined in a local theory and is sensitive to the geometry away from the GUT brane.

\section{F-theory fibrations with two \texorpdfstring{$U(1)$}{U(1)}s}

\subsection{\texorpdfstring{$\textmd{Bl}_{2}\mathbb P^2[3]$}{Bl2 P\{1,1,1\}[3]}-fibrations}
\label{sec:fthe2u1s}


In this section we discuss F-theory compactifications with two $U(1)$ gauge groups based on elliptic fibrations which are described as  generic hypersurfaces in an ambient space.
This amounts to  constructing elliptic fibrations with Mordell-Weil group of rank 2. We will identify the fibration as a ${\rm Bl}_2 \mathbb P^2[3]$-fibration,
whose form and most important properties  we have already presented in  \cite{Borchmann:2013jwa}.
The logic behind our derivation underlying the results presented in  \cite{Borchmann:2013jwa} is a generalisation of a well-known procedure in algebraic  geometry to construct the Weierstra\ss{}  model, i.e. an elliptic fibration with a section. It was applied in~\cite{Morrison:2012ei} to construct 
2-section fibrations as the ${\rm Bl}_1\mathbb P_{1,1,2}[4]$-fibrations \eqref{eq:211-hse-ng}. Note that the independent work  \cite{Cvetic:2013nia,Cvetic:2013uta} also studies the ${\rm Bl}_2\mathbb P^2[3]$-fibrations analysed here and in  \cite{Borchmann:2013jwa}.

As recalled in the introduction, a section of the fibration gives a copy of the base space. For the fibration to possess (several independent) sections given by a divisors $D_i$ in the fibred Calabi-Yau 4-fold $Y_4$, the intersection of $D_i$ with the fibre must be points and these points must not be interchanged by monodromies. This is guaranteed if the intersection points are given by rational points on the elliptic curve such that no branch cuts arise.
Applied to the present situation we are thus interested in an elliptic curve with three such rational points, which we then fibre over $B_3$.

Let us denote the rational points as $P$, $Q$ and $R$. Given the equivalence between points  and line bundles on elliptic curves we can rephrase the statement that these lie on the elliptic curve as the statement that the degree-three line bundle $L = {\cal O}(P + Q + R)$ over the elliptic curve have a section which vanishes precisely at $P$, $Q$ and $R$. 

In the first step, notice that the line bundle $L$, being of degree three, must have three independent sections which we denote by $\tu,\tw,\tv$.
Then the degree-six line bundle $L^2$ has six sections given the six monomials $\tu^2, \tv^2, \tw^2, \tu \tv, \tu \tw, \tv \tw$. The degree-nine line bundle $L^3$ has nine sections, but since one can form ten monomials $\tu^3$, $\tv^3$, $\tw^3$, $\tu^2 \tv$, $\tu^2 \tw$, $\tu \tv^2$, $\tu \tw^2$, $\tu \tv \tw$, $\tw^3$, $\tv^3$ these must satisfy one relation, which can be viewed as a generic cubic equation in $\mathbb P^2$. This leads to the representation of the elliptic curve as $\mathbb P^2[3]$ with $[\tu\,:\,\tv\,:\,\tw]$ homogeneous coordinates of the fibre ambient space $\mathbb P^2$.

\begin{figure}[h]
    \centering
\def\svgwidth{0.6 \textwidth}
 \executeiffilenewer{polygon5.svg}{polygon5.pdf}%
 {inkscape -z -D --file=polygon5.svg %
  --export-pdf=polygon5.pdf --export-latex}%
   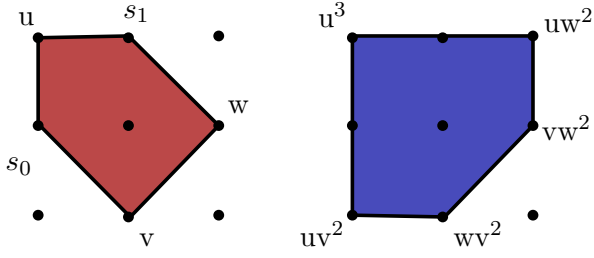%

      \caption{Polygon for $\textmd{Bl}_{2}\mathbb{P}^{2}$ }\label{fig:polygon5}
\end{figure}

In the final step one finds restrictions on this  cubic which ensure that one of the sections of $L = {\cal O}(P + Q + R)$ vanishes precisely at $P, Q, R$.
Suppose that $\tu =0$ is the section that vanishes precisely at these three points.
The locus $\tu=0$ in a generic $\mathbb P^2$ can be represented as the equation
\begin{eqnarray}
\tilde c_0  \tw^3 + \tilde c_1 \tw^2 \tv + \tilde c_2 \tw \tv^2 + \tilde c_3 \tv^3 =0
\end{eqnarray}
for some coefficients $\tilde c_i$. For $\tu$ to vanish at three distinct points, this equation must factorise as
\begin{eqnarray}
(\alpha_1 \tw + \beta_1 \tv) (\alpha_2 \tw + \beta_2 \tv)  (\alpha_3 \tw + \beta_3 \tv) = 0,
\end{eqnarray}
where the coefficients $\alpha_i$, $\beta_i$ must not vanish simultaneously and the three vanishing points must be distinct.
One can then relabel $(\alpha_1 \tw + \beta_1 \tv) \rightarrow \tw$ and $(\alpha_2 \tw + \beta_2 \tv) \rightarrow \tv$ so that $\tu=0$ becomes
\begin{eqnarray}
\tw \tv (c_1\,  \tw   + c_2\, \tv ) =0
\end{eqnarray}
for some new coefficients $c_1$ and $c_2$.
As a result, the elliptic fibre can be represented as the vanishing of the cubic  \cite{Borchmann:2013jwa} (see also \cite{Cvetic:2013nia,Cvetic:2013uta})
\begin{equation} \label{eq:hyper1}
\begin{split}
 P_T = \tv\, \tw (c_1\,  \tw + c_2\, \tv )  &+ \tu\, (b_0\, \tv^2 + b_1\, \tv\, \tw  + b_2\, \tw^2) +  \tu^2 (d_0\, \tv  + d_1\, \tw  + d_2\, \tu).
\end{split}
\end{equation}
Note that \eqref{eq:hyper1} represents the most general cubic in $\mathbb P^2$ except that the coefficients of $\tw^3$ and $\tv^3$ have been set to zero, cf.\ Figure~\ref{fig:polygon5}.
This ensures that the elliptic fibre possesses  three independent rational points, which we henceforth call 
\begin{equation}\label{eq:sections_2u1s_cubic}
\begin{split}
 \T{Sec}_0:\qquad [\tu\,:\,\tv\,:\,\tw] &= [0\,:\,0\,:\,\tw],\\
 \T{Sec}_1:\qquad [\tu\,:\,\tv\,:\,\tw] &= [0\,:\,\tv\,:\,0],\\
 \T{Sec}_2:\qquad [\tu\,:\,\tv\,:\,\tw] &= [0\,:\,-c_1\,:\,c_2].
\end{split}
\end{equation}

With $c_i$, $b_i$, $d_i$ suitable sections of some line bundle over the base $B_3$ this describes a (singular) Calabi-Yau 4-fold $Y_4: T^2 \rightarrow B_3$ as a hypersurface in the ambient space $X_5$ given by a $\mathbb P^2$-fibration over $B_3$. In the \emph{generic} case,  none of the bundles corresponding to sections $c_i$, $b_i$, $d_i$ is trivial. In this situation we have the freedom to choose  $\tw$ to be a section of $\alpha\otimes \mathcal{L}$ and $\tv$ to be a section of  $\beta\otimes \mathcal L$ where $\alpha$ and $\beta$ are some line bundles over the base and $\mathcal L$ is the line bundle corresponding to the hyperplane class of the fibre $\mathbb P^2$, cf.\ Table~\ref{tab:scaling}.\footnote{Note that this is different from the Weierstra\ss{} model, where no such freedom arises because the monomials $y^2$ and $x^3$ have constant coefficients. Also, for the $\mathbb P_{1,1,2}[4]$ fibration \eqref{eq:211-hse-ng} one can choose only one line bundle because $\tw^2$ appears with a constant.}
\begin{table}
\centering
\begin{tabular}{c||ccccc}
 & \tu & \tv & \tw \\
\hline
\hline
$\alpha$ & $\cdot$ & $\cdot$ & 1  \\
$\beta$ & $\cdot$ &1 &  $\cdot$ \\
\hline
U & 1 & 1 & 1 
\end{tabular}
\caption{Divisor classes and coordinates of the fibre ambient space of $X_{5}$;}\label{tab:scaling}
\end{table}

The bundles $\alpha$ and $\beta$ are not totally arbitrary. They are bounded because $b_i$, $c_i$, $d_i$ have to be global sections of some bundles over the base and the hypersurface \eqref{eq:hyper1} has to be Calabi-Yau. From the Calabi-Yau condition we deduce the scalings for the coefficients as given in Table~\ref{coeff}. Again, note that our analysis here and in \cite{Borchmann:2013jwa} has some overlap with the independent analysis of \cite{Cvetic:2013nia,Cvetic:2013uta}.
\begin{table} 
\centering
\begin{tabular}{c|c|c|c|c|c|c|c}
 $b_{0}$ & $b_{1}$ & $b_{2}$ & $c_{1}$ & $c_{2}$&$d_{0}$&$d_{1}$&$d_{2}$\\
\hline
 $\alpha-\beta+ \bar {\cal K}$&$\bar {\cal K}$&$-\alpha +\beta+\bar {\cal K}$&$-\alpha+\bar {\cal K}$&$-\beta + \bar {\cal K}$&$\alpha + \bar {\cal K}$&$\beta+\bar {\cal K}$&$\alpha+\beta+\bar {\cal K}$\\
\end{tabular}
\caption{Classes of the coefficients with $\alpha$ and $\beta$ `arbitrary' classes of $B_3$ and $\bar {\cal K}$ the anti-canonical class of $B_3$.}\label{coeff}
\end{table}
We can decompose $\alpha$, $\beta$ and $\bar{\mathcal{K}}$ into generators of $H^2(B_3,\mathbb Z)$\footnote{Note that we assume here that $H^2(B_3,\mathbb Z)$ is equivalent to the Picard group.},
\[
\alpha= \alpha^l H_l,\qquad \beta=\beta^l H_l\quad \textmd{and}\quad \bar{\mathcal{K}}=k^l H_l 
\]
with $\alpha^l$, $\beta^l$, $k^l$ $\in\, \mathbb Z$ and $H_l$ generators of $H^2(B_3,\mathbb Z)$. For $b_i$, $c_i$, $d_i$ to be globally well-defined we obtain for  each $H_l$ a set of inequalities,
\begin{equation}\label{eq:inequal-coeff}
\begin{split}
 &\qquad\qquad\left\{\frac{1}{k^l}\,\vec v^j\cdot(\alpha^l,\beta^l)\ge-1\right\}_{j=1,\ldots,7} \quad\textmd{with}\\
 & \left\{\vec v^j\right\}_{j=1,\ldots,7}=\left\{(1,-1)^T,\,(-1,1)^T,\,(-1,0)^T,\,(0,-1)^T,\,(1,0)^T,\,(0,1)^T,\,(1,1)^T\right\}\,.
\end{split}
\end{equation}
The polygon spanned by the vectors $\vec v^j$ in \eqref{eq:inequal-coeff} is the right-hand polygon of Figure~\ref{fig:polygon5} reflected along the $y$-axis polygon. Therefore, the solutions to the inequalities of \eqref{eq:inequal-coeff} are given by the interior of the right-hand polygon of Figure~\ref{fig:polygon5} reflected along the $y$-axis, but with a refined lattice or a scaled polygon to take account of the $\frac{1}{k^l}$ factor. Hence, for every $H_i$ we obtain the same shape for the bounded region of allowed values of $\alpha^l$ and $\beta^l$, only the size depends on $k^i$, cf.\ Table~\ref{tab:alphabeta}.
\begin{table}
\centering
\begin{tabular}{c|ccccccccc}
$\alpha\backslash\beta$&-4&-3&-2&-1&0&1&2&3&4\\
\hline
4&x&x&x&x&{\color{purple}$\checkmark$}&{\color{blue}$\checkmark$}&{\color{blue}$\checkmark$}&{\color{blue}$\checkmark$}&{\color{cyan}$\checkmark$}\\
3&x&x&x&{\color{red}$\checkmark$}&$\checkmark$&$\checkmark$&$\checkmark$&$\checkmark$&{\color{yellow}$\checkmark$}\\
2&x&x&{\color{red}$\checkmark$}&$\checkmark$&$\checkmark$&$\checkmark$&$\checkmark$&$\checkmark$&{\color{yellow}$\checkmark$}\\
1&x&{\color{red}$\checkmark$}&$\checkmark$&$\checkmark$&$\checkmark$&$\checkmark$&$\checkmark$&$\checkmark$&{\color{yellow}$\checkmark$}\\
0&{\color{red}$\checkmark$}&$\checkmark$&$\checkmark$&$\checkmark$&$\checkmark$&$\checkmark$&$\checkmark$&$\checkmark$&{\color{green}$\checkmark$}\\
-1&x&$\checkmark$&$\checkmark$&$\checkmark$&$\checkmark$&$\checkmark$&$\checkmark$&{\color{Orange}$\checkmark$}&x\\
-2&x&x&$\checkmark$&$\checkmark$&$\checkmark$&$\checkmark$&{\color{Orange}$\checkmark$}&x&x\\ 
-3&x&x&x&$\checkmark$&$\checkmark$&{\color{Orange}$\checkmark$}&x&x&x\\
-4&x&x&x&x&{\color{Orange}$\checkmark$}&x&x&x&x\\
\end{tabular}
\caption{Allowed divisor classes $\alpha$ and $\beta$ for the fibre coordinates $\text{w}$ and $\text{v}$ for the choice of $k^1 = 4$, e.g.\ if the base is a $\mathbb P^3$ or a blow-up thereof. Coloured checkmarks indicate that one or two sections are sections of the trivial line bundle; orange indicates {\color{Orange}$b_{0}$}, red {\color{red}$b_{2}$}, blue {\color{blue}$c_{1}$}, yellow {\color{yellow}$c_{2}$}, purple {\color{purple}$b_{2}$} and {\color{purple}$c_{1}$}, cyan {\color{cyan}$c_{1}$} and {\color{cyan}$c_{2}$} and green indicates {\color{green}$b_{0}$} and {\color{green}$c_{2}$}. }\label{tab:alphabeta}
\end{table}

At the boundary of the allowed region some of the sections become constant and we are not in the generic case anymore. For instance, at the dual edge (for all $H_i$) to $(-1,1)^T$ and $(-1,0)^T$ $b_2$ and $c_1$, respectively, are sections of the trivial line bundle. At these points the section which we will define as the zero section will become holomorphic. For fibrations over $\mathbb{P}^3$ this was also observed by \cite{Cvetic:2013uta}. For the other sections the points dual to the edges are $(-1,0)^T$, $(0,-1)^T$ and $(1,-1)^T$, $(0,-1)^T$.


The appearance of codimension-two singularities in \eqref{eq:hyper1} at
\begin{equation}\label{eq:co-dim-2_singularities}
\tu=\tv=c_1=b_2=0 \qquad \textmd{and} \qquad \tu=\tw=c_2=b_0=0
\end{equation}
for which  the cubic $P_T$ given in \eqref{eq:hyper1} and $d P_T$ vanish simultaneously necessitates a resolution process. 
The first singularity in  \eqref{eq:co-dim-2_singularities} can be resolved by a blow-up in the fibre ambient space $\mathbb P^2$. This introduces the new homogeneous coordinate $s_0$ via the blow-up of the point $\tu=\tv=0$,
\begin{eqnarray} \label{blowjob2}
\tu \rightarrow \tu \, s_0, \qquad \tv \rightarrow \tv \, s_0
\end{eqnarray}
with the equivalence relation
\begin{eqnarray} \label{scaling2}
(\tu, \tw, \tv, s_0) \simeq (\lambda_0^{-1} \tu,   \tw, \lambda_0^{-1} \tv, \lambda_0 s_0).
\end{eqnarray}
The second singularity is resolved similarly by blowing up the point $\tu = \tw =0$ in ${\rm Bl}_1 \mathbb P^2$,
\begin{eqnarray} \label{blowjob1}
\tu \rightarrow \tu \, s_1, \qquad \tw \rightarrow \tw \, s_1,
\end{eqnarray}
together with an extra scaling relation 
\begin{eqnarray} \label{scaling1}
(\tu, \tw, \tv, s_0, s_1) \simeq (\lambda_1^{-1} \tu,  \lambda_1^{-1} \tw, \tv, s_0, \lambda_1 s_1).
\end{eqnarray}

The proper transform of the equation \eqref{eq:hyper1}---i.e.\ \eqref{eq:hyper1}  with \eqref{blowjob2} and \eqref{blowjob1} plugged in and factorising of overall powers of $s_0$ and $s_1$---takes the form
\begin{equation}
\begin{split}
P_{T^2}:= & \tv\,  \tw (c_1\,  \tw\,   s_1 +     c_2\, \tv\,s_0)  +        \tu\, (b_0\, \tv^2\,s_0^2 + b_1\, \tv\, \tw\,s_0\,s_1 +  b_2\, \tw^2\,s_1^2) + \\
&   \tu^2 (d_0\, \tv \,s_0^2\,s_1 + d_1\, \tw \,s_0\,s_1^2 + d_2\, \tu\,s_0^2\,s_1^2)=0
\end{split} \label{eq:hyper1-res}
\end{equation}
and identifies the smooth elliptic fibration as a ${\rm Bl}_2 \mathbb P^2[3]$-fibration. We will henceforth denote the smooth elliptic fibration as $\hat Y_4$ and its ambient space as $\hat X_5$.

The equivalence relations of the five homogeneous coordinates $\tu$, $\tv$, $\tw$, $s_0$, $s_1$ of the blown-up ambient space can be expressed as in  Table~\ref{tab:scaling_hs} by taking suitable linear combinations of \eqref{scaling1},  \eqref{scaling2} and the original scaling of $\mathbb P^2$. 
\begin{table}
\centering
\begin{tabular}{c||ccccc}
 & \tu & \tv & \tw & $s_{0}$ & $s_{1}$\\
\hline
\hline
$\alpha$ & $\cdot$ & $\cdot$ & 1 & $\cdot$ & $\cdot$\\
$\beta$ & $\cdot$ &1 &  $\cdot$ & $\cdot$ & $\cdot$\\
\hline
U & 1 & 1 & 1 & $\cdot$ & $\cdot$ \\
$S_{0}$ & $\cdot$ & $\cdot$ & 1 & 1 & $\cdot$ \\
$S_{1}$ & $\cdot$ & 1 & $\cdot$ & $\cdot$ & 1 \\
\end{tabular}
\caption{Divisor classes and coordinates of the blown-up fibre ambient space of $\hat X_{5}$;}\label{tab:scaling_hs}
\end{table}
The blow-ups \eqref{blowjob2} and \eqref{blowjob1} also change the Stanley-Reisner ideal, i.e.\ the set of coordinates which are not allowed to vanish simultaneously. Prior to resolution only the simultaneous vanishing of all three homogeneous coordinates was forbidden, but now the Stanley-Reisner ideal takes the form
\begin{eqnarray}
 \{ \T w \, s_0, \T w \, \T u,  \T v \, s_1,  s_0 \, s_1,\T v \, \T u\}. \label{SR1}
\end{eqnarray}

The next step is to analyse  the points $\T{Sec}_0$, $\T{Sec}_1$ and $\T{Sec}_2$ as we blow up $\mathbb P^2$ and fibre it over $B_3$. The blow-ups replace the points $\T{Sec}_0$, $\T{Sec}_1$ by two $\mathbb P^1$s, $\{s_0=0\}$ and $\{s_1=0\}$. Both of these rational curves intersect $P_{T^2}=0$ in one point because the edges dual to $s_0$ and $s_1$ are both of length one~\cite{Kreuzer:1997zg}, cf.\ Figure~\ref{fig:polygon5}. The two intersection points are given by
\begin{equation}
 \widetilde{\T{Sec}}_0:\quad[-c_1\,:\,b_2\,:\,1\,:\,0\,:\,1] \qquad\textmd{and}\qquad \widetilde{\T{Sec}}_1:\quad [-c_2\,:\,1\,:\,b_0\,:\,1\,:\,0].
\end{equation}
For $\T{Sec}_2$ the resolutions do not change much and the point is still the intersection of the rational curve $\{\tu=0\}$ with $P_{T^2}=0$,
\begin{equation}
\widetilde{\T{Sec}}_2:\quad[0\,:\,1\,:\,1\,:\,-c_1\,:\,c_2]\,.
\end{equation}

When we consider now the fibration of ${\rm Bl}_2 \mathbb P^2[3]$ over a base $B_3$, these three points are promoted to sections of the fibration. However, due to the Stanley-Reisner ideal~\eqref{SR1} $\widetilde{\T{Sec}}_0$, $\widetilde{\T{Sec}}_1$ and $\widetilde{\T{Sec}}_2$ are ill-defined over the three curves 
\begin{equation}\label{eq:problematic-curves}
 c_1=b_2=0\,,\qquad c_2=b_0=0\quad\textmd{and}\quad c_1=c_2=0\,
\end{equation}
and, therefore, none of them defines a well-defined section on $\hat Y_4$ (the ${\rm Bl}_2 \mathbb P^2[3]$ fibration over $B_3$). To circumvent this problem, we take the divisors
\begin{equation}
 {S}_0:\,\, \{s_0 = 0\},\,\qquad {S}_1:\,\, \{s_1 = 0\},\quad\textmd{and}\quad {S}_2:\,\,\{\tu = 0\}
\end{equation}
as our sections. These agree with $\widetilde{\T{Sec}}_i$ away from the curves \eqref{eq:problematic-curves} but give a $\mathbb P^1$ instead of a point over these curves because
\[
 \left.P_{T^2}\right|_{s_0= c_1=b_2=0}=\left.P_{T^2}\right|_{s_1= c_2=b_0=0} = \left.P_{T^2}\right|_{\tu= c_1=c_2=0}\equiv 0\,.
\]
In this sense these sections are \emph{rational} sections and not \emph{holomorphic} ones  - unlike the divisor $Z$ defining the holomorphic zero-section in the Weierstra\ss{} model. 
Note however that as we choose $\alpha$ and $\beta$ such that one of the sections $c_1$, $c_2$, $b_0$, $b_2$ becomes constant, i.e.\ going to the boundary of the allowed values, one or two of the rational sections become holomorphic. As can be seen from the divisor classes of ${\rm Bl}_2 \mathbb P^2$ the three sections are independent and, therefore, generate a Mordell-Weil group of rank two.

For later purposes we need to compute the mutual intersection numbers of the sections. Our general treatment of the fibration allows us to express all intersection numbers involving any number of $S_i$ in terms of intersections entirely on $B_3$, which can then be conveniently evaluated for specific choices of $B_3$. We now list those intersection numbers which will be needed explicitly in the subsequent computations.
First, being sections, the $S_i$ satisfy
\begin{align}
\begin{split}
\int_{\hat Y_{4}}S_{k} \wedge  \pi^* \omega_6 = \int_{B_3} \omega_6, \qquad k=0,1,2,  \label{IntSec1a}
\end{split}
\end{align}
where $\omega_6$ is the volume form  of $B_3$.
Furthermore, we will need the following intersection numbers between the $S_i$ with any four-form $\omega_4 \in H^4(B_3)$,
\begin{eqnarray}
&& \int_{\hat Y_4} S_0 \wedge S_0 \wedge \pi^*\omega_4 = - \int_{B_3} \bar{\cal K} \wedge \omega_4, \qquad  \int_{\hat Y_4} S_0 \wedge S_1 \wedge \pi^*\omega_4 = 0,     \label{IntSec1} \\
&&  \int_{\hat Y_4} S_0 \wedge S_2 \wedge \pi^*\omega_4 =  \int_{B_3} [c_1] \wedge \omega_4, \qquad   \int_{\hat Y_4} S_1 \wedge S_2 \wedge \pi^*\omega_4 =  \int_{B_3} [c_2] \wedge \omega_4 \label{IntSec2},
\end{eqnarray}
and, for any  two-form $\omega_2 \in H^2(B_3)$, 
\begin{eqnarray} 
&& \int_{\hat Y_4} S_0 \wedge S_0 \wedge S_2 \wedge  \pi^*\omega_2 = \int_{B_3} (\beta - \alpha) \wedge c_1 \wedge \omega_2, \label{eq:intnuma}  \\
&& \int_{\hat Y_4} S_0^3 \wedge  \pi^*\omega_2 = \int_{B_3} \omega_2 \wedge \Big( (\bar{\cal K} -  \alpha) \wedge (2 \alpha - \beta) + \alpha \wedge \alpha \Big). \label{eq:intnumB}
\end{eqnarray}
The first equation in \eqref{IntSec1} follows from the linear relations and the SR-ideal of the divisors, cf.\ Table~\ref{tab:scaling_hs} and equation \eqref{SR1}, from which one can show that on the ambient space $\hat X_5$
\begin{equation}
[P_{T^2}] \, (S_0+\bar\cK)\,S_0=(-\alpha+\beta+\bar\cK) \, (-\alpha+\bar\cK)\,S_0\,.
\end{equation}
Together with the fact that there are no basis eight-forms this gives the first part of \eqref{IntSec1}. The second equation of \eqref{IntSec1} is a consequence of $s_0\, s_1$ being in the SR-ideal. To compute the intersection numbers in \eqref{IntSec2}, we rewrite them in the ambient five-fold $\hat X_{5}$, e.g.\ in the first case by evaluating  $P_{T^2}$ for $s_{0}=s_2=0$ as
\begin{align}
\begin{split}
[P_{T^2}] \, S_2 \,S_0 = [ c_{1}\tv \tw^{2} s_{1}]  \, \tU  \, S_0 = [c_1] \, \tU \, S_0
\end{split}
\end{align}
because $\tu\, \tv$, $\tu\, \tw$ and $s_{0}\,s_{1}$ are in the SR-ideal. Wedging this with a base four-form gives the result stated.
The same logic leads to the triple intersection numbers \eqref{eq:intnuma} and eventually allows one to deduce all possible intersections if needed.

\subsection{Holomorphic zero-section, base embedding and \texorpdfstring{$U(1)$}{U(1)} generators} \label{sec:holzerosec}

The behaviour of rational sections in fibrations with non-trivial Mordell-Weil group, i.e. the fact that the generators of the Mordell-Weil group wrap entire fibre components over certain curves on $B_3$, plays an important role in F-theory compactifications with $U(1)$ gauge groups, as has been stressed in the recent F-theory literature \cite{Morrison:2012ei,Mayrhofer:2012zy,Braun:2013yti,Borchmann:2013jwa,Cvetic:2013nia,Braun:2013nqa,Cvetic:2013uta}.
The  ${\rm Bl}_2 \mathbb P^2$-fibration under consideration here and in \cite{Borchmann:2013jwa} (see also \cite{Cvetic:2013nia,Cvetic:2013uta}), however, appears at first sight to have not even a holomorphic zero-section because all three of $S_0, S_1, S_2$ degenerate over curves. The appearance of non-holomorphic zero-sections has been pointed out recently in \cite{Braun:2013nqa,Grimm:2013oga,Cvetic:2013nia,Cvetic:2013uta}.
In the sequel, we show that the non-holomorphicity of the zero-section in the ${\rm Bl}_2 \mathbb P^2$-fibration under consideration is merely an artefact of the special resolution procedure and prove that an alternative resolution can be chosen which does admit a holomorphic zero-section. This in particular allows for a holomorphic embedding of the base $B_3$ into the fibration $\hat Y_4$, as will be crucial for our construction of $U(1)$ generators and gauge fluxes.

That we have found no holomorphic section after the resolution is due to the Stanley-Reisner ideal after resolution and, therefore, a result of how  the small resolution was performed in detail. It turns out, however, that a holomorphic zero-section can be defined  if  the first resolution is performed via the alternative procedure applied in \cite{Braun:2011zm}. We write $P_{T^2}$---with only the second blow-up implemented---as
\begin{equation}\label{eq:hyper-equ-rewritten}
 \tv\,P_1=\tu\,P_2
\end{equation}
with
\begin{equation}
\begin{aligned}
&P_1= -\tw (c_1\,  \tw\,   s_1 +     c_2\, \tv) \quad\textmd{and}\quad \\
&P_2=(b_0\, \tv^2 + b_1\, \tv\, \tw\,s_1 +  b_2\, \tw^2\,s_1^2) + \tu (d_0\, \tv \,s_1 + d_1\, \tw \,s_1^2 + d_2\, \tu\,s_1^2)\,.
\end{aligned}
\end{equation}
To resolve the conifold singularity at $\tu=\tv=c_1=b_2=0$ we paste in a $\mathbb P^1$ by defining the fibration as the complete intersection \begin{equation}\label{eq:hyper-cicy}
 \begin{aligned}
  \tv\,\lambda_1&=\lambda_2\,P_2,\\
  \tu\,\lambda_1&=\lambda_2\,P_1,\,
 \end{aligned}
\end{equation}
where $\lambda_1$ and $\lambda_2$ are the homogeneous coordinates of the $\mathbb P^1$. Hence, we obtain a complete intersection in an ambient space $\hat X_6$ of one dimension higher than in the hypersurface case, cf.\ Table~\ref{tab:scalings_ci}.
\begin{table}
\centering
\begin{tabular}{c||cccccc}
 & \tu & \tv & \tw  & $s_{1}$   & $\lambda_{1}$  & $\lambda_{2}$  \\
\hline
\hline
$\bar{\mathcal{K}}$ & $\cdot$ & $\cdot$ & $\cdot$ & $\cdot$ & 1 & $\cdot$\\
$\alpha$ & $\cdot$ & $\cdot$ & 1 & $\cdot$ & 1 & $\cdot$\\
$\beta$ & $\cdot$ &1 &  $\cdot$ & $\cdot$ & $\cdot$ & $\cdot$\\
\hline
U & 1 & 1 & 1 & $\cdot$ &   1 & $\cdot$ \\
$S_{1}$ & $\cdot$ & 1 & $\cdot$  & 1 & 1 & $\cdot$ \\
$\Lambda_{2}$ & $\cdot$ & $\cdot$ & $\cdot$ & $\cdot$ & 1 & 1  \\
\end{tabular}
\caption{Divisor classes and coordinates of the blown-up fibre ambient space of $\hat X_{6}$.}\label{tab:scalings_ci}
\end{table}
The fibre part of the Stanley-Reisner ideal is
\begin{eqnarray}
\{ \tu \,  \tw, \tv \, s_1, \lambda_1\,  \lambda_2 \}.
\end{eqnarray}
The advantage of this more involved but equivalent resolution is that $\T{Sec}_0$, which becomes $[0\,:\,0\,:\, 1 :\, 1\,:\,1\,:\,0]$ after the resolution \eqref{eq:hyper-cicy}, is a holomorphic section of the fibration. The  divisor corresponding to this section is 
\begin{eqnarray}
\Lambda_2:\,\,\{\lambda_2=0\}.
\end{eqnarray}
 The equivalent to $S_0$ of the hypersurface case is $\tu=\tv=0$. The other two sections remain, as expected, rational. 

We will not use the complete intersection description in the sequel, but it serves as a proof of principle that there does exist a holomorphic  section for our fibration even though, in the hypersurface description, this holomorphic section cannot be realised via divisors pulled back from the ambient space.  The reason why we stick to the description as a hypersurface is that the description of matter surfaces and fluxes is less involved than for of a complete intersection. Nevertheless, we need at least an object which behaves for all intersections like $\Lambda_2$ such that we have a `semi-embedding' of $B_3$ into $\hat Y_4$ to define fluxes which do not break Poincar\'e invariance of the four-dimensional spacetime. Therefore we define for the hypersurface at least a point set which has the same properties as $\Lambda_2$ in the complete intersection.

We define such a substitute for a holomorphic section by considering the following point set on $\hat Y_4$ as a complete intersection in the ambient space $\hat X_5$ of $\hat Y_4$,
\begin{eqnarray} \label{H-sec}
{\rm H} = S_0 \cap \{P_{T^2}=0\} - S_0 \cap \{b_2=0\} \cap \{c_1=0\} + S_0 \cap \{\tv=0\} \cap \{b_2=0\} \cap \{c_1=0\}.
\end{eqnarray}
The second term subtracts from $S_0 \subset \hat Y_4$ the degenerate locus, given by a $\mathbb P^1$-fibre over $b_2 = c_1=0$, and the third adds a point in the fibre over $ b_2 = c_1=0$. Thus ${\rm H}$ coincides with $S_0$ everywhere in $\hat Y_4$  except  over the degeneration curve $ b_2 = c_1=0$, where ${\rm H}$ is given by a single point. 

Our proposal for dealing with elliptic fibrations without a holomorphic zero-section that can be pulled back from the ambient space is to define the `embedding' of $B_3$ into $\hat Y_4$ via the object ${\rm H}$.
In particular, we propose to define the generators of  $U(1)$ symmetries and the $G_4$ gauge fluxes by demanding the usual transversality condition \eqref{cond1} with respect to ${\rm H}$ instead of the holomorphic section. This will also provide us with a very clear geometric interpretation of the allowed fluxes. 

With this understood we now derive the form of the two $U(1)$ generators $\tw_i$ presented already in \cite{Borchmann:2013jwa}.
Abelian gauge potentials $A_i$ arise via M/F-theory duality by dimensional reduction of the M-theory 3-form field $C_3$ as \cite{Morrison:1996na,Morrison:1996pp}
\begin{eqnarray}
C_3 = A_i \wedge \tw_i,
\end{eqnarray}
where the element $\tw_i \in H^{1,1}(\hat Y_4)$ must satisfy the transversality conditions
\begin{align} \label{cond1}
\begin{split}
\int_{\hat Y_{4}}\text{w}_{i}\wedge \pi^* \omega_6=0,\quad \int_{\hat Y_{4}}\text{w}_{i}\wedge {\rm H } \wedge \pi^* \omega_4=0 .
\end{split}
\end{align}
Here  $\omega_6$ and  $\omega_4$ denote forms of the indicated rank on $B_3$. As discussed above this condition is the well-known transversality condition from the Weierstra\ss{}  model, but with the holomorphic zero-section $Z$ of the Weierstra\ss{}  model replaced by ${\rm H }$, which serves as the substitute for our holomorphic section.
This construction of $U(1)$ generators is known, in the mathematics literature, as the Shioda map.

In order to meet the first transversality condition \eqref{cond1}, we conclude from \eqref{IntSec1a} that  
\begin{align}
\begin{split}
\int_{\hat Y_{4}}(S_0 - S_1)\wedge \pi^* \omega_6   =\int_{\hat Y_{4}}(S_{0}-S_{2}) \wedge \pi^* \omega_6 =0.
\end{split}
\end{align}
To tackle the second condition note that ${\rm H }$, as a point set, differs from $S_0$ only in codimension-two. Therefore, we obtain
\begin{eqnarray}
\int_{\hat Y_4} {\rm H} \wedge S_i \wedge \pi^*\omega_4 = \int_{\hat Y_4} S_0 \wedge S_i \wedge \pi^*\omega_4, \qquad \quad i=0,1,2.
\end{eqnarray}
With the help of \eqref{IntSec2} we conclude that \cite{Borchmann:2013jwa}
\begin{align}
\begin{split} \label{twi-1}
\text{w}_{1} = 5 (S_{1} - S_{0} - \mathcal{\bar K}), \quad \text{w}_{2} = 5 (S_2 - S_{0} - \mathcal{\bar K} - [c_{1}])
\end{split}
\end{align}
satisfy \eqref{cond1}, where we have chosen the overall normalisation such as to arrive at integral charges when generalising the construction to models with additional $SU(5)$ gauge symmetry.

\subsection{Massless charged singlets} \label{eq:U1singlets}

We now discuss the appearance of massless matter states charged under the gauge group $U(1)_1 \times U(1)_2$. 
Such states are in 1-1 correspondence with a factorisation of the fibre into two $\mathbb P^1$s over certain curves on the base $B_3$. In fact, massless charged states arise from M2-branes wrapping one of the two fibre components over such curves as these become massless in the F-theory limit of vanishing fibre volume. 
In order to identify the curves over which the fibre splits, i.e. the loci $C_i \subset B_3$ such that $P_{T^2}|_{C_i}$ factorises, it turns out more convenient to start not from the hypersurface \eqref{eq:hyper1-res}, but instead to analyse the birationally equivalent Weierstra\ss{} model prior to resolution.
This motivates us to rewrite \eqref{eq:hyper1} as a Weierstra\ss{}  model 
\begin{eqnarray} \label{Weier2}
y^2 = x^3 + f x z^4 + g z^6
\end{eqnarray}
by working out the birational map which relates the fibre coordinates $[\tu\,:\,\tv\,:\,\tw]$ to the Weierstra\ss{} coordinates $[x\,:\,y\,:\,z]$. A general algorithm for obtaining this birational map is given by the Nagell transformation~\cite{nagell1929proprietes,Connell:1999,Cassels:1991}.
We find that the transformation 
\begin{equation}\label{eq:map_P2_Weierstrass}
\begin{aligned}
 x & =  -4\,\tw^2\,(b_2\,\tu + c_1\,\tv)\left(b_0\,b_2^2\,\tu + b_2^2\,c_2\,\tw - b_2\,c_1\,(d_0\, \tu + b_1\,\tw) + c_1^2\,(d_2\,\tu + d_1\,\tw)\right)+\\
       & +\tfrac13\,\tw^2\,(b_2\,\tu + c_1\,\tv)^2\left(b_1^2 + 8\, b_0\, b_2 - 4\, (c_1\, d_0 + c_2\, d_1)\right),\\
 y & =  -4 \tw^2 (b_2\,\tu + c_1\,\tv) (2\,c_1\,(b_0\,b_2^2 + c_1\,(-b_2\,d_0 + c_1\,d_2))\, \tu\,(d_2\,\tu^2 + \tv\,(d_0\,\tu + b_0\,\tv)) +\\
      &+ ((-b_2^3\,c_2\,d_0 - 2\,b_2\,c_1^2\,d_0\,d_1 +  4\,c_1^3\,d_1\,d_2 + b_1\,b_2 c_1\,(b_2\,d_0 - 3\,c_1\,d_2) +\\
      &+  b_2^2\,c_1\,(b_0\,d_1 + 3\,c_2\,d_2))\,\tu^2 + ((b_0\,b_2 -  c_1\,d_0) (3\,b_1\,b_2\,c_1 - 2 (b_2^2\,c_2 + c_1^2\,d_1)) + b_1\,c_1^3\,d_2)\,\tu\,\tv + \\
      & +  c_1\,(b_0\,(-b_1\,b_2\,c_1 + 2\,b_2^2\,c_2 + c_1^2\,d_1) +   c_1\,c_2\,(-b_2\,d_0 + c_1\,d_2))\,\tv^2)\,\tw +\\
      & + (-b_1\,b_2\,c_1 + b_2^2\,c_2 +  c_1^2\,d_1) (-b_1\,b_2\,\tu + 2\,c_1\,d_1\,\tu + b_1\,c_1\,\tv - 2\,b_2\,c_2\,\tv)\,\tw^2), \\
z & =  \tw\,(b_2\,\tu + c_1\,\tv)
\end{aligned}
\end{equation}
maps our 3-section fibration \eqref{eq:hyper1} to a 
Weierstra\ss{} model with $f$ and $g$ given by 
\begin{equation} \label{sec-map1}
 f=-\tfrac13\T d^2 + \T c\, \T e\qquad\textmd{and}\qquad g=-  f\,\left(\tfrac13\T d\right) -\left(\tfrac{1}{3}\T d\right)^3  + \T c^2\, \T k,
\end{equation}
where
\begin{equation}
 \begin{split} \label{sec-map2}
  \T d &=b_1^2 + 8\,b_0\,b_2 - 4\,c_1\,d_0 - 4\,c_2\,d_1,\\
  \T c &=-\frac{4}{c_1}(b_0\,b_2^2 - b_2\,c_1\,d_0 + c_1^2\,d_2),\\
  \T e &=\frac{2  c_1 \left( b_0 \left( b_1  c_1  d_1-b_1^2 b_2+2  b_2  c_1  d_0+2  b_2  c_2
    d_1-2  c_1^2  d_2\right)\right)}{ b_0  b_2^2+ c_1 ( c_1  d_2- b_2  d_0)}+\\&\qquad\qquad+\frac{2  c_1 \left(-2  b_0^2  b_2^2+ c_2 ( b_1  b_2  d_0+ b_1
    c_1  d_2-2  b_2  c_2  d_2-2  c_1  d_0
    d_1)\right)}{ b_0  b_2^2+ c_1 ( c_1  d_2- b_2  d_0)},\\
  \T k &=\frac{ c_1^2 ( b_0  b_1  b_2- b_0  c_1  d_1- b_2  c_2
    d_0+ c_1  c_2  d_2)^2}{\left( b_0  b_2^2+ c_1 ( c_1
    d_2- b_2 d_0)\right)^2}.
 \end{split}
\end{equation}
The results of our application of Nagell's algorithm,  eq.\ \eqref{sec-map1} and \eqref{sec-map2},  had already been presented in \cite{Borchmann:2013jwa}. Here, for completeness, we also include the corresponding expressions \eqref{eq:map_P2_Weierstrass} for the fibre coordinates. Note that a similar analysis appears in the independent work \cite{Cvetic:2013nia,Cvetic:2013uta}.

Under the map \eqref{eq:map_P2_Weierstrass} only the last section Sec$_2$ \eqref{eq:sections_2u1s_cubic} does not map to the exceptional set of $\mathbb P_{2,3,1}$ on the Weierstra\ss{} side. Via the Nagell transformation the zero section Sec$_0$ of the cubic goes to the zero section $[\lambda^2\,:\,\lambda^3\,:\,0]$. To find the counterpart to Sec$_1$ we note that $\T c\,\T k=\T p^2$ is a complete square. Therefore, $[ \tfrac{1}{3}\T d\,:\, \T p\,:\,1]$ is the last section we were missing.

We can now search for the loci where the Weierstra\ss{} equation becomes singular. This happens at the loci
\begin{equation}
 B_1=3\,A_1^2+f =0\label{eq:conifold_locus_1}
\end{equation}
and
\begin{equation}
 B_2=3\,A_2^2+f  =  0,\label{eq:conifold_locus_2} 
\end{equation}
where $A_i$ and $B_i$ are the affine $x$ and $y$ coordinates, respectively, of the sections $\T{Sec}_1$ and $\T{Sec}_2$ in the Weierstra\ss{} model.
The solutions to \eqref{eq:conifold_locus_1} are given by
\begin{equation}\label{eq:conifold_locus_1a_explicit} 
\begin{split}
d_0\, c_2^2 & = (-b_0^2 \,c_1 + b_0\, b_1\, c_2)\,,\\
d_1\, b_0\, c_2 & = (b_0^2\, b_2 + c_2^2\, d_2)
\end{split}
\end{equation}
and
\begin{equation}\label{eq:conifold_locus_1b_explicit} 
\begin{split}
d_0\, b_2\, c_1 & = (b_0\, b_2^2 + c_1^2\, d_2)\,,\\
d_1\, c_1^2 & = (b_1\, b_2\, c_1 - b_2^2\, c_2)\,.
\end{split}
\end{equation}
For \eqref{eq:conifold_locus_2} we obtain in addition to \eqref{eq:conifold_locus_1b_explicit} the solution
\begin{equation}\label{eq:conifold_locus_2_explicit} 
\begin{split}
d_0\, c_1^3\, c_2^2 & =  (-b_0^2\, c_1^4 + b_0\, b_1\, c_1^3\, c_2 + 
  c_2^3 (-b_1\, b_2\, c_1 + b_2^2\, c_2 + c_1^2 d_1))\,,\\
d_2\, c_1^4\, c_2^2 & = -(b_0\, c_1^2 + c_2\, (-b_1\, c_1 + b_2\, c_2)) (b_0\, b_2\, c_1^2 + c_2 (-b_1\, b_2\, c_1 + b_2^2\, c_2 + c_1^2\, d_1))\,.
\end{split}
\end{equation}
Like in the $\textmd{Bl}_1\mathbb P_{1,1,2}[4]$-case~\cite{Morrison:2012ei}, over these loci \eqref{eq:hyper1} factorises in different ways. In this case there are six different loci, which we denote, as in  \cite{Borchmann:2013jwa}, as
\begin{enumerate}
\item $C_{\mathbf 1^{(1)}}$: $b_0=c_2=0$; \label{b0c2}
\item $C_{\mathbf 1^{(2)}}$: \eqref{eq:conifold_locus_1a_explicit} with $(b_0, c_2) \neq (0,0)$;
\item $C_{\mathbf 1^{(3)}}$: $b_2=c_1=0$; \label{b2c1}
\item $C_{\mathbf 1^{(4)}}$: \eqref{eq:conifold_locus_1b_explicit} with $(b_2, c_1) \neq (0,0)$;
\item $C_{\mathbf 1^{(5)}}$:   $c_1=c_2=0$;
\item $C_{\mathbf 1^{(6)}}$:  \eqref{eq:conifold_locus_2_explicit} with $(c_1,c_2) \neq (0,0)$, %
$(b_0, c_2) \neq (0,0)$ and $(b_2, c_1) \neq (0,0)$. \label{C16a} %
\end{enumerate}

\begin{figure}[h!]
\begin{center}
\def\svgwidth{0.8\textwidth}
\hspace*{3.5cm} %
 \executeiffilenewer{singlets_polygon5.svg}{singlets_polygon5.pdf}%
 {inkscape -z -D --file=singlets_polygon5.svg %
  --export-pdf=singlets_polygon5.pdf --export-latex}%
   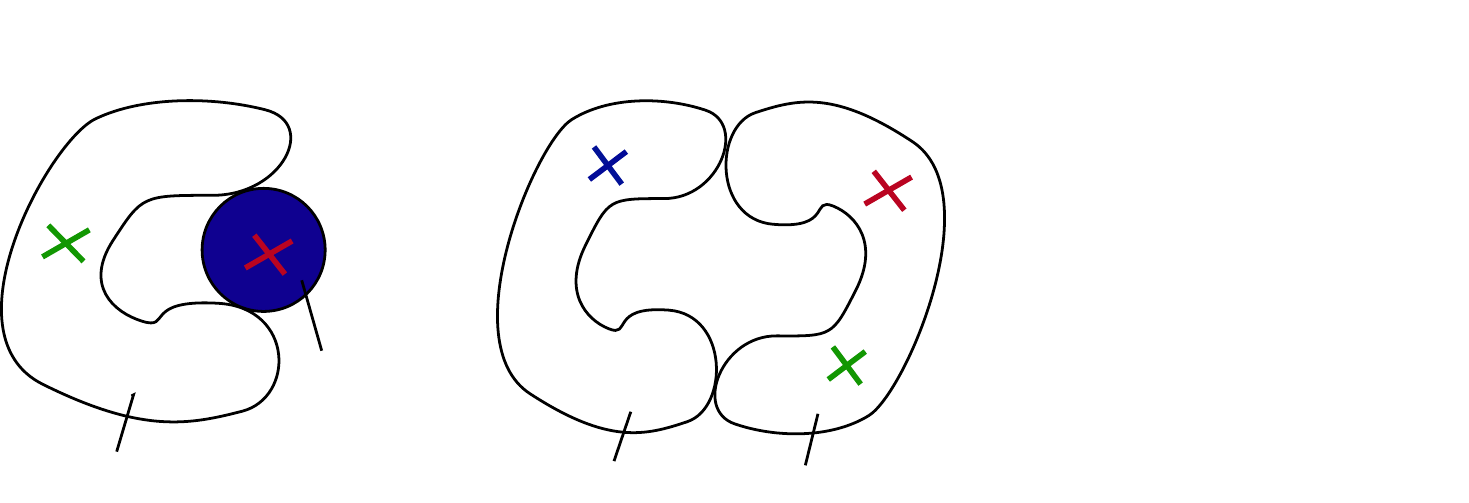%
  
 \end{center}
 \begin{center}
 \def\svgwidth{0.8\textwidth}
 \hspace*{3.5cm} %
 \executeiffilenewer{singlets_polygon5c3c4.svg}{singlets_polygon5c3c4.pdf}%
 {inkscape -z -D --file=singlets_polygon5c3c4.svg %
  --export-pdf=singlets_polygon5c3c4.pdf --export-latex}%
   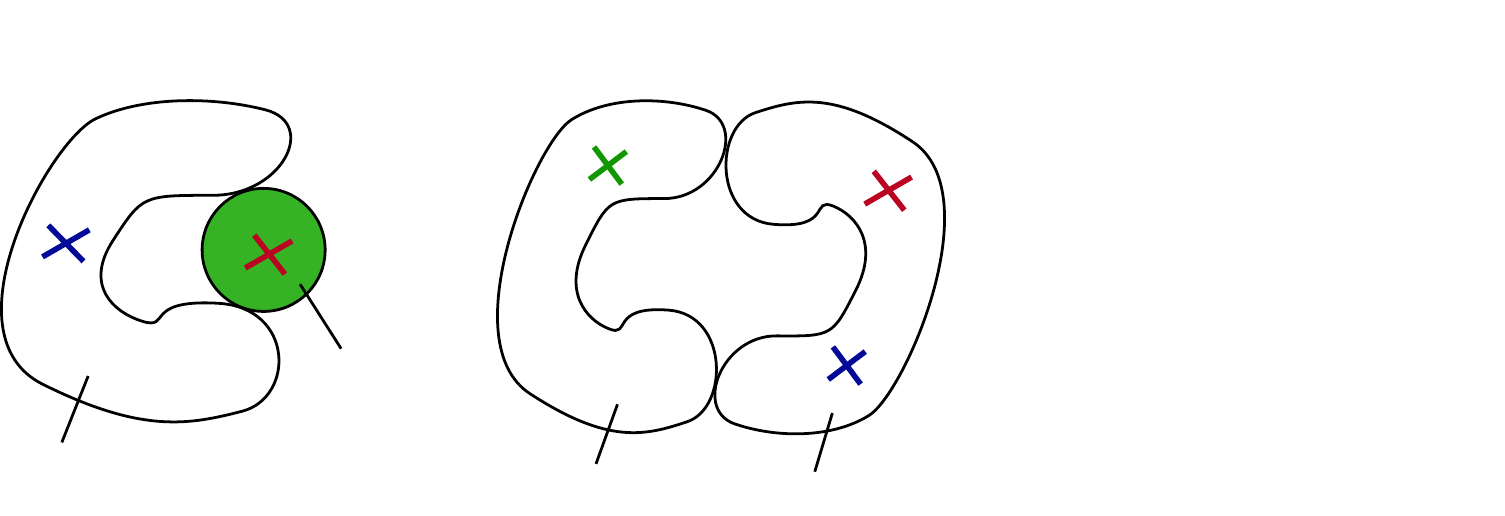%

 \end{center}
  \begin{center}
 \def\svgwidth{0.8\textwidth}
 \hspace*{3.5cm}%
 \executeiffilenewer{singlets_polygon5c5c6.svg}{singlets_polygon5c5c6.pdf}%
 {inkscape -z -D --file=singlets_polygon5c5c6.svg %
  --export-pdf=singlets_polygon5c5c6.pdf --export-latex}%
   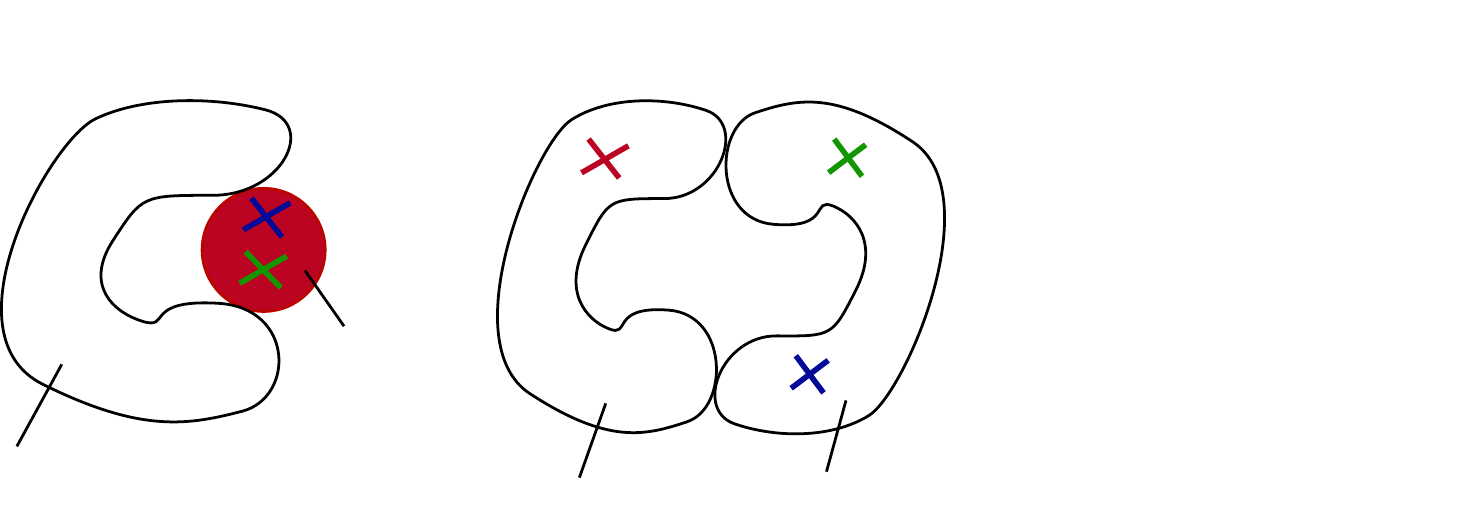%

      \caption{Topology of the fibre over the six singlet curves $C_{\mathbf 1^{(i)}}$, $i=1,\ldots,6$. Green denotes the zero section $S_0$, blue denotes $S_1$ and red corresponds to $S_2$.}\label{fig:singlets_polygon5}
\end{center}
\end{figure}

The factorisation of the hypersurface \eqref{eq:hyper1}  prior to resolution implies that also the resolved fibration \eqref{eq:hyper1-res} degenerates in the following way:
Over each of the six loci $C_{\bf 1^{(i)}}$  on $B_3$ the fibre of \eqref{eq:hyper1-res}  splits into two ${\mathbb P}^1$-components $\mathbb P^1_{A_i}$ and $\mathbb P^1_{B_i}$ intersecting like the affine Dynkin diagram of $A_1$. This can be worked out explicitly by plugging the respective defining equations for $C_{\bf 1^{(i)}}$ into \eqref{eq:hyper1-res} and observing a factorisation of the hypersurface equation. 
In the fibre over $C_{\mathbf 1^{(1)}}, C_{\mathbf 1^{(3)}}, C_{\mathbf 1^{(5)}}$, one of the factors is given respectively by the coordinates $s_1$, $s_0$ and $\tu$.
This is precisely the statement that the corresponding section $S_0, S_1$ or $S_2$ wraps that fibre part, as discussed in detail around \eqref{eq:problematic-curves}. 
The sections which do not factor out intersect one of the components in a point. 
For example, from $P_{T^2}|_{b_0=c_2=0} = s_1 \, p_1$  with $p_1$ a complicated polynomial one deduces that the fibre over $C_{\bf 1^{(1)}}$ splits into the two components
\begin{eqnarray}
&& \mathbb P^1_{A_1}= \{s_1 = 0\} \, \cap \, \{b_0=0\} \, \cap \, \{c_2=0\} \, \cap \,  D \subset \hat X_5, \\
&&  \mathbb P^1_{B_1}= \{ p_1 =0\} \,  \cap\,  \{b_0=0\} \,  \cap \,  \{c_2=0\} \,  \cap D \subset \hat X_5,
\end{eqnarray}
where we have added  an arbitrary divisor $D$ on $B_3$ that intersects the curve $\{c_2 = 0\} \cap \{b_0=0\}$ on $B_3$ in one point in order to isolate the fibre.
In particular the section $S_1$ becomes the entire $\mathbb P^1_{A_1}$ over  $\{c_2 = 0\} \,  \cap \, \{b_0=0\}$. 
The section $S_2$ intersects $\mathbb P^1_{A_1}$ precisely in one point, given by the solution to
\begin{eqnarray}
\{\tu = 0\} \, \cap \,  \{s_1 = 0\} \, \cap \, \{b_0=0\} \, \cap \, \{c_2=0\} \, \cap \,  D \subset \hat X_5,
\end{eqnarray}
while $S_0$ does not intersect $\mathbb P^1_{A_1}$  since $s_0 \, s_1$ is in the Stanley-Reisner ideal. Rather $S_0$ has one intersection point with $\mathbb P^1_{B_1}$, as it must since the intersection with the total fibre $\mathbb P^1_{A_1} + \mathbb P^1_{B_1}$ is one. 
This behaviour is depicted for all six loci in Figure~\ref{fig:singlets_polygon5}.

Let us now discuss in more detail the appearance of localised massless matter states.
Massless ${\cal N}=1$ chiral multiplets charged under the two $U(1)$ gauge groups arise from M2-branes wrapping the fibre components $\mathbb P^1_{A_i}$ and  $\mathbb P^1_{B_i}$ in the fibre over the six curves $C_{\bf 1^{(i)}}$, $i=1, \ldots,6$. 
The $U(1)$ charges of states from wrapped M2-branes along $\mathbb P^1_{A_i}$ or $\mathbb P^1_{B_i}$ are given by the integral of the U(1) generators  $\tw_1$ and $\tw_2$ determined in \eqref{twi-1} over the respective fibre component. 
As a consequence of the first condition in $\eqref{cond1}$, $\tw_1$ and $\tw_2$ integrate to zero over the full fibre $\mathbb P^1_{A_i} + \mathbb P^1_{B_i}$.
Thus M2-branes wrapping $\mathbb P^1_{A_i}$ and $\mathbb P^1_{B_i}$ give rise to oppositely charged ${\cal N}=1$ chiral multiplets which we are to be interpreted as charge conjugate to each other. 

The integrals can be performed in an elementary way. E.g. for ${\mathbb P}^1_{A_1}$ we read off from Figure~\ref{fig:singlets_polygon5} that $\int_{{\mathbb P}^1_{A_1}} S_2 =1$ while $\int_{{\mathbb P}^1_{A_1}} S_0 =0$. Since none of $\bar {\cal K}$ and $[c_1]$ can contribute either,  $\int_{{\mathbb P}^1_{A_1}} \tw_2 =5$.
On the other hand, $\int_{{\mathbb P}^1_{A_1}} \tw_1 = - \int_{{\mathbb P}^1_{{ B}_1}} \tw_1 = -5$  because $\int_{{\mathbb P}^1_{B_1}} S_2 =0$ and $\int_{{\mathbb P}^1_{B_1}} S_0 =1$ as depicted likewise in Figure~\ref{fig:singlets_polygon5}. 
Altogether we find the following singlet charges, where we list the M2 branes wrapping ${\mathbb P}^1_{A_i}$ as the states and the ones wrapping ${\mathbb P}^1_{B_i}$ as their charge conjugates \cite{Borchmann:2013jwa},
\begin{equation}\label{Singlet-charges}
\begin{aligned}
C_{\mathbf 1^{(1)}}: \mathbf 1_{5,-5} + c.c. ,\,\,\,\,\,  && C_{\mathbf 1^{(2)}}: \mathbf 1_{5,0} + c.c., \quad && C_{\mathbf 1^{(3)}}: \mathbf 1_{-5,-10} + c.c., \\
 C_{\mathbf 1^{(4)}}: \mathbf 1_{-5,-5}+ c.c. ,\, && C_{\mathbf 1^{(5)}}: \mathbf 1_{0,10}+ c.c., && C_{\mathbf 1^{(6)}}: \mathbf 1_{0,5}+ c.c. \,\,
\end{aligned}
\end{equation}

\begin{figure}[h]
\begin{center}
\def\svgwidth{0.75\textwidth}
 \hspace*{2cm}%
 \executeiffilenewer{singlets_Yukawas_polygon5all.svg}{singlets_Yukawas_polygon5all.pdf}%
 {inkscape -z -D --file=singlets_Yukawas_polygon5all.svg %
  --export-pdf=singlets_Yukawas_polygon5all.pdf --export-latex}%
   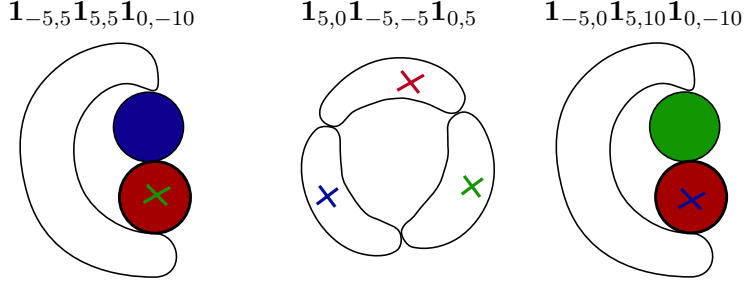%

       \caption{The fibre structure over three Yukawa points. Green corresponds to the zero section $S_0$, blue to $S_1$ and red to $S_2$.}\label{fig:singlets_Yukawas_polygon5}
\end{center}
\end{figure}

Further fibre degenerations occur in codimension-three at the intersection loci of the curves $C_{\bf 1^{(i)}}$. 
It is at these triple intersections that the Yukawa couplings between the associated charged singlets are localised.
By counting common points one straightforwardly confirms the existence of the intersection loci
\begin{equation}
\begin{split} \label{Yukawas1}
 C_{\mathbf 1^{(1)}} \cap C_{\mathbf 1^{(4)}}  \cap C_{\mathbf 1^{(5)}} & = \{ b_0 = c_2 = c_1 = 0 \}, \\
 C_{\mathbf 1^{(2)}}  \cap C_{\mathbf 1^{(3)}} \cap C_{\mathbf 1^{(5)}} & = \{ b_2 = c_1 = c_2 = 0 \}, \\ 
 C_{\mathbf 1^{(2)}} \cap C_{\mathbf 1^{(4)}}  \cap C_{\mathbf 1^{(6)}} & = \{ \ldots \},
\end{split}
\end{equation}  
where the last equation is a bit more lengthy and will not be displayed explicitly.
Over these points the fibre splits into three $\mathbb P^1$s intersecting as the affine Dynkin diagram of $A_2$ \cite{Borchmann:2013jwa}. These are depicted in Figure~\ref{fig:singlets_Yukawas_polygon5}, where we also indicate the associated Yukawa couplings, which, of course, are consistent with the $U(1)$ charges of the states.   

The Yukawa couplings \eqref{Yukawas1}, which had been presented already in \cite{Borchmann:2013jwa}, are not the only ones which are in principle allowed by the $U(1)$ charges of the states. In addition
\begin{eqnarray} \label{Yukawas2}
 C_{\mathbf 1^{(1)}} \cap C_{\mathbf 1^{(2)}}  \cap C_{\mathbf 1^{(6)}},  \qquad \quad C_{\mathbf 1^{(3)}}  \cap C_{\mathbf 1^{(4)}} \cap C_{\mathbf 1^{(6)}}  
\end{eqnarray}
could give rise to gauge invariant triple couplings, and in fact even  $C_{\mathbf 1^{(5)}} \cap C_{\mathbf 1^{(6)}}  \cap C_{\mathbf 1^{(6)}}$.
The analysis of these three geometric loci is more involved. However, our computation of chiralities in Section~\ref{sec:U(1)Fluxes} and the geometric interpretation of a certain gauge flux \emph{proves} that the second intersection is present and therefore leads to the corresponding Yukawa coupling - see the discussion at the end of Section~\ref{sec:U(1)Fluxes}. By symmetry, the same conclusion applies to the first coupling in \eqref{Yukawas2} because $(b_0,c_2)$ and $(b_2, c_1)$ are related to each other by interchanging $\alpha$ and $\beta$, see Table~\ref{coeff}. The existence of these couplings has also recently been suggested in \cite{Cvetic:2013uta}, albeit based on very different arguments. The fate of the remaining coupling, which, from the perspective of the four-dimensional effective theory, is certainly expected to be present, can also be determined by a modification of our flux computation of Section~\ref{sec:U(1)Fluxes}, but we are not presenting this analysis here.

\subsection{Gauge fluxes and chirality} \label{sec:U(1)Fluxes}

As is well-familiar \cite{Becker:1996gj}, gauge fluxes are described by 4-form flux $G_4 \in H^{2,2}(\hat Y_4)$ subject to the two transversality conditions
\begin{eqnarray} \label{transverse-flux}
\int_{\hat Y_4} G_4 \wedge \pi^* w_4 = 0,  \qquad \quad    \int_{\hat Y_4} G_4 \wedge {\rm H} \wedge \pi^* w_2 = 0
\end{eqnarray}
for $w_4$ and $w_2$ harmonic forms on $B_3$ of indicated rank. 
Again we have substituted the conventionally appearing holomorphic zero section by ${\rm H}$, see \eqref{H-sec}, which defines the embedding of $B_3$ into $\hat Y_4$.

First, as in all models with $U(1)$ gauge symmetries, the generators $\tw_i$ of the two Abelian gauge groups provide us with the corresponding Abelian gauge fluxes
\begin{eqnarray} \label{U1fluxesa}
G^{(i)}_4 = \pi^*F_i \wedge \tw_i  \qquad  \quad F_i \in H^{1,1}(B_3).
\end{eqnarray}
The analogue of these fluxes in elliptically fibred 4-folds with Mordell-Weil group of rank one has been studied intensively in the recent F-theory literature \cite{Grimm:2010ez,Braun:2011zm,Krause:2011xj,Grimm:2011fx,Krause:2012yh}.

Interestingly, the non-holomorphicity of $S_0$ and the specific form of the embedding section ${\rm H}$ allow for another simple solution to the constraints \eqref{transverse-flux} which is most easily described in terms of its dual 4-cycle. 
Consider the fibration restricted to the curve $C_{{\bf 1}^{(3)} } = b_2 \cap c_1 \subset B_3$. As depicted in Figure~\ref{fig:singlets_polygon5}, the fibre splits into $\mathbb P^1_{B_3}$ and $\mathbb P^1_{A_3}$, where the first factor is wrapped by the section $S_0$. By definition, ${\rm H}$ intersects the fibre  $\mathbb P^1_{B_3}$ in a point and therefore has no intersection with $\mathbb P^1_{A_3}$.
The fibration of $\mathbb P^1_{A_3}$ over $C_{{\bf 1}^{(3)} }$ defines a 4-cycle $\gamma$ in $\hat Y_4$. Its dual class, denoted by abuse of notation again by $ \gamma \in H^{2,2}(\hat Y_4)$,  satisfies both transversality constraints \eqref{transverse-flux} and thus represents an independent, well-defined 4-form flux. 

The 4-cycle $\gamma$ can be described very concretely as the  complete intersection on $\hat X_5$ (not on $\hat Y_4$)
\begin{eqnarray} \label{gamma-cycle}
\gamma = b_2 \cap c_1 \cap \tilde P \subset \hat X_5 \quad {\rm with} \quad P_{T^2} |_{b_2 = c_1 = 0} = s_0 \tilde P.
\end{eqnarray}
Likewise, the fibration of $\mathbb P^1_{B_1}$ over $C_{ {\bf 1}^{(1)} }$ and of  $\mathbb P^1_{A_5}$ over $C_{ {\bf 1}^{(5)} }$ give rise to well-defined transversal gauge fluxes. As can be seen from Figure~\ref{fig:singlets_polygon5} these are related to $S_1$ and $S_2$, which already appear in the $U(1)$ fluxes \eqref{U1fluxesa}. Therefore these fluxes are not independent of the fluxes we have already computed and can be discarded. 

We conclude that a simple class of 4-form flux is given by
\begin{eqnarray} \label{gammapres1}
G_4 =  G_4^{\gamma} + G_4^{(1)} +  G_4^{(2)}  
\end{eqnarray}
with 
\begin{eqnarray}
 G_4^{\gamma} = a  \gamma, \quad \qquad G_4^{(1)} =\pi^*F_1 \wedge \tw_1, \quad \qquad G_4^{(2)} =\pi^*F_2 \wedge \tw_2. 
\end{eqnarray}
Here the coefficient $a$ and  and the classes $F_i \in H^{1,1}(B_3)$ must be chosen such as to satisfy the flux quantisation condition $G_4 + \frac{1}{2} c_2(\hat Y_4) \in H^{4}(\hat Y_4, \mathbb Z)$.

The 4-cycle $\gamma$ and thus also the corresponding flux have been described as a complete intersection inside $\hat X_5$, not inside $\hat Y_4$.
For completeness we now give an equivalent presentation of $\gamma$ directly in $\hat Y_4$. 
The 4-cycle $\gamma$ is the complement of  $\mathbb P^1_{B_3}$ inside the total fibre class over the curve $C_{{\bf 1}^{(3)}}$. The total fibre over $C_{{\bf 1}^{(3)}}$ is simply the complete intersection $b_2 \cap c_1$ inside $\hat Y_4$. We would like to subtract from this the restriction of the section $S_0$ to $C_{{\bf 1}^{(3)}}$ because this is precisely what 
$\mathbb P^1_{B_3}$ fibred over $b_2 \cap c_1$ gives. To this end recall that $S_0$ intersects the fibre in a point except over  $C_{{\bf 1}^{(3)}}$. Were it not for this latter degeneration, $S_0$ would satisfy the constraint $S_0^2 + S_0 {\bar{\cal K}} =0$ (as is the case for the holomorphic zero-section in the Weierstra\ss{}  model). Thus the 4-cycle $S_0 \cap S_0 + S_0 \cap \bar {\cal K}$ is localised entirely over $b_2 \cap c_1$, and therefore the class $S_0 \wedge (S_0 + \bar{\cal K})$ is proportional to the class of  $\mathbb P^1_{B_3}$ fibred over $b_2 \cap c_1$. 
To fix the normalisation we compute    $\int_{\hat Y_4} S_0 \wedge (S_0 + \bar{\cal K}) \wedge S_2 \wedge \pi^*D_a$ with $D_a \in H^{1,1}(B_3)$ and compare this with our geometric expectation. 
First, as a consequence of the intersection numbers \eqref{IntSec2} and \eqref{eq:intnuma}
the result is $\int_{\hat Y_4} S_0 \wedge (S_0 + \bar{\cal K}) \wedge S_2 \wedge \pi^*D_a = \int_{B_3} c_1 \wedge b_2 \wedge D_a$. On the other hand this precisely matches the expected intersection between the fibration of $\mathbb P^1_{B_3}$ over $b_2 \cap c_1$ with the 4-cycle $S_2 \cap D_a$ because $S_2$ intersects $\mathbb P^1_{B}$ in one point in the fibre over $b_2 \cap c_1$, see Figure~\ref{fig:singlets_polygon5}. This identifies, as expected, $S_0 \wedge (S_0 + \bar{\cal K})$ as the class associated with $\mathbb P^1_{B_3}$ over $b_2 \cap c_1$ and in particular
\begin{eqnarray} \label{gammapres2}
\gamma = \pi^*b_2 \wedge \pi^*c_1 - S_0 \wedge (S_0 + \pi^*\bar{\cal K}) \in H^{2,2}(\hat Y_4).
\end{eqnarray}
Note that our flux $\gamma$ agrees with the flux recently presented, albeit with a rather different derivation, in \cite{Cvetic:2013uta}, which in particular classifies the primary vertical cohomology $H^{2,2}_{\rm vert.}(\hat Y_4)$ for ${\rm Bl}_2 \mathbb P^3[3]$-fibrations over $\mathbb P^3$.

Switching on gauge flux induces a D-term for the two Abelian gauge symmetries, which in the F-theory limit takes the form \cite{Grimm:2010ks,Grimm:2011tb}
\begin{eqnarray}
D_i \simeq \int_{\hat Y_4} \pi^* J \wedge G_4 \wedge \tw_i, \qquad \quad i=1,2
\end{eqnarray}
with $J$ the K\"ahler form of the base $B_3$.
To find the explicit form of the D-terms we plug in expression \eqref{twi-1} for $\tw_i$ and \eqref{gammapres1} for the flux, where it is most convenient to work directly with the presentation \eqref{gammapres2} of $\gamma$. The intersections are evaluated with the help of  the intersection numbers \eqref{IntSec1}, \eqref{IntSec2}, \eqref{eq:intnuma}, \eqref{eq:intnumB}. 
Since we will make heavy use of it momentarily we display the result
\begin{eqnarray}
D_1 &\simeq&   \int_{B_3} J \wedge \Big(  F_1 \wedge (-2 \bar {\cal K}) + F_2 \wedge  (  \alpha - \beta - \bar {\cal K})  + a (-\bar {\cal K}^2 +\alpha^2 + (2 \alpha - \beta)\wedge(\bar {\cal K} - \alpha) )   \Big), \nonumber \\
D_2 &\simeq&   \int_{B_3} J \wedge \Big( F_1 \wedge (\alpha - \bar {\cal K} - \beta)  +  F_2 \wedge (-4 \bar {\cal K} + 2 \alpha)   -a (\alpha -  \bar {\cal K} ) \wedge (\alpha - \beta - \bar {\cal K})\Big). \label{eq:Dterm}
\end{eqnarray}

Gauge fluxes induce a chiral matter spectrum, with the chiral index given by the topological intersection of $G_4$ with the matter surface associated with the specific matter state. 
In the sequel we will compute the chirality for the states wrapping the matter surfaces ${\cal C}^{(i)}$ given by the fibration of ${\mathbb P}^1_{A_i}$ over the matter curve $C_{ {\bf 1}^{(i)} }$ in the base,
\begin{eqnarray}
\chi_i = \int_{{\cal C}^{(i)}} G_4.
\end{eqnarray}

As discussed in detail in \cite{Krause:2011xj}, for the $G_4$-flux associated with the $U(1)$ symmetries this integral factorises in a simple manner and  takes the form
\begin{eqnarray}
\int_{{\cal C}^{(i)}}  G_4^{(1)} =   \int_{{\cal C}^{(i)}} \pi^*F_1 \wedge \tw_1 = q_1^{(i)} \int_{{ C}_{{\bf 1}^{(i)}}}     F_1, \qquad  \int_{{\cal C}^{(i)}}  G_4^{(2)} = \int_{{\cal C}^{(i)}} \pi^*F_2 \wedge \tw_2 = q_2^{(i)} \int_{{ C}_{{\bf 1}^{(i)}}} F_2, \nonumber \\
\end{eqnarray}
where $q^{(i)}_1$ and $q^{(i)}_2$  denote the $U(1)$ charges of the corresponding states. 
Thus the computation of the $U(1)$-fluxed induced chiralities boils down to evaluating the topological intersection number of a given class $F_i \in H^{1,1}(B_3)$ with the classes of the curves ${ C}^{(i)} \subset B_3$. For the curves $C_{ {\bf 1}^{(1)} }, C_{ {\bf 1}^{(3)} }, C_{ {\bf 1}^{(5)} }$, which are given by a complete intersection inside $B_3$, this immediately gives
\begin{eqnarray}
 \int_{{\cal C}^{(1)}}  G_4^{(1)} + G_4^{(2)} &=& \int_{B_3} (5 F_1 - 5 F_2) \wedge b_0 \wedge c_2 =  \int_{B_3} ( 5 F_1 - 5 F_2) \wedge (\bar{\cal K} - \beta) \wedge (\bar{\cal K} + \alpha - \beta), \nonumber\\
 \int_{{\cal C}^{(3)}}  G_4^{(1)} + G_4^{(2)} &=& \int_{B_3} (-5 F_1 -  10 F_2) \wedge b_2 \wedge c_1 \nonumber \\
 &=&  \int_{B_3} (-5 F_1 -  10 F_2) \wedge (\bar{\cal K} - \alpha) \wedge (\bar{\cal K} + \beta - \alpha), \nonumber \\
  \int_{{\cal C}^{(5)}}  G_4^{(1)} + G_4^{(2)} &=& \int_{B_3} (   10 F_2) \wedge c_1 \wedge c_2 
= \int_{B_3} (   10 F_2) \wedge  (\bar{\cal K} - \alpha) \wedge  (\bar{\cal K} - \beta). \label{eq:U1Fchi1}
\end{eqnarray}
To determine the class of  $C_{ {\bf 1}^{(2)} }$ and  $C_{ {\bf 1}^{(4)} }$ we recall that the complete intersections \eqref{eq:conifold_locus_1a_explicit} and \eqref{eq:conifold_locus_1b_explicit} factor into the components $C_{ {\bf 1}^{(1)} }$ and $C_{ {\bf 1}^{(2)} }$ and, respectively, $C_{ {\bf 1}^{(3)} }$ and $C_{ {\bf 1}^{(4)} }$. The multiplicity of the classes $C_{ {\bf 1}^{(1)} }$ and $C_{ {\bf 1}^{(3)} }$ within the original complete intersections can be computed to be $2\times 2$ in both cases by noting that the sections $b_0$ and $c_2$ both appear quadratic in \eqref{eq:conifold_locus_1a_explicit} (and likewise for $b_2$ and $c_1$ in \eqref{eq:conifold_locus_1b_explicit}. This leads to the classes
\begin{eqnarray}
[C_{ {\bf 1}^{(2)} }] = (2 c_2 + d_0)\wedge (2 b_0 + b_2) - 4 b_0 \wedge c_2, \quad [C_{ {\bf 1}^{(4)} }] = (2 c_1 + d_2)\wedge (2 c_1 + d_1) - 4 b_2 \wedge c_1 \label{eq:c2c4}
\end{eqnarray}
and thus, with the help of Table~\ref{coeff},
\begin{eqnarray}
&& \int_{{\cal C}^{(2)}}  G_4^{(1)} + G_4^{(2)}  =  \int_{B_3} (5 F_1) \wedge  (\alpha^2 + \alpha \beta - 2 \beta^2 + 2 \alpha \bar{\cal K} - \beta \bar{\cal K} + 5 \bar{\cal K}^2 ), \label{eq:U1Fchi2} \\
&& \int_{{\cal C}^{(4)}}  G_4^{(1)} + G_4^{(2)}  =  \int_{B_3} (-5 F_1 - 5 F_2)\wedge  (\beta^2 + \alpha \beta - 2 \alpha^2 + 2 \beta \bar{\cal K} - \alpha \bar{\cal K} + 5\bar{\cal K}^2   ). \nonumber
\end{eqnarray}

The computation of $\int_{{\cal C}^{(6)}}  G_4^{(1)} + G_4^{(2)} $ can likewise be performed geometrically by carefully determining the class of ${{ C}^{(6)}}$. Alternatively one can deduce the chirality of the associated matter states by exploiting 4-dimensional anomalies as follows:
Consider switching on only the fluxes $F_1 \wedge \tw_1$ and $F_2 \wedge \tw_2$, and demand that the D-term $D_2$ of $U(1)_2$ vanishes identically, i.e. for every class $J$.
This condition can be read as a constraint on $F_1$, while $F_2$ is taken to be completely general. More precisely, from \eqref{eq:Dterm} we deduce that $F_1$ must satisfy, in cohomology on $B_3$,
\begin{eqnarray} \label{eq:ancostr1}
F_1 \wedge (\alpha - \bar {\cal K} - \beta)  = -   F_2 \wedge (-4 \bar {\cal K} + 2 \alpha). 
\end{eqnarray}
If we switch on a flux combination with this property, $U(1)_2$ is guaranteed to remain massless since no D-term is induced. In particular, for such fluxes $U(1)_2$ is non-anomalous. Therefore we can conveniently read off the chiral index of the matter states localised on ${{\cal C}^{(6)}} $ by exploiting that the cubic $U(1)^{\rm cub}_2$ anomaly
\begin{eqnarray}
{\cal A}\Big(U(1)^{\rm cub}_2 \Big) \simeq \sum_{i=1}^{6} \Big(q^{(2)}_i\Big)^3   \int_{{\cal C}^{(i)}} G_4^{(1)} + G_4^{(2)}  
\end{eqnarray}
vanishes. Indeed, plugging in our results \eqref{eq:U1Fchi1}, \eqref{eq:U1Fchi2} and imposing the constraint \eqref{eq:ancostr1}
uniquely fixes
\begin{eqnarray}
\int_{{\cal C}^{(6)}} G_4^{(1)} + G_4^{(2)}  = - \int_{B_3} F_2 \wedge (2 \alpha^2 + 2 \beta^2 - 10 {\bar{\cal K} }^2).
\end{eqnarray}
Note that this result can be brought into the form
\begin{eqnarray}
\int_{{\cal C}^{(6)}} G_4^{(1)} + G_4^{(2)}  = - \int_{B_3} F_2 \wedge [C_{ {\bf 1}^{(6)} }],
\end{eqnarray}
where $[C_{ {\bf 1}^{(6)} }]$ is given, similarly to \eqref{eq:c2c4}, by subtracting from the class of the complete intersection \eqref{eq:conifold_locus_2_explicit} the three components $4 \, [C_{ {\bf 1}^{(1)} }], 8\,  [C_{ {\bf 1}^{(3)} }], 20 \, [C_{ {\bf 1}^{(5)} }]$ including their multiplicities.

The computation of $\int_{{\cal C}^{(j)}} \gamma$ 
 can be performed geometrically by analysing the intersection properties of the 4-cycle $\gamma$ defined in \eqref{gamma-cycle}, which is just the fibre component $\mathbb P^1_{A_3}$ depicted in Figure \ref{fig:singlets_polygon5} fibred over the curve ${{ C}_{{\bf 1}^{(3)}}}$,  and the 4-cycles  ${{\cal C}^{(j)}}$.
 This approach can be easiest performed for 
 \begin{eqnarray}
\int_{{\cal C}^{(1)}}  \gamma &=& 0, \label{chi1}\\
 \int_{{\cal C}^{(2)}}   \gamma &=& \int_{B_3} b_2 \wedge c_1 \wedge c_2   =  \int_{B_3} (\bar{\cal K} - \alpha)\wedge (-\alpha + \beta + \bar{\cal K}) \wedge (- \beta + \bar{\cal K})    \label{chi2}, \\
  \int_{{\cal C}^{(3)}}    \gamma &=& \int_{B_3} b_2 \wedge c_1 \wedge (c_2 - 2 c_1)   = \int_{B_3} (\alpha- \bar{\cal K}) \wedge (\alpha-\beta-\bar{\cal K}) \wedge (2 \alpha - \beta - \bar{\cal K})     \label{chi3}, \\
 \int_{{\cal C}^{(5)}}      \gamma &=&  \int_{B_3} b_2 \wedge c_1 \wedge c_2 =  \int_{B_3} (\bar{\cal K} - \alpha) \wedge (\alpha-\beta-\bar{\cal K}) \wedge (\beta-\bar{\cal K})  \label{chi5}.
 \end{eqnarray} 
Eq.~\eqref{chi1} follows from the fact that  the curves ${ C}_{{\bf 1}^{(1)}}$ and ${ C}_{{\bf 1}^{(3)}}$  do not intersect on $B_3$ and consequently the geometric intersection of the 4-cycle $\gamma$ with ${\cal C}^{(1)}$ is empty. 
To derive \eqref{chi5} we inspect the intersection $\gamma \cap {\cal C}^{(5)}$ and recall that both 4-cycles are fibred over the curves ${ C}_{{\bf 1}^{(3)}}$ and ${ C}_{{\bf 1}^{(5)}}$, respectively, which intersect at $\int_{B_3} b_2 \wedge c_1 \wedge c_2$ common points.
The fibre topology at these points is depicted in the third diagram of Figure~\ref{fig:singlets_Yukawas_polygon5}.
The fibre part of $\gamma$ over generic points on ${ C}_{{\bf 1}^{(3)}}$ is given by $\mathbb P^1_{A_3}$.
Over the intersection points with ${ C}_{{\bf 1}^{(5)}}$ this component splits into  $S_2$ and the component  $\tilde P'$ (depicted white in Figure~\ref{fig:singlets_Yukawas_polygon5}). Instead of working with these fibre components, we can work with their complement $S_0$ and include a minus sign.
The fibre $\mathbb P^1_{A_5}$ of ${\cal C}^{(5)} $ is the complement of $S_2$, which itself does not split over the intersection with ${ C}_{{\bf 1}^{(3)}}$. Again we choose to work not with $\mathbb P^1_{A_5}$, but with its complement $S_2$ and include another minus sign.
Thus the intersection of $\gamma$ and ${\cal C}^{(5)}$ in the fibre is
$(-1)(-1) (S_0) \cdot S_2 = (-1)(-1)1 = 1$. Similar reasoning leads to  \eqref{chi2}.
For \eqref{chi3} we observe that this is just $\int_\gamma \gamma = \int_{\hat Y_4} \gamma^2$, 
which can be evaluated by a trick used in \cite{Braun:2011zm} as the integral $\int_\gamma c_2(N_{\gamma \subset \hat Y_4})$. The Chern class of the  normal bundle $N_{\gamma \subset \hat Y_4}$ is
\begin{eqnarray} \label{eq:tricka}
c(N_{\gamma \subset \hat Y_4}) = \frac{c(N_{\gamma \subset X_5})}{c(N_{\hat Y_4 \subset X_5})} = \frac{(1+b_2)(1 + c_1) (1 + 2 \tv + \tw + c_2)}{1 + 2 \tv + \tw + c_2 + S_0}.
\end{eqnarray}
The terms in brackets follow from the divisor classes appearing in the definition of $\gamma$ in  \eqref{gamma-cycle} and, in the denominator, the hypersurface class of $P_{T^2}$, see \eqref{eq:hyper1-res}.
Expanding this and performing the integral with the help of general intersection properties results in
\begin{eqnarray}
a^2 \int_\gamma \gamma = \int_{\hat Y_4} G_4^{\gamma} \wedge G_4^{\gamma}  = a^2 \int_{B_3} b_2 \wedge c_1 \wedge (c_2 - 2 c_1),
\end{eqnarray}
where again $G_4^\gamma = a \gamma$ for a suitable normalization factor $a$.
The remaining integrals can be determined either geometrically or, since this is slightly more involved than in the above cases, again be deduced from 4-dimensional anomalies:
To this end, we consider a flux $G_4 = G_4^{(1)} +G_4^{(2)} +  G_4^{\gamma}$ and impose that either $D_1 \equiv 0$ or $D_2 \equiv 0$ (where the expressions for the D-terms are given in \eqref{eq:Dterm}). Each time this  implies a cohomological relation similar to \eqref{eq:ancostr1} for $F_1$  in terms of $F_2$ and $G_4^{\gamma}$. Since all other chiralities have been determined already, the vanishing of the cubic $U(1)_1$ and, respectively, $U(1)_2$ anomaly gives two independent relations for $ \int_{{\cal C}^{(4)}} \gamma$ and  $ \int_{{\cal C}^{(6)}} \gamma$ with the unique solution
\begin{eqnarray}
  \int_{{\cal C}^{(4)}}    \gamma & =&  \int_{B_3}  2 \,  (\bar{\cal K} + \beta - \alpha)  \wedge (\alpha^2 + \alpha \wedge  \bar{\cal K} - 2 \bar{\cal K}^2  ),  \\
    \int_{{\cal C}^{(6)}}    \gamma &=&  - \int_{B_3}  2 \,  (\bar{\cal K} + \beta - \alpha)  \wedge (\alpha^2 + \alpha \wedge  \bar{\cal K} - 2 \bar{\cal K}^2  ).
\end{eqnarray}

Finally, let us take up the discussion of Yukawa couplings at the end of Section~\ref{eq:U1singlets}. Since the matter surface ${\cal C}^{(3)}$ is simply the complement to the 4-cycle $\gamma$, the expression $  \int_{{\cal C}^{(4)}}    \gamma$ is $(-1)$ times the intersection number of ${\cal C}^{(3)}$ and ${\cal C}^{(4)}$. This proves that indeed the fibre components over the curves $C_{ {\bf 1}^{(3)} }$ and $C_{ {\bf 1}^{(4)} }$ intersect, and this intersection is automatically over the points which also lie on $C_{ {\bf 1}^{(6)} }$. Thus the second Yukawa coupling \eqref{Yukawas2} exists, and by symmetry also the first one. To check for the Yukawa $C_{\mathbf 1^{(5)}} \cap C_{\mathbf 1^{(6)}}  \cap C_{\mathbf 1^{(6)}}$ we could repeat the above chirality computation for the flux given by the fibration of $\mathbb P^1_{A_5}$ over $C_{ {\bf 1}^{(5)} }$. A non-vanishing chirality of the states ${\bf 1}^{(6)}$ with respect to this flux would indicate the presence of the Yukawa coupling.

\section{Construction of the tops over the \texorpdfstring{$\textmd{Bl}_{2}\mathbb{P}^{2}$}{Bl2 P2} fibration} \label{sec:Tops-all}
In this section we  implement  extra non-Abelian gauge symmetry in the $\textmd{Bl}_{2}\mathbb{P}^{2}$-fibration introduced previously, with special emphasis on the construction of GUT models with gauge group $SU(5) \times U(1) \times U(1)$.
This amounts to specialising further the sections $b_i, c_i, d_i$ appearing in the hypersurface equation \eqref{eq:hyper1-res} such as to create an $A_4$ singularity in the fibre over a divisor $S: w=0$ on $B_3$, and resolving the induced singularity. 
In general, there are many possible choices of $b_i, c_i, d_i$ that induce an $SU(5)$ singularity at $w=0$. A special subclass of such enhancements is given by models where the sections $b_i, c_i, d_i$ merely factor out suitable powers of $w$, but remain otherwise generic. These types of fibrations are naturally described in the language of toric geometry, with the help of the notion of tops introduced in \cite{Candelas:1996su} and classified in \cite{Bouchard:2003bu}. 

In this article we make use of these toric methods to classify the possible $SU(5)$ enhancements of the above type for our $\textmd{Bl}_{2}\mathbb{P}^{2}$-fibration.
In fact, the $\textmd{Bl}_{2}\mathbb{P}^{2}$-fibration described by hypersurface \eqref{eq:hyper1-res} is one out of 16 possible polygons analysed in  \cite{Bouchard:2003bu} which describe a torus fibration as a hypersurface (as opposed to complete intersection) in a toric ambient space.
For this polygon, which is polygon $5$ in the list of \cite{Bouchard:2003bu}, we find four possible tops which lead to inequivalent such $SU(5) \times U(1) \times U(1)$-enhancements.\footnote{A priori, there are five such inequivalent tops, but the fifth leads to a non-flat fibre in codimension-two. We therefore do not include this one in the sequel.}
The results of this analysis have been presented already in \cite{Borchmann:2013jwa} in a manner that allows for the construction of $SU(5) \times U(1) \times U(1)$ without reference to a specific base manifold $B_3$. The list of \cite{Borchmann:2013jwa} and of this article also includes the specific model presented in \cite{Cvetic:2013uta}.
A survey of the $SU(5)$ enhancements of the 16 polygons of \cite{Bouchard:2003bu} has been given in \cite{Braun:2013nqa}. 

After describing the main logic behind this toric construction we exemplify the procedure for one $SU(5) \times U(1) \times U(1)$ model. The reader not interested in the details of the toric construction can jump right away to the discussion following eq.~\eqref{eq:top4-proper-transform}, which gives the vanishing orders of the sections $b_i, c_i, d_i$ of the hypersurface  inducing an $SU(5)$ singularity. In Appendix~\ref{app-SU5} we summarise the details of the remaining three tops with $SU(5) \times U(1) \times U(1)$ gauge symmetry and give further $SU(5)$ tops for other elliptic fibres. In Appendix~\ref{app-SU4} we apply this to $SU(4)$, which, as discussed in \cite{Mayrhofer:2012zy}, can be a starting point  for the construction of fibrations with several ${\bf 10}$-curves.

Since in the sequel we will make use of toric methods in order to describe elliptically fibred Calabi-Yau fourfolds $\hat Y_{4}$ as hypersurfaces in a five complex-dimensional ambient space $\hat X_{5}$ we refer to \cite{Kreuzer:2006ax} for an introductory review of these methods. The most important points to recall about toric varieties are that they are generally described by fans of rational, polyhedral cones in a lattice $N$. 
Furthermore, there is a correspondence connecting one-dimensional cones, also called rays, with homogeneous coordinates  \cite{Cox:1993fz}. This generates a connection between toric varieties and weighted projective spaces. The coordinate corresponding to a ray $v_{i}$ will be called $x_{i}$ in the following. The lattice points of $M$, the dual lattice to $N$, correspond to monomials made out of the homogeneous coordinates $x_i$.

\subsection{The tops construction of \texorpdfstring{\cite{Bouchard:2003bu}}{[1]}} \label{sec:topsintro}

Tops were first discussed by Candelas and Font \cite{Candelas:1996su}, who observed that the intersection of certain reflexive polytopes with a plane was reflexive itself. The notion of a top was then generalised by Candelas and Skarke \cite{Candelas:1997pq}; a top $\Diamond$ is defined as
\begin{align}
\begin{split}
\Diamond = \{ v \in N_{\mathbb{R}}: \langle u_{i},v\rangle \ge -1\ \wedge \ \langle u_{0},v\rangle \ge 0 \}
\end{split}
\end{align} 
for some $u_{i} \in M$, where $M$ and $N$ are dual lattices with $N \simeq \mathbb{Z}^{3}$ and $N_{\mathbb{R}} = N \otimes \mathbb{R}$. We will say two tops are isomorphic if there exists a $GL(3,\mathbb{Z})$-transformation mapping one top to the other. This enables us to set $u_{0}=(0,0,1)$ because for $u_{0} \neq (0,0,1)$ we can always find a $GL(3,\mathbb{Z})$-transformation that maps it to $(0,0,1)$.

Analogously to reflexive polygons, we can define the dual $\Diamond^{*} \subset M_{\mathbb{R}}$ to be the polyhedron\footnote{It is important to note that the dual of a top is not a top itself.}
\begin{align}
\begin{split}
\Diamond^{*} = \{p \in M_{\mathbb{R}}: \langle p,v_{i}\rangle \ge -1  ,\ v_{i} \text{ vertices of } \Diamond \}.
\end{split}
\end{align}
To further investigate the form of the dual $\Diamond^{*}$, we note that the inequality due to $u_{0}$ singles out the hyperplane $F_{0} = \{ v \in \Diamond: \langle u_{0},v\rangle =0 \}$, which is a reflexive polygon. Written in local coordinates $(\bar x, \bar y, \bar z)$, $F_{0}$ can be denoted as the hypersurface $\bar z = 0$. Due to this, the vertices of $F_{0}$ only lead to inequalities for $\Diamond^{*}$ in the $x$- and $y$-coordinates,
\begin{align}
\begin{split}
x\bar x_{i} + y\bar y_{i} \ge -1 \ \forall i,
\end{split}
\end{align}
where the index $i$ runs over all vertices of $F_{0}$. Therefore, the dual to $F_{0}$ is just the two dimensional dual $F_{0}^{*}$. The remaining vertices of $\Diamond$ give inequalities of the form
\begin{align}
\begin{split}
z\bar z_{i} \ge -1 - x\bar x_{i} - y\bar y_{i},\label{poly}
\end{split}
\end{align}
showing that for fixed $x$ and $y$, the $z$-coordinate is only bounded by below, as $\bar z > 0$. This lower bound will be called $z_{\text{min}}(x,y)$ in the following. Thus, $\Diamond^{*}$ has the form of a prism, where the cross section is given by $F_{0}^{*}$ and extends to infinity in positive $z$-direction. This makes it natural to define a projection $\pi: \Diamond^{*} \rightarrow F_{0}^{*}, \  (x,y,z) \mapsto (x,y)$, mapping lattice points of $\Diamond^{*}$ to lattice points of $F_{0}^{*}$. In fact, every (finite) vertex of the dual top $\Diamond^{*}$ is of the form $(x,y,z_{\text{min}}(x,y))$ implying that specifying $F_{0}^{*}$ and all the $z_{\text{min}}$ describes the dual $\Diamond^{*}$ completely.

To see why tops are related to torus fibrations and, in the case of tops over $F_0$s with a dual edge of length one~\cite{Kreuzer:1997zg}\footnote{To be precise, in \cite{Kreuzer:1997zg} it was noted that for fibred toric varieties described as hypersurfaces with fibre $F_0$ one of the 16 two-dimensional reflexive polygons, cf.\ page 216 of \cite{Bouchard:2003bu}, every edge of $F_0^*$ gives rise to a section. However, the rank of the Mordell-Weil group is not necessarily the number of length one edges of $F_0^*$ minus one because, as was also realised in \cite{Braun:2013nqa}, the sections might be dependent. We also refer to \cite{GrassiPerduca} for a classification of toric and non-toric sections in elliptic fibrations realised as hypersurfaces.}, to elliptic fibrations, we have to look at the hypersurface $P$ given by $\Diamond^*$. The polynomial $P$ is defined, like in the case of reflexive polytopes, as
\begin{equation}\label{eq:definition_of_P}
P:=\sum_{p_j\in\Diamond^*} a_j \prod_{v_i\in \Diamond} x_{i}^{\langle p_j,v_{i}\rangle+1}=\sum_{p_j\in\Diamond^*} a_j \Big(\prod_{\{v_k\in \Diamond: \bar z>0\}} x_{k}^{\langle p_j,v_{k}\rangle+1}\Big)\Big(\prod_{v_l\in F_0}x_{l}^{\langle p_j,v_{l}\rangle+1}\Big)\,.
\end{equation} 
Firstly, we note that due to the definition of $\Diamond$ and $\Diamond^*$ all of the exponents in \eqref{eq:definition_of_P} are non-negative. Furthermore, \eqref{eq:definition_of_P} is a one-dimensional Calabi-Yau equation in the homogeneous coordinates corresponding to $v_l\in F_0$ with coefficients as power series in the coordinates $x_k$ with $\{v_k\in \Diamond: \bar z>0\}$. In the sum over all the lattice points $p_j$ of $\Diamond^*$ the second factor is independent of the $z$-coordinate of the $p_j$ and therefore the same for all the points along a ray, i.e.\ the preimage of a point $(x,y)\in F_0^*$ under the above define projection. This now   naturally defines a torus or---depending on $F_0$---an elliptic fibration  over $\mathbb C$. The projection to the base is given by\footnote{Since all the cones of $\Diamond$ only contain points with positive $z$-coordinates, the projection along the plan spanned by $F_0$ gives a well-defined toric morphism.}
\begin{align}\label{eq:projection-top}
\begin{split}
\pi: (x_1,\ldots,x_n)\mapsto w= \prod_{v_i\in\Diamond} x_{i}^{\langle u_{0},v_{i}\rangle} = \prod_{v_i\in\Diamond}x_{i}^{\bar z_{i}} = \prod_{\{v_i:\bar z_{i}\neq 0\}}x_{i}^{\bar z_{i}} \,,
\end{split}
\end{align}
where $w\in\mathbb C$ is the base coordinate.

In order to see how one can obtain a non-Abelian symmetry with this construction, we start with a top with only one point at height one, $\bar z_{j} =1$. 
We get a hypersurface equation in the $x_{i},i \neq j$ with coefficients being power series in $x_{j}$, all starting with a constant term. Each of these coefficients corresponds to a ray of $\Diamond^{*}$, as already explained above,  which is described by an inequality of type \eqref{poly}. If we now restrict the power series of the coefficients to start at higher powers, singularities at $w=0$ will occur. In toric terms this means that the monomials with non-vanishing coefficients only correspond to a subset $\Diamond'^{*}$ of $\Diamond^{*}$. The dual $\Diamond'$ will now include additional points at $\bar z \ge 1$, which correspond to blow-ups of the singularity. Thus, we can interpret every top $\Diamond$ as the smooth resolution of a singularity at $w=0$. The fibre over $w=0$ will contain rational curves $C_{i}$ that intersect each other. Note that in \eqref{eq:projection-top} we have, after the resolution, a product of the exceptional divisors.  Each of the curves $C_{i}$ has self-intersection minus 
two, whereas intersections of different curves are 
governed by the geometry of the top. Two curves can have intersection one if the corresponding points lie next to each other on an edge, whereas otherwise the intersection is zero. Curves corresponding to lattice points in the interior of a facet of $\Diamond$ do not intersect the hyperplane while points in the interior of an edge correspond to $l$ curves. Here, $l$ is the length of the dual edge in $\Diamond$ corresponding to the point in question.  This was proven for K3 hypersurfaces in \cite{Perevalov:1997vw} but the results obviously extend to tops \cite{Bouchard:2003bu}. In all the cases considered in this paper, the dual length of the points in question will be one, as we are only interested in $SU(N)$-symmetries, for which this reproduces an intersection pattern $C_{i}\cdot C_{j} = C_{ij}$, where $C_{ij}$ is the Cartan matrix of the $SU(N)$-algebra (in conventions with minus two on the diagonal).
In this article, however, we are interested in fibrations over a three-dimensional base $B_3$ where $w$ is some Cartier divisor on $B_3$. Therefore, the above intersection pattern, where one of the two curves must always be replaced by its corresponding divisor, will only by valid in co-dimension one. Furthermore,  in higher co-dimension one finds loci where some of the rational curves become further reducible. Additionally, divisors coming from the interior points of facets of $\Diamond$ can intersect the hypersurface and render it, already in co-dimension two, non-flat. This means that the dimension of the fibre jumps in that co-dimension.

The starting point for the construction of a top with an $SU(5)$-singularity is the choice of polygon $F_{0}$ and consequently $F_{0}^{*}$.  In order to determine the $z_{i} \equiv z_{\text{min}}(r_{i})$ for every point $r_{i}\in F_{0}^{*}$, we require local convexity,
\begin{align}
\begin{split}
z_{i-1} + (l_{i} - 2)z_{i} + z_{i+1} + l_{i} \ge 0 \label{ineq},
\end{split}
\end{align}
to get a system of inequalities which can be used to solve for the $z_{i}$. Here, $l_{i}$ is the length of the dual edge in $\Diamond$ corresponding to $r_{i}$. Note that for an interior point we have $l_{i}=0$. Additionally, we have the equation
\begin{align}
\begin{split}
n+1 = \sum_{i}l_{i}(z_{i} + 1),
\end{split}
\end{align} 
which determines the type of $A_{n}$-singularity we encounter---in our case $n=4$. As can be seen from the classification \cite{Bouchard:2003bu}, for every $A_{n}$-singularity $z_{0}=z_{1}=z_{2}=-1$ and we only have to solve for the remaining $z_{i}$. Constructing the top $\Diamond$ via inequalities of the type \eqref{poly} is now straightforward, as we have already determined all the vertices of $\Diamond^{*}$. 
In general, there will be multiple solutions to the inequalities \eqref{ineq}, resulting in different tops with different charged matter content.

\begin{figure}[h]
    \centering
\def\svgwidth{0.8 \textwidth}
 \executeiffilenewer{su5-tops_polygon5.svg}{su5-tops_polygon5.pdf}%
 {inkscape -z -D --file=su5-tops_polygon5.svg %
  --export-pdf=su5-tops_polygon5.pdf --export-latex}%
   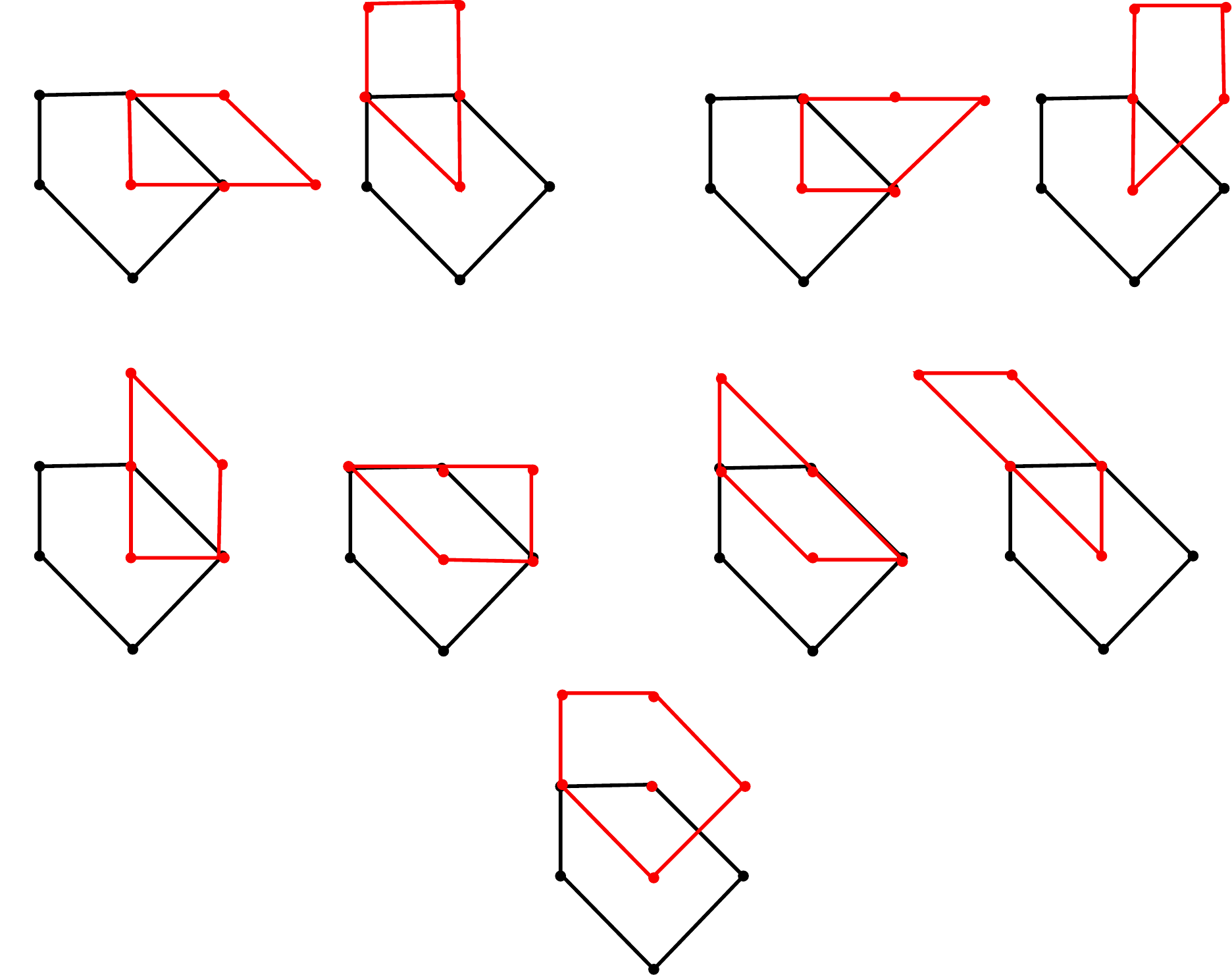%

      \caption{The tops over polygon 5 \cite{Bouchard:2003bu}, where black describes points at $\bar z=0$ and red at $\bar z=1$.}\label{fig:su5-tops_polygon5}
\end{figure}

In the sequel, we calculate the $SU(5)$ or $A_4$ tops for the elliptic fibre described by the hypersurface equation \eqref{eq:hyper1-res}. The corresponding polygon  $F_{0}$ is polygon 5 from \cite{Bouchard:2003bu}, the red polygon as depicted in Figure \ref{fig:polygon5}, with its dual $F_{0}^{*}$ given by polygon 12, the blue one in Figure \ref{fig:polygon5}. Applying the conditions \eqref{ineq}, we get a system of inequalities,
\begin{align}
\begin{split}
z_{7} + 1 \ge 0,
\quad z_{3} + 1 \ge 0, 
\quad z_{4} - z_{3} \ge 0, 
\quad z_{3} - z_{4} + z_{5} +1 \ge 0, \\
z_{4} - z_{5} + z_{6} +1 \ge 0, 
\quad z_{5} - 2\,z_{6} + z_7 + 1 \ge 0, 
\quad z_{6} - z_{7} \ge 0,
\end{split}
\end{align}
plus the equation
\begin{align}
\begin{split}
z_{3} + z_{4} + z_{5} + z_{7} = 1,
\end{split}
\end{align}
for which the possible solutions are given by
\begin{align}
\begin{split}
(z_{3},z_{4}&,z_{5},z_{6},z_{7}) \in { }\{ (-1,0,1,1,1),(-1,0,2,1,0),(-1,1,1,0,0),(-1,1,2,0,-1),\\&(0,0,1,0,0),(0,1,0,0,0),(0,1,1,-1,-1),(0,1,1,0,-1),(1,1,0,-1,-1)\}.\label{tops5}
\end{split}
\end{align}
This specifies the dual tops $\Diamond^{*}$ completely and we are in a position to construct $\Diamond$ for each element of \eqref{tops5}. The inequalities defining the tops read
\begin{align}
\begin{split}
\bar z \ge 0, \quad -\bar z + 1 \ge 0, \quad -\bar z+\bar y+1 \ge 0, \quad -\bar z+\bar x+\bar y+1 \ge 0, \quad z_{3}\, \bar z + \bar x + 1 \ge 0, \\
z_{4}\,\bar z - \bar y + 1 \ge 0, \quad z_{5}\,\bar z -\bar x -\bar y +1 \ge 0, \quad z_{6}\,\bar z - \bar x + 1 \ge 0, \quad z_{7}\,\bar z - \bar x + \bar y + \ge 0.
\end{split}
\end{align}
The resulting tops are sketched in Figure \ref{fig:su5-tops_polygon5} where we group the tops according to their $GL(3,\mathbb Z)$-equivalence. From the discussion above it is clear that the additional points at $\bar z = 1$ correspond to the resolution divisors of the singularity and their intersection pattern in co-dimension one forms the affine Dynkin diagram of $SU(5)$. 


\subsection{The \texorpdfstring{$SU(5) \times U(1) \times U(1)$}{SU(5) x U(1) x U(1)} model based on Top 4: \texorpdfstring{$(z_{3},z_{4},z_{5},z_{6},z_{7})= (0,1,0,0,0)$}{(z3,z4,z5,z6,z7)= (0,1,0,0,0)}} \label{sec_Top1}


From a top $\Diamond$ and its dual $\Diamond^*$ we can read off via equation \eqref{eq:definition_of_P} the proper transform of the resolved hypersurface equation. 
In this section we exemplify this procedure for 
 top 4 over polygon 5 in the list of   \cite{Bouchard:2003bu}. This is depicted in Figure~\ref{fig:su5-tops_polygon5} and corresponds to $(z_{3},z_{4},z_{5},z_{6},z_{7})= (0,1,0,0,0)$. The associated hypersurface equation takes the form
\begin{equation}\label{eq:top4-proper-transform}
\begin{aligned}
\tilde P_{T^2}=&\,\tilde b_1s_0 s_1 \tu\,\tv\,\tw\,\Big(\prod_{\{v_k\in \Diamond: \bar z>0\}} x_{k}^{\langle (0,0,z_0),v_{k}\rangle+1}\Big)+\tilde d_1 s_0s_1^2\tw\,\tu^2\times\\
&\times\Big(\prod_{\{v_k\in \Diamond: \bar z>0\}} x_{k}^{\langle (0,1,z_1),v_{k}\rangle+1}\Big)+\ldots+\tilde d_2 s_0^2s_1^2\tu^3\,\Big(\prod_{\{v_k\in \Diamond: \bar z>0\}} x_{k}^{\langle (-1,1,z_7),v_{k}\rangle+1}\Big)=\\
=&\,
\tilde b_{1}\,s_{0} s_{1}\tu\, \tv\, \tw  + 
\tilde d_{1}e_{2}e_{3}^{2}e_{4}\,s_{0}s_{1}^{2}\tu^{2}\tw  + 
\tilde b_{2}e_{1}e_{2}e_{3}\,s_{1}^{2} \tu\, \tw^{2}  +
\tilde c_{1}e_{0}e_{1}^{2}e_{2}\,s_{1}\tv\, \tw^{2} +\\&+
\tilde c_{2}e_{0}^{2}e_{1}^{2}e_{2}e_{4}\,s_{0}\tv^{2} \tw  + 
\tilde b_0 e_{0}e_{4}\,s_{0}^{2}\tu\, \tv^{2}  + 
\tilde d_{0}e_{0}e_{2}e_{3}^{2}e_{4}^{2}\,s_{0}^{2}s_{1}\tu^{2}\tv  + 
\tilde d_{2}e_{0}e_{2}^{2}e_{3}^{4}e_{4}^{3}\,s_{0}^{2}s_{1}^{2}\tu^{3},
\end{aligned}
\end{equation} 
where the homogeneous coordinates $x_i$ associated with $v_i\in F_0$ are renamed to those appearing in \eqref{eq:hyper1-res} and the $x_i$s coming from $\{v_k\in \Diamond: \bar z>0\}$ are relabelled, starting from $(0,0,1)^T$ going clockwise 
 to $e_i$ with $i=0,\,1,\ldots,4$.

Note that for hypersurfaces in which no monomial appears with a constant coefficient the identification of  $e_0$ is not unique due to the projection \eqref{eq:projection-top}. This subtlety is relevant for the present fibration, but not  for the Weierstra\ss{} model in which two monomials, $x^3$ and $y^2$, appear with constant coefficients. The proper transforms of these monomials become $e_3 e_4\,y^2$ and $e_1 e_2^2 e_3\,x^3$, and  out of all the $e_i$s, $e_0$ is the only homogeneous coordinate not appearing in these terms. Therefore, in agreement with Tate's algorithm, $e_0$ is the unique choice for assigning the vanishing orders.
For the ${\rm Bl}_1 {\mathbb P}^1_{1,1,2}[4]$ fibration \eqref{eq:211-hse-ng} only the monomial $\tw^2$ comes with a constant coefficient. Since the proper transform of this term is multiplied with some $e_i$s, but not all, the choice of assigning the vanishing orders is reduced but not unique.
Our identification of $e_0$ with $(0,0,1)^T$ ensures that the powers of $e_0$ are directly given by $z_i+1$. Different choices can give rise, for  the same top, to  different patterns of vanishing orders, which nonetheless describe the same elliptic fibration.
This is possible because the classes  $\alpha$, $\beta$ can be adjusted in such a way that for different choices we nevertheless obtain the same line bundles for the coefficients $\tilde b_{i}$, $\tilde c_{i}$, $\tilde d_{i}$ which define the fibration after resolution.\footnote{Taking this into account, one can show that the model discussed in \cite{Cvetic:2013uta} is top 3 of Appendix~\ref{app-SU5} (cf.\ Figure~\ref{fig:su5-tops_polygon5} and Table~\ref{tab:top3_poly5}), presented also in \cite{Borchmann:2013jwa}, where the vanishing orders are taken with respect to $e_2$ instead of $e_0$.}

To read off the higher co-dimension enhancements it is most convenient to consider the blow-down of the hypersurface \eqref{eq:top4-proper-transform} to the singular fibration with $SU(5)$ singularity over the divisor $S: \{w=0\} \subset B_3$. 
To this end we simply set the homogeneous coordinates of the $SU(5)$ resolution divisors $e_i$ with $i=1, \ldots,4$ in \eqref{eq:top4-proper-transform} to one and identify $e_0$ with $w$, obtaining
\begin{equation}
 \begin{aligned}
  0 =& \,b_{0,1} w \tu \tv^{2} s_{0}^{2} + d_{0,1} w \tu^{2}\tv s_{0}^{2}s_{1}  + d_{2,1} w \tu^{3}s_{0}^{2}s_{1}^{2} + c_{2,2} w^{2} \tv^{2} \tw s_{0} + b_{1}\tu \tv \tw s_{0} s_{1} \nonumber\\
&+ d_{1}  \tu^{2}\tw s_{0}s_{1}^{2} + c_{1,1} w \tv \tw^{2}s_{1} +b_{2} \tu \tw^{2} s_{1}^{2} \,.
 \end{aligned}
\end{equation}
Here we are using the subscript $i,j$ to denote the sections after factoring out the powers of $w$, e.g.\ by replacing $d_2 \rightarrow d_{2,2} w^2$; the $b_{i,j}$, $ c_{i,j}$, $ d_{i,j}$ are the  $\tilde b_i$, $\tilde c_i$, $\tilde d_i$ of eq.~\eqref{eq:top4-proper-transform}.
In particular the classes of, for example, $d_{i,j}$ are given by
$$[d_{i,j}] = [d_i] - j [w] $$
with $[d_i]$ as in Table~\ref{coeff}. We emphasise once more that the vanishing orders can be different for different labellings of the points at height one.

\begin{table}
\centering
\begin{tabular}{c||ccccc|c|cccc}
 & \tu & \tv & \tw & $s_{0}$ & $s_{1}$&$e_{0}$&$e_{1}$&$e_{2}$&$e_{3}$&$e_{4}$\\
\hline
\hline
[W] & $\cdot$ & $\cdot$ & $\cdot$ & $\cdot$ & $\cdot$ & 1 & $\cdot$& $\cdot$& $\cdot$& $\cdot$\\
$\alpha$ & $\cdot$ & $\cdot$ & 1 & $\cdot$ & $\cdot$ & $\cdot$& $\cdot$& $\cdot$& $\cdot$& $\cdot$\\
$\beta$ & $\cdot$ &1 &  $\cdot$ & $\cdot$ & $\cdot$ & $\cdot$& $\cdot$& $\cdot$& $\cdot$& $\cdot$\\
\hline
[U] & 1 & 1 & 1 & $\cdot$ & $\cdot$& $\cdot$ & $\cdot$& $\cdot$& $\cdot$& $\cdot$\\
$[S_{0}]$ & $\cdot$ & $\cdot$ & 1 & 1 & $\cdot$ & $\cdot$& $\cdot$& $\cdot$& $\cdot$& $\cdot$ \\
$[S_{1}]$ & $\cdot$ & 1 & $\cdot$ & $\cdot$ & 1 & $\cdot$ & $\cdot$& $\cdot$& $\cdot$& $\cdot$\\
\hline
$[E_{1}]$& $\cdot$& $\cdot$ & -1 & $\cdot$& $\cdot$ & -1& 1& $\cdot$& $\cdot$& $\cdot$ \\
$[E_{2}]$& -1& $\cdot$ & -1 & $\cdot$& $\cdot$ & -1& $\cdot$& 1 & $\cdot$& $\cdot$ \\
$[E_{3}]$& -2 & $\cdot$ & -1 & $\cdot$& $\cdot$ & -1& $\cdot$& $\cdot$& 1& $\cdot$ \\
$[E_{4}]$& -1& $\cdot$ & $\cdot$ & $\cdot$& $\cdot$ & -1& $\cdot$& $\cdot$& $\cdot$& 1 \\
\hline
\hline
 & -1 & 0 & 1 & -1 & 0 & 0 & 1 & 0 & -1 & -1\\
 & 1 & -1 & 0 & 0 & 1 & 0 & 0 & 1 & 2 & 1\\
 & $\underline{0}$ & $\underline{0}$ & $\underline{0}$ & $\underline{0}$ & $\underline{0}$ & $\underline{v}$& $\underline{v}$& $\underline{v}$& $\underline{v}$& $\underline{v}$ \\
\end{tabular}
\caption{Divisor classes and coordinates of the ambient space $\hat X_{5}$.}\label{scaling}
\end{table}

With the help of the birational map \eqref{eq:map_P2_Weierstrass} one can bring this hypersurface equation in Weierstra\ss{} form, identify $f$ and $g$ and compute the discriminant via the usual formula $\Delta = 4 f^3 + 27 g^2$. The result is
\begin{align}
\begin{split}
\Delta = w^{5}(R + Qw + \mathcal{O}(w^{2}))
\end{split}
\end{align}
with
\begin{align}
\begin{split}
R = { } &\frac{1}{16}b_{0,1}b_{1}^{4}b_{2}c_{1,1}(b_{0,1}c_{1,1} - b_{1}c_{2,2})(b_{0,1}d_{1}^{2} - b_{1}d_{0,1}d_{1} + b_{1}^{2}d_{2,1}), \\
Q = { }& \frac{1}{16} b_{0,1} b_{1}^2 (8 b_{0,1}^3 b_{2}^2 c_{1,1}^2 d_{1}^2 + b_{0,1}^2 b_{1} c_{1,1} (c_{1,1}^2 d_{1}^3 - 4 b_{2}^2 (2 d_{0,1} (c_{1,1} d_{0,1} + c_{2,2} d_{1}) - 3 b_{1} c_{1,1} d_{2,1})) \\  
   &{} + b_{1}^3 c_{2,2} (c_{1,1}^2 d_{1} (d_{0,1} d_{1} - b_{1} d_{2,1}) + 
      b_{2}^2 (c_{1,1} d_{0,1}^2 + c_{2,2} d_{0,1} d_{1} - b_{1} c_{2,2} d_{2,1})) -  \\
 & {} -  b_{0,1} b_{1}^2 (c_{1,1}^2 d_{1} (c_{1,1} d_{0,1} d_{1} + c_{2,2} d_{1}^2 - b_{1} c_{1,1} d_{2,1})  \\
  &{}  +  b_{2}^2 (c_{1,1}^2 d_{0,1}^2 - 8 c_{1,1} c_{2,2} d_{0,1} d_{1} + c_{2,2}^2 d_{1}^2 + 12 b_{1} c_{1,1} c_{2,2} d_{2,1})))\,.
      \end{split}
      \end{align}
From the form of the discriminant we expect, following the usual logic of Weierstra\ss{}  models, a single ${\bf 10}$-matter at
\begin{eqnarray}
C_{\bf 10} = \{b_1 = 0\} \cap \{w =0\}
\end{eqnarray}
as well as five distinct ${\bf 5}$-curves at
\begin{eqnarray}
&& C_{\bf 5^{(1)}} = \{b_{0,1} = 0\} \cap \{w =0\}, \qquad C_{\bf 5^{(2)}} = \{b_{2}= 0\} \cap \{w =0\},\nonumber \\
&& C_{\bf 5^{(3)}} = \{ c_{1,1}= 0\} \cap \{w =0\}, \qquad C_{\bf 5^{(4)}} = \{b_{0,1}c_{1,1} - b_{1}c_{2,2}=0\} \cap \{w =0\}, \\
&&  C_{\bf 5^{(5)}} = \{b_{0,1}d_{1}^{2} - b_{1}d_{0,1}d_{1} + b_{1}^{2}d_{2,1}=0\} \cap \{w =0\}.\nonumber
\end{eqnarray}

To actually prove the existence of such matter states, one must analyse the fibre structure over the respective curves in the base by working out the splittings of the resolution $\mathbb P^1_i$. We use the methods discussed in great detail in \cite{Krause:2011xj} and \cite{Mayrhofer:2012zy} for $SU(5) \times U(1)$, to which we refer for more details. This analysis confirms that the fibres over the curves $C_{\bf 10}$ and $C_{\bf 5^{(i)}}$  form the affine Dynkin diagram of $SO(10)$ and $SU(6)$, respectively.  
Note that the details of the computations depend on the specific triangulation via the Stanley-Reisner ideal of the fibre coordinates. 
For definiteness we work with a triangulation for which the Stanley-Reisner ideal includes the elements
\begin{align}
\begin{split}
\{  & \T w \, s_0, \T w \, \T u,\T w  \, e_{0}, \T w \, e_{2}, \T w \, e_{3}, \T w \, e_{4},  \T v \, s_1,  s_0 \, s_1,  s_1 \, e_{0}, s_1 \, e_{4}, s_0 \, e_{1}, \\ &  s_0 \, e_{2},  s_0 \, e_{3}, \T u \, e_{0}, \T u \, e_{1} ,\T u \, e_{2}, e_{0} \, e_{2}, e_{1} \, e_{3},  \T v \, \T u,\T v  \, e_{2}, \T v  \, e_{3}, \T v  \, e_{4}, e_{0} \, e_{3}\} .
\end{split}
\end{align}

As a next step we evaluate the generators of the two Abelian gauge group factors $U(1)$.
To this end it is important to determine the intersection numbers of the resolution divisors with the sections $S_0$, $S_1$ and $S_2$.
These can be determined in the same manner as the various intersection numbers of the fibration without $SU(5)$ enhancement as detailed in Section~\ref{sec:fthe2u1s} by evaluating intersections on the ambient space $\hat X_5$ with the help of the Stanley-Reisner ideal.
In particular this analysis implies
\begin{align}
\begin{split}
\int_{\hat Y_{4}} S_0\wedge E_{j} \wedge \pi^*\omega_4 &{ } = \delta_{j4} \int_{B_3} w \wedge \omega_4, \\
\int_{\hat Y_{4}} S_1\wedge E_{j} \wedge  \pi^*\omega_4  &{ } = \delta_{j3} \int_{B_3} w \wedge \omega_4, \\
\int_{\hat Y_{4}} S_2 \wedge E_{j} \wedge  \pi^*\omega_4  &{ } = \delta_{j3} \int_{B_3} w \wedge \omega_4. \label{int}
\end{split}
\end{align}
We will examine the first relation closely in order to explain how one arrives at these equations. First note that since $s_{0} \, e_{1}, s_{0}e_{2}$ and $s_{0} \, e_{3}$ are in the Stanley-Reisner ideal, those intersections vanish. Furthermore $\int_{\hat Y_{4}} S_{0}\wedge E_{0} \wedge \pi^*\omega_4$ is given as an intersection in the ambient space $\hat X_{5}$ as
\begin{align}
\begin{split}
\{ b_{2}=0\}\cap \{ e_{0}=0\}\cap \{ s_{0} =0\}\cap \{ D_{a}=0\} \cap \{ D_{b}=0\},
\end{split}
\end{align}
where the Stanley-Reisner ideal was used and the 4-form $\omega_4 \in H^4(B_3)$ is written as the wedge product of two divisors $D_a, D_b$. This vanishes since the Poincar\'e-dual two forms of the resolution divisors only have `one leg along the fibre'. From $E_{0} + \sum_{i=1}^{4}E_{i} = \pi^*w$ we obtain the intersection of $S_{0}$ with $E_{4}$ as  in \eqref{int}. The same kind of reasoning gives us the intersections of $S_{1}$ and $S_2$ with the $E_{i}$. By repeated use of the homological relations displayed in Table~\ref{scaling} one can equally determine intersection numbers involving self-intersections of sections or resolution divisors if needed.

What remains is to analyse the modification of the $U(1)$ generators $\tw_i$ in \eqref{twi-1} of the two $U(1)$ groups such as to ensure
\begin{eqnarray} \label{cond2}
\int_{\hat Y_{4}} \tw_{i}\wedge E_j \wedge  \pi^*\omega_4=0, \qquad j=1,2,3,4, \qquad \quad \forall \,  \omega_4 \in H^4(B_3). 
\end{eqnarray}
This guarantees that the $SU(5)$ gauge bosons are uncharged under $U(1)_i$ and therefore normalises the Abelian gauge groups as orthogonal to $SU(5)$.
In view of  \eqref{int} we add to both $\tw_1$ and $\tw_2$ a linear combination $\sum_{i=1}^{4}l_{i}E_{i}$ with $l_i = k_{i}+m_{i}$ such that 
\begin{eqnarray}
\sum_{i=1}^{4}\frac{k_{i}}{5}\int_{\hat Y_{4}}E_{i}\wedge E_{j} \wedge  \pi^*\omega_4  { } = \sum_{i}\frac{k_{i}}{5}C_{ij}\int_{B_3} w \wedge  \omega_4  = - \delta_{j3} \int_{B_3} w \wedge  \omega_4,
\end{eqnarray}
and
\begin{eqnarray}
\sum_{i=1}^{4}\frac{m_{i}}{5}\int_{\hat Y_{4}}E_{i}\wedge E_{j} \wedge  \pi^*\omega_4  { } = \sum_{i}\frac{m_{i}}{5}C_{ij}\int_{B_3}w \wedge  \omega_4  = \delta_{j4} \int_{B_3}w \wedge \omega_4.
\end{eqnarray}
The solution to these two equations is $k_{i}=(2,4,6,3)$ and $m_{i}=(-1,-2,-3,-4)$. At this stage we would therefore take
\begin{align}
\begin{split}
\text{w}_{1} &= 5(S_{1} - S_{0} - \mathcal{\bar K})+ \sum_{i}l_{i}E_{i}, \quad l_{i}=k_{i}+m_{i}=(1,2,3,-1), \\ \text{w}_{2} &= 5(S_2  - S_{0}- \mathcal{\bar K} - c_{1,1}) + \sum_{i}l_{i}E_{i}.
\end{split}
\end{align}
The normalisation has been chosen such as there do not occur fractional charges. Due to the non-vanishing intersection of $S_0$ with $E_{i}$, the correction term just introduced gives a contribution to the second equation of \eqref{cond1} and has to be corrected by yet another term. Altogether we arrive at 
\begin{align}
\begin{split} \label{eq:twiwithSU5}
\text{w}_{1} &= 5(S_{1} - S_{0} - \mathcal{\bar K})+ \sum_{i}l_{i}E_{i} - l_{4} \pi^*w, \quad l_{i}=k_{i}+m_{i}=(1,2,3,-1), \\ \text{w}_{2} &= 5(S_2 - S_{0} - \mathcal{\bar K} - c_{1,1}) + \sum_{i}l_{i}E_{i} - l_{4} \pi^*w .
\end{split}
\end{align}

The computation of the $U(1)$ charges of the $SU(5)$ matter states proceeds completely analogously to the charge computation of the singlet states in Section~\ref{eq:U1singlets}. For further details we also refer to \cite{Krause:2011xj} and \cite{Mayrhofer:2012zy}, where the same methods have been applied extensively.
In the fibre over each of the matter curves we choose one linear combination of $\mathbb P^1$s corresponding to a weight vector of the representation and evaluate the integral of the 2-forms $\tw_i$ over this combination of $\mathbb P^1$s. This requires in particular the geometric intersection properties of the sections $S_0, S_1, S_2$ with the fibre over the matter curves. Since the method is clear by now we merely state the result of this computation:
\begin{center}
\begin{tabular}{c|c}
Curve (on $w=0$)& matter representation\\
\hline
$\{b_{1}=0\}$ & $\mathbf{10_{2,2}}$ + $\mathbf{\overline{10}_{-2,-2}}$\\
$\{b_{0,1}=0\}$ & $\mathbf{5_{-4,1}}$ + $\mathbf{\overline{5}_{4,-1}}$\\
$\{b_{2}=0\}$ & $\mathbf{5_{-4,-4}}$ + $\mathbf{\overline{5}_{4,4}}$\\
$\{c_{1,1}=0\}$ & $\mathbf{5_{1,6}}$ + $\mathbf{\overline{5}_{-1,-6}}$\\
$\{b_{0,1}c_{1,1} - b_{1}c_{2,2}=0\}$ & $\mathbf{5_{1,-4}}$ + $\mathbf{\overline{5}_{-1,4}}$\\
$\{b_{0,1}d_{1}^{2} - b_{1}d_{0,1}d_{1} + b_{1}^{2}d_{2,1}=0\}$ & $\mathbf{5_{1,1}}$ + $\mathbf{\overline{5}_{-1,-1}}$\\
\end{tabular}
\end{center}
By further inspection of the curve intersections in codimension three, we have checked the existence of the following Yukawa couplings involving $SU(5)$ charged matter states:
\begin{center}
\begin{tabular}{c|c}
Point (on $w=0$)& Yukawa coupling\\
\hline
$\{b_{1}=b_{2}=0\}$ & $\mathbf{10_{2,2}} \mathbf{10_{2,2}}\mathbf{5_{-4,-4}}$ \\
$\{b_{1}=c_{1,1}=0\}$ & $\mathbf{10_{2,2}} \mathbf{\overline{5}_{-1,4}}\mathbf{\overline{5}_{-1,-6}}$ \\
$\{ b_{1} = d_{1} =0\}$ & $\mathbf{10_{2,2}} \mathbf{\overline{5}_{-1,-1}}\mathbf{\overline{5}_{-1,-1}}$ \\
$\{b_{1}=b_{0,1}=0\}$ & non-flat fibre \\
\end{tabular}
\end{center}
The fibre forms a non-extended $E_6$ Dynkin diagram over $\{b_{1}=b_{2}=0\}$ and an extended $D_6$ Dynkin diagram over $\{b_{1}=c_{1,1}=0\}$. Over $\{ b_{1} = d_{1} =0\}$, the fibre structure is that of a non-extended $D_6$ Dynkin diagram, as a consequence of a non-trivial monodromy at that point.  

In addition ${\bf 1}^{(i)} \, {\bf 5}^{(j)} \,  {\bf \bar 5}^{(k)}$-type Yukawa couplings appear at the pairwise intersection of the ${\bf 5}^{(i)}$-curves. For reasons that will be discussed in detail in section \eqref{sec:Flatness} we analyse these for the special case that the section $b_{0,1}$ is constant such that the non-flat point is absent. Since this removes the matter curve at 
$\{b_{0,1}=0\}$ there are six possible intersections of the ${\bf 5}$-curves. It turns out that these points intersect one of the singlet curves $C_{{\bf 1}^{(i)}}$ in precisely the right pattern such that a Yukawa coupling with suitably charged singlet exists. Concretely, the Yukawa couplings are the following:
\begin{center}
\begin{tabular}{c|c}
Point (on $w=0$) & Yukawa coupling\\
\hline
$\{c_{1,1}=b_{2}=0\}$ & $\mathbf{5_{-4,-4}} \mathbf{\bar 5_{-1,-6}}  \mathbf{ 1_{5,10}}     $ \\
$\{b_{0,1}c_{1,1} - b_{1}c_{2,2}=b_{2}=0\}$ & $\mathbf{5_{-4,-4}} \mathbf{\bar 5_{-1,4}}  \mathbf{ 1_{5,0}}     $ \\
$\{b_{0,1}d_{1}^{2} - b_{1}d_{0,1}d_{1} + b_{1}^{2}d_{2,1} =b_{2}=0\}$ & $\mathbf{5_{-4,-4}} \mathbf{\bar 5_{-1,-1}}  \mathbf{ 1_{5,5}}     $ \\
$\{c_{1,1}=c_{2,2}=0\}$ & $\mathbf{5_{1,6}} \mathbf{\bar 5_{-1,4}}  \mathbf{ 1_{0,-10}}     $ \\
$\{c_{1,1}=b_{0,1}d_{1}^{2} - b_{1}d_{0,1}d_{1} + b_{1}^{2}d_{2,1} = 0\}$ & $\mathbf{5_{1,6}} \mathbf{\bar 5_{-1,-1}}  \mathbf{ 1_{0,-5}}     $ \\
$\{b_{0,1}c_{1,1} - b_{1}c_{2,2}  =b_{0,1}d_{1}^{2} - b_{1}d_{0,1}d_{1} + b_{1}^{2}d_{2,1} =  0\}$ & $\mathbf{5_{1,-4}} \mathbf{\bar 5_{-1,-1}}  \mathbf{ 1_{0,5}}     $ \\

\end{tabular}
\end{center}
The fibre topology is in each case that of an extended $SU(7)$ Dynkin diagram, as expected.

\subsection{Flatness of the fibration} \label{sec:Flatness}

We now address a complication of the types of models discussed so far that arises in complex codimension three. 
This complication is related to the fact that the fibration becomes non-flat over special points on the base $B_3$, meaning that the dimension of the fibre jumps over these points. The non-flat points are given by the intersection of the $SU(5)$ divisor $w=0$ with two more divisor classes $D_1$ and $D_2$. For the example studied in Section~\ref{sec_Top1}, $D_1$ and $D_2$ are given by $b_1$ and $b_{0,1}$. 
Indeed the hypersurface $\tilde P_{T^2}$ displayed in \eqref{eq:top4-proper-transform} splits off a factor of $e_2$ at $b_1=b_{0,1}=0$. Therefore the locus $\{\tilde P_{T^2}=0\} \cap \{e_2=0\} \cap \{b_1=0\} \cap \{b_{0,1}=0\} \subset \hat X_5$, which would usually describe a $\mathbb P^1$ in the fibre, is really of dimension 2, as opposed to one. For the remaining $SU(5)$ models studied in this paper the non-flat points are indicated in the tables in the appendices.


In the presence of non-flat fibres tensionless strings appear in the effective action describing the F-theory compactification. A safe way to arrive at a globally well-defined Calabi-Yau fibration suitable for F-theory compactifications is to avoid these non-flat points. In principle there are two alternative strategies. First, for a  specific fibration over a concrete base space $B_3$ it may happen that the set $\{ w=0\} \cap \{ D_1 =0 \} \cap \{D_2 =0 \}$ is simply empty.
A necessary condition for this is that the topological intersection number  of the $SU(5)$ divisor class $w$ with the classes $D_a$ and $D_b$ is non-positive,
\begin{eqnarray}
\int_{B_3} w \wedge D_1 \wedge D_2 \leq 0.
\end{eqnarray}
Indeed if $\int_{B_3} w \wedge D_1 \wedge D_2 <  0$,  the triple intersection does not correspond to a geometric point set. If $\int_{B_3} w \wedge D_1 \wedge D_2 =0$, then
generic representatives of the three divisor classes will have no intersection points.

A second, and more drastic, way to forbid the non-flat points is to exploit the freedom in the definition of the sections appearing in the hypersurface equation such as to identify one of the two divisor classes $D_i$ with the trivial class. Whether or not this is possible consistently must again be checked for concrete examples following the logic detailed below. Note that in contrast to the first approach,  this strategy removes not only the non-flat points, but also the matter curve defined by the intersection of the divisor $D_i$ in question and the $SU(5)$ divisor, thereby restricting the model further.

We exemplify this latter strategy by constructing a concrete Calabi-Yau 4-fold without non-flat points realising the $SU(5) \times U(1) \times U(1)$ model studied in Section~\ref{sec_Top1}.
 Let us collectively denote any of the sections $c_i, d_i, b_i$ by $k_i$ and recall that in $k_{i,j}$ a power of $w^j$ has been factored out as $k_i = k_{i,j} w^j$. The classes of the sections $k_i$  are collected in Table~\ref{coeff} and we furthermore recall that $\alpha$ and $\beta$ are in principle arbitrary classes on $B_3$ which must be chosen such that all the sections $k_{i,j}$ appearing in the hypersurface equation $\tilde P_{T^2}$ displayed in \eqref{eq:top4-proper-transform} exist. This means that all classes $k_{i,j}$ must be effective.

 We investigate the possibility of setting the section $b_{0,1}$ constant by choosing its corresponding class to be trivial. 
 This implies the relation
 \begin{eqnarray}
 \alpha = \beta - \bar{\cal K} + w \label{eq:nfcondition}
 \end{eqnarray} 
in homology, where  $\bar{\cal K}$ and $w$ are positive classes. Demanding existence of all the remaining sections $k_{i,j}$ appearing in the hypersurface equation $\tilde P_{T^2}$ in \eqref{eq:top4-proper-transform}  yields the non-trivial constraints
 \begin{eqnarray}
 \beta \geq 0, \qquad \bar{\cal K} -  2 w - \beta \geq 0, \qquad 2 \bar{\cal K} - w \geq 0,
 \end{eqnarray} 
 where equality will remove further matter curves whose associated classes will then be trivial.
 Note that setting $b_{0,1}$ constant removes one of the five ${\bf 5}$-curves given by $\{b_{0,1} = 0 \} \cap \{ w=0 \}$.

 It is not hard to find explicit base spaces $B_3$ and well-defined fibrations over them where these conditions 
 can be met.
 The probably simplest example is to consider $B_3 = \mathbb P^3$ with $\bar{\cal K}={\cal O}(4)$, i.e. $c_1(B) = 4 H$ in terms of the hyperplane class $H$ which spans $H^{1,1}(B_3)$. 
 For instance, for the choice of classes
 \begin{eqnarray}
 \beta = H, \qquad w= H, 
 \end{eqnarray}
$b_{0,1}$ is constant and all remaining sections $k_{i,j}$ appearing in $\tilde P_{T^2}$ exist and are in classes 
\begin{eqnarray}
&& d_{0,1} =H, \quad d_1 = 5H, \quad d_{2,1} = 2H, \quad b_0=H, \quad b_1=4H, \\
&& b_2 = 7H, \quad c_{1,1}=5H, \quad c_{2,2} = H. \nonumber 
\end{eqnarray}

Indeed one can check that the fibration \eqref{eq:top4-proper-transform} over $\mathbb P^3$ with these class assignments leads to a smooth Calabi-Yau 4-fold $\hat Y_4$ given by a flat elliptic fibration. 
The topological data of $\hat Y_4$ can be computed torically by describing $\hat Y_4$ via the reflexive polytope given in Table \ref{tab:polyY4}.
\begin{table}
\centering
\begin{tabular}{ccccccccccccc}
 \tu & \tv & \tw & $s_{0}$ & $s_{1}$ & $e_0$ & $e_1$ & $e_2$ & $e_3$ & $e_4$ & $z_1$ & $z_2$ & $z_3$\\
\hline
 0 & 0 & 0 & 0 & 0 & 0 & 0 & 0 & 0 & 0 &-1 & 1 & 0\\
 0 & 0 & 0 & 0 & 0 & 1 & 1 & 1 & 1 & 1 &-1 & 0 & 0\\
 0 & 0 & 0 & 0 & 0 & 0 & 0 & 0 & 0 & 0 &-1 & 0 & 1\\
-1 & 0 & 1 &-1 & 0 & 0 & 1 & 0 &-1 &-1 & 0 & 0 & 2\\
 1 &-1 & 0 & 0 & 1 & 0 & 0 & 1 & 2 & 1 & 0 & 0 & 1
\end{tabular}
\caption{Points of the toric ambient space $\hat X_{5}$ of $\hat Y_4$.}\label{tab:polyY4}
\end{table}
In particular the Hodge numbers of $\hat Y_4$ are
\begin{eqnarray}
h^{1,1} = 8, \quad h^{2,1} =  0,  \qquad h^{3,1}= 267
\end{eqnarray}
and the Euler characteristic is $\chi(\hat Y_4) = 1698$.

\subsection{Gauge fluxes}

We now describe the construction of a class of chirality inducing $G_4$ gauge fluxes for the $SU(5) \times U(1) \times U(1)$ fibrations presented in this work.

We are interested in those gauge fluxes which, in addition to satisfying the transversality conditions \eqref{transverse-flux}, do not break gauge invariance.
To analyse this condition we note that the operation of integrating a flux $G_4$ over one of the resolution ${\mathbb P}^1_i$ in the fibre over the $SU(5)$ divisor $S: w=0 \subset B_3$ gives us an element in $H^{1,1}(S)$ which corresponds to the first Chern class of a line bundle $L_i$ to which the $SU(5)$ gauge bosons couple,
\begin{eqnarray}
c_1(L_i) = \int_{\mathbb P^1_i} G_4.
\end{eqnarray}
For $SU(5)$ to be unbroken these bundles $L_i$ must be trivial for $i=1,2,3,4$. If the $SU(5)$ surface has $H^{1}(S)=0$, which we assume in the sequel, this is equivalent to stating that $\int_S c_1(L_i) \wedge \omega_a=0$ for a basis $\omega_a$ of $H^{1,1}(S)$. 
The fluxes that we are going to construct will have the property that $c_1(L_i) \in \iota^* H^{1,1}(B_3)$. For these fluxes it suffices to check that $\int_S c_1(L_i) \wedge \iota^*\omega_2 =0$ for every $\omega_2 \in H^{1,1}(B_3)$, where $\iota: S\rightarrow B_3$ is the embedding of the divisor $S$ into $B_3$. Thus we will impose the constraint
\begin{eqnarray}
\int_{\hat Y_4} G_4 \wedge E_i \wedge \pi^*\omega_2  = 0 \quad \forall \, \, \omega_2 \in H^{1,1}(B_3), \qquad \quad \quad i=1,2,3,4. \label{eq:orth1}
\end{eqnarray}

The gauge fluxes associated with the two $U(1)$ symmetries, normalised as explained in the previous sections, are guaranteed to meet these requirements.
Thus  the $U(1)$ gauge fluxes take the form
\begin{eqnarray}
G_4^{(i)} = F_i \wedge \tw_i, \qquad F_i \in H^{1,1}(B_3)
\end{eqnarray}
for suitably normalised generators $\tw_i$. For instance, for the $SU(5) \times U(1) \times U(1)$ fibration of Section~\ref{sec_Top1}, the $\tw_i$ must be chosen as in \eqref{eq:twiwithSU5}. The extra flux \eqref{gamma-cycle} is not modified by the $SU(5)$ singularity and its resolution: The 4-cycle  \eqref{gamma-cycle} is simply the matter surface associated with an $SU(5)$ singlet, and since the  $SU(5)$ weights of this singlet state vanish, the constraint \eqref{eq:orth1} holds automatically.

In addition one can construct a simple type of $G_4$ flux from each of the $SU(5)$ charged matter surfaces. We first describe the general logic behind  these extra  fluxes and then exemplify our construction for the $SU(5) \times U(1) \times U(1)$ fibration of Section~\ref{sec_Top1}.
Let $R$ be one of the $SU(5)$ representations present in the model, i.e.  $R={\bf 10}$ or  $R={\bf 5}^{(i)}$ for one of the ${\bf 5}$ representations.
In the fibre over the matter curve $C_R$ one or several of the $\mathbb P^1$s split such as to form the extended Dynkin diagram of an enhanced symmetry group (e.g. $SO(10)$ for $R={\bf 10}$). Choose one of these, denoted by ${\mathbb P}^1_R$ in the sequel, with the property that its intersections with the resolution divisors $E_i$ correspond to a weight vector $\beta_R$ of the representation $R$.  
More precisely, if we define the 4-cycle $\gamma_R$ as the fibration of this ${\mathbb P}^1_R$ over $C_R$, then
\begin{eqnarray} \label{eq:weightsveca}
\int_{\gamma_R} E_i \wedge \pi^*\omega_2 = [\beta_R]_i \int_{C_R} \omega_2 =  [\beta_R]_i \int_{B_3} w \wedge D_{C_R} \wedge \omega_2,
\end{eqnarray}
where $C_R$ is the intersection of $w$ with the divisor $D_{C_R}$ in the base, e.g. $D_{\bf 10} = b_1$ for the model in Section~\ref{sec_Top1}.

Based on this 4-cycle $\gamma_R$ we can now construct a $G_4$-flux. 
The 4-cycle $\gamma_R$ is easily described as the complete intersection of three divisors within $\hat X_5$, similarly to the flux \eqref{gamma-cycle}.
As a consequence of \eqref{eq:weightsveca} the linear combination
\begin{eqnarray}
\gamma_R - [\beta_R]_i  \, C_{ij}^{-1}  \,  E_j \wedge \pi^*D_{C_R}
\end{eqnarray}
 fulfills the constraint \eqref{eq:orth1}. Here we abuse notation and assign the symbol $\gamma_R$ also to the dual class $H^{2,2}(\hat Y_4)$ and $C_{ij}$ denotes the $SU(5)$ Cartan matrix. Moreover, both contributions separately satisfy the first of the transversality constraints \eqref{transverse-flux}. It is then always possible to add a suitable multiple of the total fibre class over $C_R$, i.e. to add a multiple of $\pi^*w \wedge \pi^*D_R$ such as to ensure also the second constraint in \eqref{transverse-flux}.
 As a result,
 \begin{eqnarray} \label{eq:Rflux}
 G^R_4 = \lambda_R \Big( \gamma_R - [\beta_R]_i  \, C_{ij}^{-1}  \,  E_j \wedge \pi^*D_{C_R} + \Delta \, (\pi^*w \wedge \pi^*D_R) \Big)
 \end{eqnarray}
 with suitably chosen $\Delta$ represents a well-defined gauge flux. Here we have introduced an overall constant $\lambda_R$ to be chosen such that the full $G_4$ flux (i.e. the linear combination of all types of fluxes present)
satisfies the quantization condition $G_4 + \frac{1}{2} c_2(\hat Y_4) \in H^4(\hat Y_4, \mathbb Z)$.

Such a flux exists for each of the matter curves. Note that the specific representation of this flux depends on the choice of ${\mathbb P}^1_R$. However,  for fixed $C_R$ these various choices of ${\mathbb P}^1_R$ do not give rise to independent fluxes because the rank of the fibre is increased only by one compared to the generic $SU(5)$ fibre and thus only one independent new 4-cycle class exists for each $C_R$. Also the number of independent fluxes is in general smaller than the number of matter surfaces because of homological relations between the various $G_4^R$. These relations can be determined from the intersection numbers of the fluxes.

The appearance of fluxes of the type \eqref{eq:Rflux} is not new. The so-constructed $G_4^R$ with $R={\bf 10}$ gives rise to the so-called universal flux constructed in \cite{Marsano:2011hv}. 
This flux exists as an independent flux also for $SU(5)$ fibrations with no extra $U(1)$s \cite{Marsano:2011hv} and persists for the $SU(5) \times U(1)$ models described in \cite{Krause:2011xj} (see also \cite{Grimm:2011fx}). Its Type IIB interpretation was given in \cite{Krause:2012yh}: In models with a type IIB dual, the universal flux describes the generically present D5-tadpole free diagonal $U(1)_a \subset U(5)_a$ flux of class $F_a = \bar {\cal K}$, which maps to $F_a = [O7]$ in Type IIB models with an orientifold 7-plane.

In less generic fibrations, such as the $SU(5) \times U(1) \times U(1)$ at hand, also the fluxes from the {\bf 5}-curves give rise to gauge fluxes independent of 
$G_4^{{\bf 10}}$ and the $U(1)$ fluxes.    
Let us exemplify this construction for our example given by top 4 over a general basis $B_3$, where we explicitly enforce flatness of the fibration via \eqref{eq:nfcondition}, as discussed in the previous section. Indeed it turns out that absence of non-flat fibres is crucial in order to arrive at a consistent set of fluxes.

We choose $R= {\bf 5}^{(3)}$, i.e. construct the flux associated with the fibration over the curve
\begin{eqnarray}
C_{{\bf 5}^{(3)}} = \{w=0 \} \cap \{c_{1,1} = 0 \}.
\end{eqnarray}
Inspection of the fibre shows that the resolution divisor $\{e_4=0\}$ splits over $C_{{\bf 5}^{(3)}}$ because  the hypersurface equation \eqref{eq:top4-proper-transform} factorises as 
\begin{eqnarray}
\tilde P_{T^2} |_{c_{1,1}=0 \cap e_4=0} = u \, \tilde p
\end{eqnarray}
for some polynomial $\tilde p$.
We choose the 4-cycle
\begin{eqnarray}
\gamma_{{\bf 5}^{(3)}} = \{ u=0 \} \cap \{ e_4=0 \} \cap \{ c_{1,1}=0 \}, 
\end{eqnarray}
described as a complete intersection inside $\hat X_5$. It is associated with the weight vector $[0,0,1,-1]$. Following the general discussion we define  the gauge flux
\begin{eqnarray} \label{eq:53flux}
G_4^{{\bf 5}^{(3)}} = \lambda_{{\bf 5}^{(3)}} \Big( \gamma_{{\bf 5}^{(3)}} +  l_i E_i \wedge \pi^*c_{1,1} - \frac{4}{5} \, \,   (\pi^*w \wedge \pi^*c_{1,1})  \Big) , \qquad \quad l_i = \frac{1}{5} (1,2,3,-1).
\end{eqnarray}
The value of $\Delta = -\frac{4}{5}$ follows from 
\begin{eqnarray}
&& \{e_4 = 0 \} \cap \{\tu = 0 \} \cap \{s_0=0 \} \cap \{c_{1,1} = 0\}  \cap \{D_a = 0\}  = \int_{B_3} w \wedge c_{1,1} \wedge D_a, \\
&& \int_{\hat Y_4} E_i \wedge S_0 \wedge c_{1,1} \wedge D_a = \delta_{4i} \int_{B_3} w \wedge c_{1,1} \wedge D_a
\end{eqnarray}
 for every base divisor $D_a$. Again, the overall normalisation $\lambda_{{\bf 5}^{(3)}}$  is to be chosen in agreement with the quantisation condition for the full $G_4$ flux, $G_4 + \frac{1}{2} c_2 (\hat Y_4) \in H^4(\hat Y_4,\mathbb Z)$. With this understood, we will not make this overall factor explicit in the sequel.

It is now a straightforward, albeit tedious task to evaluate the intersection properties of this gauge flux. 
We only exemplify here the chirality the flux induces for the $SU(5)$ matter states. All remaining computations can be performed with the help of the methods discussed here. 

Consider first the chiral index of the ${\bf 10}$-states. To this end we first pick one of the 10 possible surfaces given by the fibration of a combination of $\mathbb P^1$s over $C_{\bf 10}$ which corresponds to a weight vector of the representation ${\bf 10}$. The chiral index of ${\bf 10}$-states is given by the integral of \eqref{eq:53flux} over this surface. Clearly the result is independent of the specific choice of matter surface. 
For example, we know that $\tilde P_{T^2}|_{b_1=0 \cap e_1=0}$ splits off a factor of $e_4$. 
We can therefore pick the 4-cycle
\begin{eqnarray}
{\cal C}_{14} = \{ e_1=0 \} \cap \{e_4 =0 \} \cap \{b_1 =0 \} \qquad \quad {\rm with} \quad \beta_{14} = [-1,1,0,-1].
\end{eqnarray} 
We then compute the 
intersection of this 4-cycle with $\gamma_{{\bf 5}^{(3)}}$. The two base 2-cycles intersect at $b_{1}=c_{1,1}=w=0$. This is just one of the ${\bf 10 \, \bar 5 \, \bar 5}$ couplings and the fibre takes the form of the extended Dynkin diagram of $SO(12)$. However, one finds that the fibres of $\gamma_{{\bf 5}^{(3)}}$ and ${\cal C}_{14}$ do not intersect because $\tu \, e_1$ is in the Stanley-Reisner ideal. Therefore $\int_{{\cal C}_{14} }   \gamma_{{\bf 5}^{(3)}} =0$, and the chiral index comes entirely from the second piece in $\eqref{eq:53flux}$,
\begin{eqnarray}
\chi_{\bf 10} &= &\int_{{\cal C}_{14} } G_4^{{\bf 5}^{(3)}}  = \sum_i l_i \int_{{\cal C}_{14} } E_i \wedge \pi^*c_{1,1} = \sum_i l_i [-1,1,0,-1]_i  \int_{B_3} w \wedge c_{1,1} \wedge b_1  \nonumber \\
&=& \frac{2}{5} \int_{B_3} w \wedge c_{1,1} \wedge b_1.
\end{eqnarray}

The chiral index of ${\bf 5}^{(2)}$, ${\bf 5}^{(4)}$ and ${\bf 5}^{(5)}$ is computed analogously by inspection of the fibre of the associated matter curves and over the intersection points with $C_{{\bf 5}^{(3)}}$. For the computation of $\chi_{{\bf 5}^{(3)}}$ we can take, as the matter surface, $\gamma_{{\bf 5}^{(3)}} $ itself, where we need to remember that its weight vector $[0,0,1,-1]$ is associated with a ${\bf \bar 5}$. The self-intersection $\int_{\gamma_{{\bf 5}^{(3)}}} \gamma_{{\bf 5}^{(3)}} $ is computed with the same trick as described around eq.~\eqref{eq:tricka}. 
Finally, the matter representation ${\bf 5}^{(1)}$ is absent since we are assuming that $b_{0,1}$ corresponds to the trivial class, i.e we assume that \eqref{eq:nfcondition} holds in order to ensure flatness of the fibration. 

As a result, the $SU(5)$ matter chiralities read
\begin{eqnarray}
\chi_{\bf 10}&=& \frac{2}{5} \int_{B_3} w \wedge c_{1,1} \wedge b_1 = -\frac{2}{5} \int_{B_3} w \wedge \bar {\cal K} \wedge (\beta - 2  \bar {\cal K}  + 2 w), \nonumber \\
\chi_{{\bf 5}^{(2)}} &=&   \frac{1}{5} \int_{B_3} w \wedge c_{1,1} \wedge b_2 =-\frac{1}{5} \int_{B_3} w \wedge (2\bar {\cal K} - w ) \wedge(\beta - 2 \bar {\cal K} + 2 w),\nonumber  \\
\chi_{{\bf 5}^{(3)}} &=&    \frac{1}{5} \int_{B_3} w \wedge  (\beta^2 + \beta \wedge  (\bar {\cal K} - w) - 6  (\bar {\cal K} - w) ^2 )  ,\\
\chi_{{\bf 5}^{(4)}} &=&   \int_{B_3} w \wedge c_{1,1} \wedge \Big(-\frac{4}{5} \, c_{1,1} + c_{2,2} \Big) =  \nonumber\\
&=&  \frac{1}{5} \int_{B_3}  w \wedge  \Big( \beta^2 - 6 \bar {\cal K}^2+ 2 \bar {\cal K} \wedge w + 4 w^2 + \beta \wedge (\bar {\cal K} + 4 w)\Big), \nonumber \\
\chi_{{\bf 5}^{(5)}} &=&   \frac{2}{5} \int_{B_3} w \wedge c_{1,1} \wedge d_1 = - \frac{2}{5} \int_{B_3}w \wedge (\beta + \bar {\cal K}) \wedge (\beta - 2 \bar {\cal K} + 2 w). \nonumber  
\end{eqnarray}
Note that for a well-quantised linear combination of fluxes such that $G_4 + \frac{1}{2} c_2(\hat Y_4) \in \mathbb Z$, the final result for the chiralities is guaranteed to be integer \cite{Collinucci:2012as}.

\section{Embedding into a local Higgsed \texorpdfstring{$E_8$}{E8}}
\label{sec:embe8}

In this section we investigate whether our models can be described by a Higgsed $E_8$ gauge theory in their local limit. Local model building in F-theory has been studied intensively (see \cite{Weigand:2010wm,Maharana:2012tu} for reviews), starting from the initial constructions of \cite{Donagi:2008ca,Beasley:2008dc,Beasley:2008kw,Donagi:2008kj}. A common feature of all local models analysed so far is that they are based on a Higgsed $E_8$ gauge theory. More precisely since we have an $SU(5)$ symmetry all over the divisor $S \subset B_3$, we can decompose $E_8 \rightarrow SU(5) \times SU(5)_{\perp}$ and study local models through the Higgsing of $SU(5)_{\perp}$.\footnote{For clarity we will henceforth refer to the visible  gauge group $SU(5)$ realized along $S$ as $SU(5)_{GUT}$ to distinguish it from $SU(5)_{\perp}$.} A useful tool, brought from Heterotic string compactifications and introduced to F-theory in \cite{Donagi:2009ra}, for doing this has been the spectral cover. In particular local models involving $U(1)$ symmetries correspond to split spectral covers \cite{Marsano:2009gv,Marsano:2009wr,Dudas:2010zb}. Having constructed global models involving $U(1)$ symmetries it 
is therefore natural to ask whether they admit an embedding into $E_8$, a necessary requirement for their local limits to fall into the class of models studied so far in the literature, and if they do not admit such an embedding what does this teach us about extending the spectrum of local theories possible in F-theory?

First we should define precisely what we mean by embeddable into $E_8$. We consider the embedding into $E_8$ as
\begin{equation}
E_8 \rightarrow SU(5)_{GUT} \times SU(5)_{\perp} \;.
\end{equation}
The commutant group $SU(5)_{\perp}$ controls the possible $U(1)$ factors and charges of the GUT states that can appear in models based on $E_8$. There are various possibilities for embedding the $U(1)$ factors into $SU(5)_{\perp}$ but the most general such embedding is usefully parametrised in terms of the embedding into its Cartan subgroup $G_{\perp} = U(1)^4$. An embedding of a $U(1)$ into $G_{\perp}$ is specified by 5 parameters $a_i$ which determine its embedding into $S\left[U(1)^5\right]$ and therefore should satisfy a tracelessness constraint $\sum_i a_i=0$. Our  notation is to write a particular $U(1)$ embedding as
\begin{equation}
U(1)_A = \sum_{i=1}^5 a^A_i t^i \;, \label{u1}
\end{equation}
where the $t^i$ are introduced to determine the $U(1)$ charges of the states as follows. If we decompose the adjoint of $E_8$ under $S\left[U(1)^5\right]$ we find the following representations of $SU(5)_{GUT}$ with $U(1)$ charges labelled by $t_i$,
\begin{eqnarray}
\te_i: t_i \;, \;\;\;
\fb_{ij}: t_i+t_j \;, \;\;\;
\un_{ij}: t_i-t_j \;, \label{u1charges}
\end{eqnarray}
where for the $\fb$s and $\un$s we have that $i \neq j$. Here the $t_i$ correspond to the $U(1)$ charges of the representations in the sense that for a given $U(1)$, specified by \eqref{u1}, the charges are simply given by the contraction of the $t_i$ and $t^i$ using $t_i t^j = \delta_i^j$.

There are two types of gauge invariant operators which can be constructed from the fields in \eqref{u1charges}. There are mesonic type (in $SU(5)_{\perp}$) operators whose charges $t_i$ sum to zero, for example $\f \, \te \, \te$ couplings, and baryonic operators, for example $\fb \, \fb \, \te$ couplings, whose $t_i$ sum to $t_1+t_2+t_3+t_4+t_5$.

We define an embedding of a global model into $E_8$  by specifying the embedding of the global $U(1)$ into the Cartan of $SU(5)_{\perp}$. The first check that an embedding must pass is that all the charges of the states in the global model appear for some of the curves in \eqref{u1charges}. Generally, for a given global model to be embeddable into $E_8$ the number of global massless $U(1)$s must not exceed 4, the rank of the $SU(5)_{\perp}$. If a  model has less than $4$ $U(1)$s, this means that once we specify an embedding, the matter states in \eqref{u1charges} will not all have different charges under the global $U(1)$s. This raises the possibility that Yukawa couplings between matter curves could be allowed by the global symmetries but forbidden by the full Cartan $U(1)$s of $E_8$. However if we identify all the matter curves which have the same charges under the global $U(1)$s then any Yukawa couplings present in the global models will be allowed by the embedding into $E_8$.

We expect that in this way, embedding the global $U(1)$s and then identifying matter curves with equal global $U(1)$ charges, all the global models can be embedded into $E_8$. However, the subtlety lies with the identification of the matter curves which have equal charges under the global $U(1)$s but different charges under the full Cartan: a global decomposition of $E_8$ over $S_{}$ does not allow for all such possible identifications but only a subset of them. This subset is the set of identifications than can be reached by identifying two $t_i$'s. Physically one can view them as giving a vev to the singlet $\un_{t_i-t_j}$ and thereby recombining the curves that have a cubic interaction $\un\,  \f \, \fb$ or $\un \,  \te \, \teb$. Suppose the embedding of the global $U(1)$s of a model into the Cartan is such that the matter curves with equal charges under the global $U(1)$s  do not differ by just $t_i-t_j$ in their Cartan charges \eqref{u1charges}. Then this model cannot be described by $E_8$ or 
spectral cover models, and we will denote such models as not embeddable into $E_8$. Physically this arises because a global decomposition of the adjoint of $E_8$ over $S_{}$ does not contain the appropriate singlets to recombine the curves. We will show that the local limit of one of our models falls into the class of such theories which are not embeddable into $E_8$, as defined above. 

\subsection{Embedding the global models in \texorpdfstring{$E_8$}{E8}}
\label{sec:embede8top6}

Before proceeding with the group theory analysis for our two $U(1)$ models there is an important restriction on the global models, discussed previously in Section~\ref{sec:Flatness}, but which is worth highlighting again. The elliptic fibrations studied in this work are non-flat at particular points in the base where three divisors intersect. To have a flat fibration this intersection point must be absent. As discussed in Section~\ref{sec:Flatness} in principle there are two ways to do this: the first is to find a base where the particular sections do not intersect, and the second is to set one of the sections that appear in the intersection to be trivial. The second method is sufficient but not necessary in principle, though in practice it is the only way we have been able to construct flat fibrations. The features of the embedding into $E_8$ are strongly dependent on the method of ensuring flatness of the fibration. If  flatness is ensured by turning off the section we find that all the charges of the 
matter in all the models have a global embedding into a Higgsed $E_8$ theory, but there is a Yukawa coupling in the model of top 4 which does not. We discuss this case in detail in this section. If flatness of the fibration could be somehow ensured by avoiding the intersection point of the sections, rather than turning the full section off, then we find that the Yukawa coupling of top 3 also can not be embedded into $E_8$ and that also the charges of the matter states in top 2 can not be embedded into $E_8$. 

Now let us proceed to study the models with one of the sections turned off. For the analysis of embedding into $E_8$ it is useful to work in a different $U(1)$ basis where the charges can be brought into the form shown in Table~\ref{tab:topcharges}. The crossed-out states correspond to matter curves that are turned off for ensuring  flatness of the fibration. The Yukawa couplings in the models are such that all the cubic couplings allowed by the $U(1)$ charges are present. The $U(1)$ charges and Yukawa couplings of the models comprise our global data which we now want to embed into a local Higgsed $E_8$ model.
\begin{table}
\center
\begin{tabular}{|c|c|c|c|c|}
\hline
State  & Top 1 & Top 2 & Top 3 & Top 4 \\
\hline
$\te$ & $\left(-1,0\right)$ & $\left(-1,0\right)$ & $\left(-1,0\right)$ &  $\left(-2,0\right)$  \\
\hline
\hline
$\fb_1$ & $\left(3,1\right)$ & $\cancel{\left(3,-2\right)}$ & $\left(3,-1\right)$ &  $\cancel{\left(-4,1\right)}$  \\
\hline
$\fb_2$ & $\left(-2,0\right)$ & $\left(3,0\right)$ & $\cancel{\left(-2,-1\right)}$ & $\left(-4,0\right)$   \\
\hline
$\fb_3$ & $\left(-2,-2\right)$ & $\left(3,-1\right)$ & $\left(-2,1\right)$ & $\left(1,1\right)$  \\
\hline
$\fb_4$ & $\left(3,2\right)$ & $\left(-2,0\right)$ & $\left(3,0\right)$ &  $\left(1,-1\right)$ \\
\hline
$\fb_5$ & $\left(-2,-1\right)$ & $\left(-2,1\right)$ & $\left(-2,0\right)$ & $\left(1,0\right)$ \\
\hline
\end{tabular}
\caption{Matter curves and charges for the four $SU(5) \times U(1) \times U(1)$ models  - see Appendix~\ref{app-SU5}. The entries are the charges under the 2 global $U(1)$ symmetries. The strike-through entries are matter curves which are turned off for ensuring  flatness of the fibration.}
\label{tab:topcharges}
\end{table}

The charges of the matter sectors in the two $U(1)$ models can be embedded into $E_8$, and for each model the embedding of the two global $U(1)$ symmetries in the Cartan of $SU(5)_{\perp}$ is unique up to permutations. In Table~\ref{tab:e8embed} we present this embedding of the two $U(1)$s and also the charges of the states of $E_8$ in \eqref{u1charges}. 
\begin{table}
\center
\begin{tabular}{|c|c|c|c|}
\hline
      & Top 1 & Top 2 / Top 3& Top 4 \\
\hline
\hline
$U(1)_1$ & $-t^1-t^2-t^3-t^4+4t^5$ & $-t^1-t^2-t^3-t^4+4t^5$  &  $-2t^1-2t^2+3t^3+3t^4-2t^5$  \\
\hline
$U(1)_2$ & $-t^3-t^4+2t^5$ & $t^4-t^5$ &  $t^4-t^5$  \\
\hline
\hline
$\te_A : t_1$ & $\left(-1,0\right)$ & $\left(-1,0\right)$ & $\left(-2,0\right)$   \\
\hline
$\te_B : t_2$ & $\left(-1,0\right)$ &  $\left(-1,0\right)$ & $\left(-2,0\right)$  \\
\hline
$\te_C : t_3$ & $\left(-1,-1\right)$ &  $\left(-1,0\right)$ & $\left(3,0\right)$  \\
\hline
$\te_D : t_4$ & $\left(-1,-1\right)$ &  $\left(-1,1\right)$ & $\left(3,1\right)$  \\
\hline
$\te_E : t_5$ & $\left(4,2\right)$ &  $\left(4,-1\right)$ & $\left(-2,-1\right)$   \\
\hline
\hline
$\fb_A: t_1 + t_2$ & $\left(-2,0\right)$ & $\left(-2,0\right)$ & $\left(-4,0\right)$    \\
\hline
$\fb_B: t_1 + t_3$ & $\left(-2,-1\right)$ & $\left(-2,0\right)$ & $\left(1,0\right)$   \\
\hline
$\fb_C: t_1 + t_4$ & $\left(-2,-1\right)$ & $\left(-2,1\right)$ & $\left(1,1\right)$   \\
\hline
$\fb_D: t_1 + t_5$ & $\left(3,2\right)$ & $\left(3,-1\right)$ & $\left(-4,-1\right)$   \\
\hline
$\fb_E: t_2 + t_3$ & $\left(-2,-1\right)$ & $\left(-2,0\right)$ &$\left(1,0\right)$   \\
\hline
$\fb_F: t_2 + t_4$ & $\left(-2,-1\right)$ & $\left(-2,1\right)$ & $\left(1,1\right)$   \\
\hline
$\fb_G: t_2 + t_5$ & $\left(3,2\right)$ & $\left(3,-1\right)$ & $\left(-4,-1\right)$   \\
\hline
$\fb_H: t_3 + t_4$ & $\left(-2,-2\right)$ & $\left(-2,1\right)$ & $\left(6,1\right)$   \\
\hline
$\fb_I: t_3 + t_5$ & $\left(3,1\right)$ & $\left(3,-1\right)$ & $\left(1,-1\right)$   \\
\hline
$\fb_J: t_4 + t_5$ & $\left(3,1\right)$ & $\left(3,0\right)$ & $\left(1,0\right)$   \\
\hline
\end{tabular}
\caption{Embedding of the four $SU(5) \times U(1) \times U(1)$ models  - see Appendix~\ref{app-SU5} -  into $E_8$. The entries are the charges under the 2 global $U(1)$ symmetries. Note that it is possible to recombine all the tops along $t_1 \leftrightarrow t_2$, top 1 also along $t_3 \leftrightarrow t_4$, and tops 2 and 3 also $t_1 \leftrightarrow t_2 \leftrightarrow t_3$.}
\label{tab:e8embed}
\end{table}
The embeddings given in Table~\ref{tab:e8embed} are appropriate for the case where the non-flat point is avoided by switching off the appropriate section and therefore also some of the matter curves as in Table~\ref{tab:topcharges}. It is important to note though that if we were not to switch off those matter representations the embeddings would be modified: the charges of top 3 would now differ from top 2, but would still be embeddable by choosing $U(1)_2=-t^1-t^2+4t^3-t^4-t^5$. The charges of top 2 on the other hand would not be embeddable into $E_8$, at least not if we require the appropriate $\f \, \te \, \te$ coupling to be present.

We now explain why the above embeddings are the appropriate choices. To be explicit, let us attempt to embed top 2.
Since we have a symmetry between the $t_i$s we can choose the $\te$ to correspond to $t_1$ in \eqref{u1charges}. Therefore its neutrality under the second $U(1)$ implies $a^2_1 = 0$. Since there is a coupling $\te \, \te \, \f_4$, we should take another $\te$ to have equal charge so that one can form such a coupling, so we take $\te$ coming from $t_2$ and impose that it has the same charges under the 2 $U(1)$s as $t_1$. This implies that it is possible to recombine the two curves $t_1$ and $t_2$ without breaking the 2 $U(1)$s. This in turn identifies $\fb_4$ as having charges $t_1+t_2$, and sets $a^2_1 = a^2_2 = 0$. Now we note that the state $\fb_2$ is also neutral under $U(1)_2$ but has different charges under $U(1)_1$ from $\fb_4$. Therefore it must be some other state in \eqref{u1charges}. There are two such candidate states: $t_4+t_5$ is neutral if we set $a_4^2=-a_5^2$, and also $t_1+t_3$ if we set $a^2_3=0$, but the tracelessness constraint forces one to imply the other, $\left\{a_4^2=-a_5^2\right\} \
implies \left\{a^2_3=0\right\}$. So there is really no choice (up to permutations of $\left\{t_3,t_4,t_5\right\}$, which is a choice of basis). Therefore we are lead to
\begin{equation}
U(1)_2 = t^4 - t^5 \;. \label{genericu12}
\end{equation}
Once $U(1)_2$ is fixed this way, the embedding of $U(1)_1$ follows straightforwardly. The argument just presented to determine $U(1)_2$ also holds unmodified for tops 3 and 4.

Having fixed $U(1)_2$ as \eqref{genericu12} for top 2, it is simple to check that it is not possible to reproduce the $-2$ charge of $\fb_1$ from the charges \eqref{u1charges}. Hence, unless we turn off that matter curve, as may be required to maintain the flatness of the fibration, the charges cannot be embedded into $E_8$.

We now consider the embeddings of the Yukawa couplings into $E_8$. At first sight it might seem that since we can embed the charges of the states in Table~\ref{tab:topcharges} into $E_8$ the Yukawa couplings should automatically also follow. However, as discussed at the start of this section, this relies on whether the appropriate recombinations can be performed to break from the Cartan of $SU(5)_{\perp}$ to the two global $U(1)$s. The important point is that a Higgsed $E_8$ theory relates the splitting of the matter curves to their charges under the Cartan of $SU(5)_{\perp}$ such that two curves with the same charges under the global $U(1)$s but different charges under the Cartan would be distinct. On the other hand, from a global perspective the splitting of the curves, as appears in the local limit of the determinant, is fixed by their charges under the global $U(1)$s, so that the additional splitting due to the differing Cartan charges should not be present if the theories are to match generally (i.e. in 
all points in moduli space). Therefore the Higgsed $E_8$ theories can only match the local limit of the global models if all the curves which have the same charges under the global $U(1)$s have been recombined.

For the cases of tops 1, 2 and 3 we see that this is the case because there are two recombinations possible for each top thereby breaking the $U(1)^4 \rightarrow U(1)^2$. Hence the number of local selection rules and global $U(1)$ charges match in those cases. Therefore all the appropriate Yukawas are reproduced. Top 4 is different because only one possible recombination within $E_8$ is possible, $t_1 \leftrightarrow t_2$. Therefore there is one additional local selection rule that is not captured by the 2 global $U(1)$ charges of the states. And this selection rule effectively forbids the Yukawa coupling $\fb_5 \,  \fb_5 \, {\bf 10}$ which is present in the global model as we show below. 

On closer inspection we see that there are two possible embeddings of the $\fb_5$ state of top 4 into $E_8$, as $\fb_B$ or as $\fb_J$ in Table~\ref{tab:e8embed} ($\fb_B$ and $\fb_E$ have been recombined already). But neither embedding has a gauge neutral, under the additional local selection rule, Yukawa coupling to match $\fb_5 \,  \fb_5 \, \te$, as can be seen from the charges under the Cartan. However the coupling $\fb_B \, \fb_J \, {\te}_B$ is allowed. Therefore in order to reproduce the correct intersection structure we should recombine the matter curves $\fb_B$ and $\fb_J$, but the crucial point is that there is no $E_8$ singlet that can do this by forming the appropriate gauge invariant cubic coupling $\un \, \f_B \, \fb_J$: all $E_8$ singlets take the form $t_i-t_j$. In other words the model can not be embedded into a global breaking of $SU(5)_{\perp}\rightarrow S\left[U(2)\times U(1)\times U(1)\times U(1)\right]$ nor into $SU(5)_{\perp}\rightarrow S\left[U(2)\times U(2)\times U(1)\right]$ or $SU(5)_
{\perp}\rightarrow S\left[U(3)\times U(1)\times U(1)\right]$. Therefore the Yukawa couplings do not have a global embedding into a Higgsed $E_8$ gauge theory.

We have also performed an analysis of single $U(1)$ models in the literature. As a simple example let us consider the model of a single $U(1)$ presented in \cite{Braun:2013yti}, to show how its charges are embedded into $E_8$. The states in the theory have charges
\begin{equation}
Q\left(\te\right)= -1\;,\;\; Q\left(\fb\right)= 8\;,\;\; Q\left(\fb\right)= 3\;,\;\; Q\left(\fb\right)= -2\;,\;\; Q\left(\fb\right)= -7\;.
\end{equation}
It is simple to find some $a_i$ in $U(1) = \sum_{i=1}^5 a_i t^i$ that reproduces this for the charges \eqref{u1charges}, for example
\begin{equation}
a_1 = -1\;,\;\; a_2 = -1\;,\;\; a_3 = -1\;,\;\; a_4 = -6\;,\;\; a_5 = 9\;. \label{embeddingbraun}
\end{equation}
Therefore the charged spectrum of the model in \cite{Braun:2013yti} can be embedded into $E_8$. It can be checked that also the Yukawas of the embedding and the model match. We have also checked that all the single $U(1)$ global models presented in \cite{Borchmann:2013jwa} can also be embedded into $E_8$, both the charges and Yukawa couplings. 

\subsection{Recombining beyond \texorpdfstring{$E_8$}{E8}}
\label{sec:recombe8}

In this section we make some general remarks about the implications of the existence of embeddings of $U(1)$s into the Cartan of $SU(5)_{\perp}$ such that the matter curves with duplicate charges do not differ just by $t_i-t_j$, top 4 in Table~\ref{tab:e8embed} being an example of such an embedding, although many more possibilities exist. First we note that any two $\te$-matter curves differ in their Cartan charges by $t_i-t_j$ and therefore this possibility can only exist for $\f$-matter curves. The relevant study is therefore that of $\f$-matter curves and their intersection structure.

We work with the formalism introduced in Section~\ref{sec:embede8top6} where we think about embeddings into $E_8$ in terms of the Cartan charges. In particular we consider the fact that all global embeddings into $E_8$, and therefore all spectral cover models, can be understood by starting from the complete breaking of $SU(5)_{\perp}$ to its Cartan and then subsequently recombining matter curves by using the $E_8$ GUT singlets. This corresponds to turning on (in a D-flat manner) off-diagonal components of the adjoint Higgs in $SU(5)_{\perp}$, eventually classifying all the possible Higgs backgrounds (where the Higgs commutes with its conjugate) by recombining all the way to no remaining $U(1)$s. In thinking this way we have seen, in Section~\ref{sec:embede8top6}, that the missing ingredient in embedding top 4 into $E_8$ is a GUT singlet that is able to recombine the two $\f$-matter curves $\fb_B=t_1+t_3$ and $\fb_J=t_4+t_5$. In this section we will argue that actually such singlets, which go beyond $E_8$, 
are present in global F-theory constructions, first examples of which have been identified already in \cite{Mayrhofer:2012zy}. Recombining matter curves using these singlets could account for the embedding into $E_8$ of the global models.

Consider the possible operators that GUT singlets can form with $\fb$-matter curves of type $\un \, \f \, \fb$. Such operators are localised at points on $S_{}$ where two $\f$-matter curves intersect and a GUT singlet, which is localised on a locus in the bulk of the Calabi-Yau, also intersects $S_{}$. This leads to the following puzzle: In view of the charges under the Cartan of $SU(5)_{\perp}$ of the $\f_i$ and GUT singlets, as given in \eqref{u1charges}, not all the possible pairs of $\f_i$ can form a gauge invariant cubic coupling with a singlet. For example there is no singlet with appropriate charges to couple $\fb_A=t_1+t_2$ and $\f_H=-t_3-t_4$. This is puzzling because we have just argued that such cubic couplings occur at points on $S_{}$ and generically two curves intersect at a point on a surface. Therefore generically $\fb_A$ and $\f_H$ will intersect on $S_{}$ and one wonders what happens at this point if no such cubic coupling is possible? Or in terms of enhancement of gauge groups, 
at this point of intersection, what gauge group do we enhance to since an enhancement to $SU(7)$ requires the presence of a singlet to complete the adjoint
\begin{equation}
\bf{48} \rightarrow \bf{24}^{(0,0)} \op \un^{(0,0)} \op \un^{(0,0)} \op \left(\f \op \fb\right)^{(-1,0)} \op \left(\f \op \fb\right)^{(0,1)} \op
\left({\bf 1} \op {\bf \overline{1}}\right)^{(1,-1)} \;.
\end{equation}

This puzzle is an artifact of a deeper aspect of $\f$-matter curves in local models. This aspect is the fact that enhancement loci at intersections of $\f$-matter curves can not be determined in a local theory defined by taking the leading terms in the coordinate $w$ normal to the $SU(5)$ divisor $S$ in the sections defining the fibration. To see this consider a Tate model
\begin{equation}
y^2 = x^3 + a_1 x y z + a_{2}  x^2 z^2 + a_{3}  y z^3 + a_{4}  x z^4 + a_{6}  z^6 \; \label{tate}
\end{equation}
with
\bea \label{aijdef}
a_2 = a_{2,1} w, \qquad a_3 = a_{3,2} w^2, \qquad a_4 = a_{4,3} w^3, \qquad a_6 = a_{6,5} w^5.
\eea
Its discriminant can be written as
\begin{eqnarray}
\Delta = - w^5 \left[P^4_{10} P_5 + w P^2_{10} \left(8a_{2,1}P_5 + P_{10} R\right) + 2 w^2\left( -8 a_{3,2}^2 a_{2,1}^3 + {\cal O}\left(P_{10}\right) \right) + {\cal O}\left(w^3\right) \right] \;, \label{discp5}
\end{eqnarray}
where we define
\begin{eqnarray}
P_{10} &=& a_1 \;, \\
P_5 &=& a_{3,2}^2 a_{2,1} - a_{4,3} a_{3,2} a_1 + a_{6,5} a_1^2 \;, \\ \nn
R &=& -a_{3,2}^3-a_{4,3}^2a_1 + 4a_{6,5}a_{2,1}a_1 \;.\label{p5p10}
\end{eqnarray}
We recall here that an enhancement to $SO(12)$ and $E_6$, which are associated to Yukawa couplings, require a vanishing order for the discriminant of 8, while an enhancement to $SU(7)$ requires a vanishing order of 7. Now we should think about the discriminant as evaluated by the leading order behaviour, in $w$, of the Tate coefficients $a_{i}$. The $\te$-matter curves are associated to a vanishing of $P_{10}$, while the $\f$-matter curves to a vanishing of $P_5$. It can be seen that a vanishing of both $P_{10}$ and $P_{5}$ implies a vanishing of the discriminant to order $8$, independent of the higher order corrections to the $a_{i}$. Therefore such points can be determined purely locally. Since the $\un\,  \te \, \teb$ coupling is also associated to an $E_6$ point it can also be determined locally. However the $\un \, \f \, \fb$ coupling associated to $SU(7)$ can not be determined locally because any term subleading in one power of $w$ in the $a_{i}$ can influence the result of whether the second term in
\eqref{discp5} vanishes or not. More precisely since $P_{10} \neq 0$ we require that $R=0$ for an $SU(7)$ enhancement, but the leading order form of $R$ receives corrections from subleading order corrections to $a_{i}$ coming from the first term in \eqref{discp5} . 

Just to be explicit let us show how this occurs in a global context using an example from this paper. For the models studied in this paper it is actually possible to write them in Tate form globally: generally, a map that takes an elliptic fibration based on a cubic in $\mathbb P^2[3]$ as given by (\ref{eq:hyper1}) to Tate form
 is
\begin{eqnarray}
{a}_1 &=& b_1 \;,  \nn\\
{a}_2 &=& -\left(b_2 b_0 + d_1 c_2 + d_0 c_1 \right) \;,  \nn\\
{a}_3 &=& -\left(b_2 d_0 c_2 + d_1 b_0 c_1 + d_2 c_2 c_1 \right) \;, \label{cubictotate} \\  
{a}_4 &=& b_2 d_1 b_0 c_2 + d_2 b_2 c_2^2 + b_2 b_0 d_0 c_1 + d_1 d_0 c_2 c_1 + d_2 b_0 c_1^2 \;,  \nn\\
{a}_6 &=& -\left(d_2 b_2^2 b_0 c_2^2 + b_2 d_1 b_0 d_0 c_2 c_1 - d_2 b_2 b_0 b_1 c_2 c_1 + \right. \nn \\ & &
   \left. d_2 b_2 d_0 c_2^2 c_1 + d_2 b_2 b_0^2 c_1^2 + d_2 d_1 b_0 c_2 c_1^2\right)\;.  \nn
\end{eqnarray}
Consider the model based on top 2 as listed in table \ref{top1and2} of the appendix. The complete form of the Tate coefficients as determined by \eqref{cubictotate} reads
\begin{eqnarray}
{a}_1 &=& b_1 \;, \\
{a}_2 &=& - d_1 w - c_1 d_{0,2} w^2 - b_{0,3} b_2 w^3 \;, \nn\\
{a}_3 &=& -c_1 d_{2,1} w^2 - b_2 d_{0,2} w^3 - b_{0,3} c_1 d_1 w^3 \;, \nn\\
{a}_4 &=& c_1 d_{0,2} d_1 w^3 + b_2 d_{2,1} w^3 + b_{0,3} b_2 d_1 w^4 + 
 b_{0,3} c_1^2 d_{2,1} w^4 + b_{0,3} b_2 c_1 d_{0,2} w^5 \;, \nn\\
{a}_6 &=& b_{0,3} b_1 b_2 c_1 d_{2,1} w^5 - b_2 c_1 d_{0,2} d_{2,1} w^5 - 
 b_{0,3} c_1^2 d_1 d_{2,1} w^5 - b_{0,3} b_2 c_1 d_{0,2} d_1 w^6 -  \\
&&  b_{0,3} b_2^2 d_{2,1} w^6 - b_{0,3}^2 b_2 c_1^2 d_{2,1} w^7 \;.\nn \label{cubictotatetop2}
\end{eqnarray}
Now consider the discriminant on the locus $d_{2,1}=0$, which is one of the $\f$-matter curves. If we evaluate the next-to-leading order piece in $w$ of the discriminant, as in \eqref{discp5}, using the leading order behaviour of the $a_{i}$ we find
\begin{equation}
\left.\Delta\right|_{d_{2,1}=0} = w^6 b_1^4 c_1^2 d_{0,2}^2 d_1^2 + {\cal O}(w^7)\;, \label{wrongsu7}
\end{equation}
while including the next order corrections to the $a_{i}$ we get
\begin{equation}
\left.\Delta\right|_{d_{2,1}=0} = w^6 b_1^4 c_1 d_{0,2} (b_{0,3} b_1 - d_{0,2}) d_1 (b_1 b_2 - c_1 d_1) + {\cal O}(w^7)\;. \label{rightsu7}
\end{equation}
The difference between the two results is crucial. Consider the intersection of the two $\f$-matter curves $\f_3\;:\;d_{2,1}=0$ and $\f_4\;:\;b_1 b_2 - c_1 d_1=0$. Using \eqref{wrongsu7} we find no $SU(7)$ enhancement at that point, while using \eqref{rightsu7} yields the correct enhancement.

Returning to the implications of the fact that $SU(7)$ points can not be determined locally, in the sense of the leading order behaviour of the $a_{i}$, this means that a local theory should not, in general, be able to split two $\f$-matter curves that have equal charges under any global $U(1)$s. Or in other words, one should have a local theory for every embedding of any number (up to 4) of $U(1)$s into the Cartan of $SU(5)_{\perp}$ where all the matter curves only factorise according to the $U(1)$ charges, with no additional factorisation coming from the additional Cartan charges. This set of local theories is larger than those that come from a global Higgsing of $E_8$ precisely by the subset of theories that correspond to identifying, or recombining, two $\f$-matter curves which do not differ by $t_i-t_j$ in their Cartan charges. The local limit of top 4 was one such example.

Thinking in terms of recombination singlets, we therefore expect that in any local limit of a global theory, for any two $\f$-matter curves with different $U(1)$ charges there should be an associated singlet at their intersections that enhances to $SU(7)$. Generally such singlets can not be embedded inside a single global decomposition of the adjoint of $E_8$ over $S_{}$ and in that sense go beyond $E_8$. Recombining with such singlets leads to a Yukawa structure in the local model which goes beyond $E_8$. One should be able to see directly the presence of such singlets in global models, and in fact examples of them have already been identified in \cite{Mayrhofer:2012zy} within global Factorised Tate models. 

Factorised Tate models, as defined in \cite{Mayrhofer:2012zy}, are a nice testing ground for these ideas in the sense that they are constructed as global extensions of local models based on Higgsing $E_8$. They therefore have the same structure as local $E_8$ theories in the $a_{i}$ (and in fact include the subset of possible models where the $a_{i}$ have no subleading corrections in $w$). Although only the models with a single additional $U(1)$ were resolved in detail in \cite{Mayrhofer:2012zy}, the singular forms of the fibrations were given for all the possible extensions of local $E_8$ models. In the singular F-theory limit we expect to be able to use Tate's algorithm to determine the enhancement loci and this is sufficient for our purposes. Using this one can explore the presence of the appropriate recombination singlets through the enhancement loci to $SU(7)$ and $SU(2)$ and confirm that the appropriate singlets are present for all the models.

In more detail one finds that generically all pairs of $\f$-matter curves intersect, and that the intersections fall into two classes: those that correspond to pairs of $\f$s that can form a gauge invariant operator with an $E_8$ GUT singlet, and those that can not. Let us call the latter points $Q_a$s, where the index runs over the number of such points. Then the $Q_a$ points in turn split into pairs $QX_a$ and $QY_a$. At the $QX_a$ points we have enhancements to $SO(12)$, which correspond to couplings of type $\fb \, \fb \, \te$, and at the $QY_a$ points we have enhancements to $SU(7)$. Now a crucial aspect is that to each $QY_a$ point we can associate a $\te$-matter curve such that the two can never coincide, otherwise one induces a non-Kodaira singularity in the fibre.

This latter property hints at an interpretation of the presence of these singlets as a non-global (over $S_{}$) embedding into $E_8$. Since the points on $S_{}$ where such singlets intersect can never coincide with the associated $\te$-matter curve (as associated above), one can not compare their embedding inside a global decomposition of the adjoint of $E_8$. Therefore their presence is associated to the fact that the embedding of the states into $E_8$ can vary over $S_{}$, so that although the full spectrum can not be embedded into $E_8$, pointwise one can always do so.

Before we outline some further results it is worth keeping in mind that Higgs backgrounds which preserve an Abelian subgroup of $SU(5)_{\perp}$ map to Tate models that factorise so that we can write the elliptic fibration as \cite{Mayrhofer:2012zy}
\begin{equation} \label{FacTate}
X Q = z \prod_i^n Y_i \;.
\end{equation}
The $Y_i$ are some holomorphic polynomials in the sections of the base, $z$, $x$ and $t\equiv y/x$ whose degrees in $t$ sum to 5. Note that $t^{-1}$ plays the role of the coordinate usually referred to as $s$ in the local spectral cover limit,
in which the spectral cover is expressed as \footnote{The spectral cover sections $\hat b_i$ are the local limit of the Tate sections $a_{i,j}$ appearing in (\ref{aijdef}), i.e. in the local limit we identify $\hat b_0 \leftrightarrow a_{6,5}$, $\hat b_2 \leftrightarrow a_{4,3}$, $\hat b_3 \leftrightarrow a_{3,2}$, $\hat b_4 \leftrightarrow a_{2,1}$, $\hat b_5 \leftrightarrow a_{1}$.}
 \begin{equation}
SC = \hat b_5 + \hat b_4 s +  \hat b_3 s^2 + \hat b_2 s^3 + \hat b_0 s^5 = 0 \;. \label{spec}
\end{equation}
The factorised Tate models map to split spectral cover constructions as studied in \cite{Marsano:2009wr}.

We have the picture that the classification of $\un \,  \f \, \fb$ couplings should not be done by looking at the gauge invariant combinations in $E_8$ but rather by considering all the possible $\f$-matter curve pairs in $E_8$ and associating to each pair a singlet which makes the coupling gauge invariant. The charges of all such singlets can always be constructed by adding the charges of two $E_8$ singlets. Indeed, it is possible to see this explicitly in the geometry as follows. In general the polynomials $Y_i$ are not linear in $t$. The points on $S_{}$ where GUT singlets intersect it are the points where a root from one polynomial $Y_1$ say, and another $Y_2$, coincide and we enhance to at least $SU(2)$. To each such collision of roots we can associate an $E_8$ singlet, whose charge is embedded in the Cartan of $SU(5)_{\perp}$ as $t_i-t_j$. The indices $i$ and $j$ can be associated to the polynomials whose roots coincide. Now one thing to consider are points where more than 2 roots coincide, and in 
particular the possibility that there are points where 4 roots, of the polynomials $Y_i$, coincide. If such points existed we should expect to find there two types of singlets, and so the point would have a 'charge' corresponding to adding the charges of the two singlets. These are exactly the type of points which we labelled $QY_a$ above. One may think of the two $E_8$ singlets as forming a new doubly charged singlet not in $E_8$, or more formally we have that the intersection of the bulk singlet loci intersect $S_{}$ at more points than $E_8$ gauge invariant operators would predict. However unlike the argument that two $\f$-matter curves generically intersect at a point, four roots of complex equations generically do not coincide at a point. The difference is resolved by the fact that the polynomials $Y_i$ are not all independent. Rather there is a tracelessness constraint, which ensures the absence of a linear term in $t$ in their product. This relation between the polynomials ensures that there are 
points where 4 roots coincide.

Let us look at an explicit example studied in detail in \cite{Mayrhofer:2012zy} which, as a Factorised Tate model, is based on a 3-2 splitting $X Q = z Y_1 Y_2$ with $Y_1$ and $Y_2$ of degree 2 and 3 in $t$. Therefore, locally it flows to the spectral cover given by\footnote{The sections $\hat c_i, \hat d_i$ are called $c_i, d_i$ in \cite{Mayrhofer:2012zy}. We include the $\hat{}$ to avoid confusion with the sections appearing in the definition of the cubic in $
{\rm Bl}_2\mathbb P^2[3]$ or $
\mathbb P^2[3]$.}
\begin{equation}
SC=\left(\hat c_2 + \hat c_1 s + \hat c_0 s^2\right)\left(\hat d_3 + \hat d_2 s + \hat d_1 s^2 + \hat d_0 s^3\right) = 0 \;. \label{locdecom}
\end{equation}
Such a local model was first studied in \cite{Marsano:2009wr}. The tracelessness constraint ensuring the absence of a linear term in $t$ can be solved in a number of ways, one possibility being the one presented in \cite{Dolan:2011iu} which imposes
\begin{eqnarray}
\hat d_1 &=& - \gamma \hat c_1 \;, \nn \\ \label{32trace}
\hat d_0 &=& \gamma \hat c_0 \;,
\end{eqnarray}
with $\gamma$ some arbitrary section (this corresponds to setting $\alpha=1$ in the solution of \cite{Mayrhofer:2012zy}). It was shown that if one extends the local sections $\hat c_i$ and $\hat d_i$ to global ones, the full global Tate fibration can be written in the form
\begin{equation}
X Q = z Y_1 Y_2 \;.
\end{equation}
$Y_1$ and $Y_2$ take the explicit form\footnote{We have set $z=e_0=1$ in the notation of \cite{Mayrhofer:2012zy} as they do not play a role.}
\begin{equation}
Y_1 = \hat c_2 t^2 + \hat c_1 t  + \hat c_0 \;,\;\; Y_2 = \hat d_3 t^3 + \hat d_2 t^2  - \hat c_1 \gamma t + \hat c_0 \gamma \;. 
\end{equation}
The matter curves are charged under a single $U(1)$ and are given by the sections
\begin{eqnarray}
\te_{t_1} &:& \hat c_2, \\
\te_{t_3} &:& \hat d_3, \\
\fb_{2t_1} &:& \hat c_1, \\
\fb_{2t_3} &:& \hat c_0 \hat d_3 + \hat d_2 \hat c_1, \\
\fb_{t_1+t_3} &:& \hat c_2 \hat d_2^2 + \hat c_0 \hat d_3^2 + \hat d_2 \hat d_3 \hat c_1 - 2 \hat c_2^2 \hat d_2 \gamma - 
\hat c_2 \hat d_3 \hat c_1 \gamma + \hat c_2^3 \gamma^2 \;.
\end{eqnarray}
We have denoted the Cartan charges in subscripts as embedded into a Higgsed $E_8$ model with $t_1 \leftrightarrow t_2 $ and $t_3 \leftrightarrow t_4 \leftrightarrow t_5$ recombined. 

We can now identify the required properties. We consider the intersection of $\fb_{2t_1}$ and $\fb_{2t_3}$ and see that it decomposes into two loci
\begin{equation}
QX = \hat c_1 \cap \hat d_3 \;\;,\;\; QY = \hat c_1 \cap \hat c_0 \;.
\end{equation}
At $QX$ we have the $E_8$ $\fb \, \fb \, \te$ coupling, while at $QY$ we would predict a $\un \, \f \, \fb$ coupling outside $E_8$. Now we note that indeed $Y_1$ and $Y_2$ have 2 roots each coinciding at that point, so that overall there are 4 roots and so a doubly charged singlet as required to make the gauge invariant operator. We also see that the $\te$-matter curve $c_2$ and QY can not coincide otherwise the full $Y_1$ vanishes signalling a non-Kodaira singularity.

Finally, an important feature of the point $QY$ is that $\hat b_0 = \hat c_0 \hat d_0 = 0$ while $\hat b_5 \neq 0$. The vanishing of $\hat b_0$ is common to the classs of points $QY_a$ and may be understood as a signal for going beyond $E_8$. It would be interesting to study the relation between $\hat b_0=0$ and $E_8$ further.

\subsection{Embedding into a local split spectral cover}
\label{sec:embedspeccov}

In this section we briefly address the issue of whether the models constructed in this work allow for a split spectral cover model in the local limit. This question arises for tops 1, 2 and 3 where an embedding into a globally Higgsed $E_8$ theory is in principle possible, and therefore their local limit could in principle correspond to a split spectral cover model. The procedure for comparing the local limit of F-theory models with a spectral cover description is a bit subtle. Directly, one can bring the model into local Tate form as in \eqref{tate}, and then map it to the spectral cover given by the equation
\begin{equation}
SC = \hat b_5 + \hat b_4 s +  \hat b_3 s^2 + \hat b_2 s^3 + \hat b_0 s^5 = 0 \;. \label{spec}
\end{equation}
For the models studied in this paper it is actually possible to write them in Tate form globally using the map \eqref{cubictotate}.
The local limit for the Tate sections $a_{i,j}$ is then extracted for each top from the vanishing orders in $w$ of the base sections in the ${\rm Bl}_2\mathbb P^2[3]$-fibration. This yields the local coefficients given in Table~\ref{tab:locb}.
\begin{table}
\center
\begin{tabular}{|c|c|}
\hline
 Local Section  & Top 1 \\
\hline
$\hat b_5$ & $b_1$ \\
\hline
$\hat b_4$ & $-c_{2,1}$  \\
\hline
$\hat b_3$ & $-b_{0,2} c_1$  \\
\hline
$\hat b_2$ & $b_{0,2} b_2 c_{2,1} + c_1 c_{2,1} d_{0,2}$\\
\hline
$\hat b_0$ & $-b_{0,2} b_2 c_1 c_{2,1} d_{0,2} + b_{0,2} b_1 b_2 c_1 c_{2,1} d_{2,2} - b_{0,2} c_1^2 c_{2,1} d_{2,2}$  \\
\hline
\hline
 & Top 2 \\
\hline
$\hat b_5$ & $b_1$   \\
\hline
$\hat b_4$ & $-d_1$  \\
\hline
$\hat b_3$ & $-c_1 d_{2,1}$  \\
\hline
$\hat b_2$ & $c_1 d_{0,2} d_1 + b_2 d_{2,1}$ \\
\hline
$\hat b_0$ & $b_{0,3} b_1 b_2 c_1 d_{2,1} - b_2 c_1 d_{0,2} d_{2,1} - 
 b_{0,3} c_1^2 d_1 d_{2,1}$ \\
\hline
\hline
 & Top 3 \\
\hline
$\hat b_5$ & $b_1$   \\
\hline
$\hat b_4$ & $-d_{0,1}$  \\
\hline
$\hat b_3$ & $-d_1 b_{0,2}$  \\
\hline
$\hat b_2$ & $b_{0,2} b_2 d_{0,1} + c_{2,2} d_{0,1} d_1 + b_{0,2} d_{2,1}$ \\
\hline
$\hat b_0$ & $-b_{0,2} b_2 c_{2,2} d_{0,1} d_1 - b_{0,2}^2 b_2 d_{2,1} + b_{0,2} b_1 b_2 c_{2,2}d_{2,1} - b_{0,2} c_{2,2}d_1 d_{2,1}$ \\
\hline
\hline
 & Top 4 \\
\hline
$\hat b_5$ & $b_1$   \\
\hline
$\hat b_4$ & $-b_2$  \\
\hline
$\hat b_3$ & $-c_{1,1} d_1$  \\
\hline
$\hat b_2$ & $b_2 c_{1,1} d_{0,1} + b_2 c_{2,2} d_1$ \\
\hline
$\hat b_0$ & $b_2 c_{1,1} c_{2,2} d_{0,1} d_1 - b_2 c_{1,1}^2 d_{2,1} + b_1 b_2c_{1,1} c_{2,2} d_{2,1}$ \\
\hline
\end{tabular}
\caption{Table showing the Tate coefficients in the local limits of global models with a $U(1)\times U(1)$ Abelian sector. The appropriate sections have been turned off in the global models to ensure the flatness of the fibration.}
\label{tab:locb}
\end{table}

It can be checked that for each of the tops, if we take the local limits for the coefficients in the ${\rm Bl}_2 \mathbb P^2[3]$-fibration, as given in Table~\ref{tab:locb}, in the expression \eqref{spec} there is no factorisation of the polynomials. Hence the particular local limit of these theories in Table~\ref{tab:locb} is not described by a split spectral cover, even though they do have $U(1)$ symmetries. Note that this is the case even for the models that do have a global embedding into $E_8$ as discussed in Section~\ref{sec:embede8top6}.

However, the relation between the local limit of Tate models and the spectral cover is subtle: the results for the local form of the $\hat b_i$ can be modified by coordinate transformations of $y$ and $x$. Explicitly the general transformation
\begin{equation}
y \rightarrow y + p^3 w^3 z^3 + q x w z \;, \;\; x \rightarrow x + p^2 w^2 z^2 \;,
\end{equation}
for some sections $p$ and $q$, modifies the leading order coefficients of the Tate form, while maintaining the $SU(5)$ singularity at $y=x=w=0$. Such a transformation can take a split spectral cover local limit to a non-split one. Therefore the fact that the particular local limit studied here does not lead to a split spectral cover model does not rule out that such a limit exists for some appropriate choice of $p$ and $q$. We have not been able to find an appropriate choice that leads to a local splitting, nor have we shown that no such choice is possible.

\section{Summary}

In this paper we have constructed F-theory compactifications with up to two Abelian gauge groups as initiated in \cite{Borchmann:2013jwa}. Following our general approach developed in \cite{Krause:2011xj,Mayrhofer:2012zy} we have focused on Abelian gauge group factors which appear \emph{generically} in the class of elliptic fibrations under consideration, for every base space $B_3$ with sufficiently many sections so that the fibration can exist. This approach allows us to make generic statements in the framework of well-defined elliptically fibred Calabi-Yau 4-folds without necessitating a scan over concrete base manifolds. 
Specifically we have analysed the implementation of two Abelian gauge group factors by describing the elliptic fibre as the hypersurface ${\rm Bl}_2 \mathbb P^2[3]$, which is the only generic hypersurface representation with two generic $U(1)$s. Different types of models can occur either by implementing the fibre as a complete intersection, or \cite{Braun:2013yti,Braun:2013nqa} by enforcing non-generic constraints on the fibration. 
Even though our fibrations can be described as hypersurfaces, we have demonstrated that an alternative, more complicated resolution of its singular loci in terms of a complete intersection makes one important aspect of these geometries manifest which is obscure in the hypersurface description, namely the existence of a holomorphic zero-section. 
The fact that one such holomorphic section does exist has allowed us to define an embedding of the base space $B_3$ and to construct the $U(1)$ generators and a class of gauge fluxes, which we have then analysed geometrically.

Based on the fact that  ${\rm Bl}_2 \mathbb P^2[3]$ corresponds to one of the 16 polygons analysed in \cite{Bouchard:2003bu} we have  implemented an additional $SU(5)$ symmetry by constructing the four\footnote{We ignore a fifth top which leads to non-flat fibres in codimension two.} possible inequivalent tops \cite{Candelas:1996su,Bouchard:2003bu} for this class of fibrations. We have analysed the matter spectrum, the Yukawa interactions and described the construction of a class of chirality inducing gauge fluxes. Moreover we have shown that it is possible to avoid the notorious points with non-flat fibres in a fully-fledged Calabi-Yau 4-fold by imposing certain restrictions on the fibration. 

Our analysis of ${\rm Bl}_2 \mathbb P^2[3]$-fibrations has some overlap with the work of \cite{Cvetic:2013nia,Cvetic:2013uta}, which also studies such 3-section fibrations. Oftentimes the methods in their approach and ours are complementary. 

The toric technology has also been applied to the implementation of $SU(5)$ symmetries for the remaining polygons which give rise to up to two generic $U(1)$ gauge groups, the results of which we have collected in the appendix. 
Furthermore we have presented the implementation of $SU(4)$-tops. This is motivated by the analysis of \cite{Mayrhofer:2012zy}, which has shown that such $SU(4)$ tops, but with further non-generic constraints on the coefficients of the fibration, can lead to fibrations with two ${\bf 10}$-curves, again provided the final resolution is performed as a complete intersection and certain restrictions on the fibrations are imposed similar to the ones avoiding the non-flat points in this work.

We have studied the local limit of our models and in particular whether it is possible to embed them into a Higgsed $E_8$ theory, a structure that forms the framework for the class of local models considered in the literature to date. We have found that one of our models, the one based on top 4, has a Yukawa coupling that requires a recombination of matter curves which goes beyond $E_8$ in the sense that a global decomposition of the adjoint of $E_8$ over $S_{}$ does not contain the appropriate singlet that can account for such recombination. Such singlets have already been encountered in \cite{Mayrhofer:2012zy} and we have argued that they are generically present in F-theory models. These allow any two $\f$-matter curves to be recombined thereby enlarging the possible local theories that can arise in F-theory beyond the class considered so far in the literature based on $E_8$. 

\subsection*{Acknowledgements}

We thank Sakura Sch{\"a}fer-Nameki and Thomas Grimm for very interesting and useful discussions regarding the local limit of the model presented in \cite{Braun:2013yti}. We also thank Harald Skarke and Roberto Valandro for important discussions. 
The research of EP is supported by a Marie Curie Intra European Fellowship within the 7th European Community Framework Programme and by the Heidelberg Graduate School for Fundamental Physics. This work is supported in part by the Collaborative Research Center TR33 "The Dark Universe".

\newpage

\appendix
\section{More \texorpdfstring{$SU(5)$}{SU(5)} models} \label{app-SU5}
This appendix lists the tops with $SU(5)$-symmetry and their matter content for those of the 16 polygons from \cite{Bouchard:2003bu} which describe elliptic fibrations with up to  one or two extra generic sections. The tops over polygon 5 as described in Section~\ref{sec:Tops-all} are the only ones with two generically present extra sections.\footnote{A non-generic version of this is polygon 9, which we do not discuss in this appendix.} In addition, we have listed all the tops over the polygons that generically possess one extra section. These polygons are number 6, 8 and 11 from \cite{Bouchard:2003bu}. Note that the tops over polygon 6 correspond to the models discussed in \cite{Morrison:2012ei}. Additionally, we have added the $SU(5)$ tops over polygon 3. Although, generically these fibrations do not have an extra section, it is possible to construct special cases in which the rank of the Mordell-Weil group is increased \cite{Braun:2013yti}. The numbering of the tops follows the order in which they are 
presented in the figures for each polygon. Each table lists the following information: 

\begin{itemize}
\item The lower bounds $z_{i}$ for the dual top $\Diamond^{*}$, which, combined with $F_{0}^{*}$, completely specify $\Diamond^{*}$.
\item The proper transform of the hypersurface equation $P_{T}$, which describes the completely resolved elliptic fibration $\hat Y_{4}$ over some base space $B_{3}$, where the fibre is specified by the choice of polygon $F_{0}$.
\item The lowest order of the discriminant $\Delta$, where, for an $A_{n-1}$-singularity over some base divisor $w=0$, the discriminant $\Delta$ is given by an expression of the form $\Delta = w^{n}(P + \mathcal{O}(w))$.
\item The intersection numbers of the sections $S_k,  k=0,1,2$ with the resolution divisors, which are in general given by
\begin{align}
\begin{split}
\int_{\hat Y_{4}}S_{k}\wedge E_{j}\wedge \pi^*\omega_4 = \delta_{ij}\int_{B_{3}}W\wedge \omega_4,
\end{split}
\end{align}
where the specific $\delta_{ij}$ are listed in the table.
\item The Shioda maps $\tw_{i}$ corresponding to $S_{i}, i=1,2$. We will oftentimes denote $S_2$ by $\tU$.
\item The positions of the matter curves including $U(1)$-charges.
\item The positions of the Yukawa points and the couplings, including the position of a possible non-flat fibre.
\end{itemize}
\clearpage

\subsection*{\texorpdfstring{$SU(5)$}{SU(5)} on Polygon 5}
\begin{table}[h] 
\centering
\begin{tabular}{|>{\centering}c|c|c@{$\mspace{6mu}$}|@{$\mspace{6mu}$}c@{$\mspace{4mu}$}|}
\hline
$(z_{3},z_{4},z_{5},z_{6},z_{7})$ & \multicolumn{3}{c|}{$(-1,0,1,0,0)$} \\
\hline
\multirow{3}{*}{$P_{T}$} & \multicolumn{3}{c|}{$0 = b_{0,2} e_{0}^2 e_{1} e_{4}  s_0^2 \T v^2  \T u +c_{2,1} e_{0} e_{1} e_{2}   s_0 \T w \T v^2$} \\
&\multicolumn{3}{c|}{$+ d_{0,2} e_{0}^2 e_{1} e_{3} e_{4}^{2}  \T v  s_0^2 s_1  \T u^{2} + b_1  s_0 s_1 \T w \T v \T u +  c_1 e_{1}e_{2}^2 e_{3}  \T w^{2} \T v s_1$}\\
&\multicolumn{3}{c|}{$ +  d_{2,2} e_{0}^2 e_{1} e_{3}^2 e_{4}^{3}  s_0^2  s_1^2 \T u^{3} + d_1 e_{3} e_{4}  s_0 s_1^2 \T w \T u^{2} + b_2 e_{1} e_{2}^2 e_{3}^2 e_{4}  s_1^{2} \T w^2 \T u$ }\\
\hline
$P$ & \multicolumn{3}{c|}{$\frac{1}{16}b_{0,2}b_1^{4}c_{1}c_{2,1}(b_{1}b_{2} - c_{1}d_{1})(b_{0,2}d_{1}^{2} - b_{1}d_{0,2}d_{1} + b_{1}^{2}d_{2,2})$} \\
\hline
Intersection numbers & $S_{0}: \delta_{j0}$ & $S_{1}:\delta_{j3}$ & $U: \delta_{j1}$ \\
\hline
\multirow{2}{*}{Shioda-map} & \multicolumn{3}{c|}{$\text{w}_{1} = 5( S_1 -  S_{0} - \mathcal{\bar K}) + \sum_{i}m_{i}E_{i},  \quad m_{i}=(2,4,6,3)$}\\
\hhline{~---}
& \multicolumn{3}{c|}{$ \text{w}_{2} = 5(\T U - S_{0} - \mathcal{\bar K} - [ c_{1}] )+ \sum_{i}l_{i}E_{i}, \quad l_{i}=(1,2,3,4)$}\\
\hline
\multirow{4}{*}{Matter curves} & $\{ b_1=0\}$ & $\{c_{2,1}=0\}$ & $\{ c_1=0\}$\\
& $\mathbf{10_{-1,2}}$ + $\mathbf{\overline{10}_{1,-2}}$ & $\mathbf{5_{2,-4}}$ + $\mathbf{\overline{5}_{-2,4}}$ & $\mathbf{5_{2,6}}$ + $\mathbf{\overline{5}_{-2,-6}}$ \\
\hhline{~---}
& $\{ b_{0,2}=0\}$ & $\{b_{1}b_{2} - c_{1}d_{1}=0\}$ & \parbox[c][10mm][c]{40mm}{\center  \vspace*{-4mm}$\{b_{0,2}d_{1}^{2}$ \\$- b_{1}d_{0,2}d_{1} + b_{1}^{2}d_{2,2}=0\}$} \\
& $\mathbf{5_{-3,1}}$ + $\mathbf{\overline{5}_{3,-1}}$ & $\mathbf{5_{-3,-4}}$ + $\mathbf{\overline{5}_{3,4}}$ & $\mathbf{5_{2,1}}$ + $\mathbf{\overline{5}_{-2,-1}}$\\
\hline
\multirow{4}{*}{Yukawa points}  & $\{ b_1= b_{0,2}=0\}$ & $\{ b_1= c_1=0\}$ & $\{ b_1= c_{2,1}=0\}$\\
& $\mathbf{\overline{10}_{1,-2}} \mathbf{5_{2,1}} \mathbf{5_{-3,1}}$ & $\mathbf{10_{-1,2}}\mathbf{\overline{5}_{3,4}}\mathbf{\overline{5}_{-2,-6}}$ & $\mathbf{\overline{10}_{1,-2}}\mathbf{\overline{10}_{1,-2}}\mathbf{\overline{5}_{-2,4}}$ \\
\hhline{~---}
& $\{ b_1= d_1=0\}$ &  & \\
& non-flat fibre &  & \\
\hline
\hline
$(z_{3},z_{4},z_{5},z_{6},z_{7})$ & \multicolumn{3}{c|}{$(-1,0,2,1,0)$} \\
\hline
\multirow{3}{*}{$P_{T}$} & \multicolumn{3}{c|}{$0 = b_{0,3} e_{0}^{3}e_{1}^{2}e_{2}e_{4}^{2}\tu \tv^{2} s_{0}^{2} + d_{0,2}e_{0}^{2} e_{1}^{2}e_{2}e_{4}\tu^{2}\tv s_{0}^{2}s_{1}$} \\
&\multicolumn{3}{c|}{$+ d_{2,1}e_{0}e_{1}^{2}e_{2}\tu^{3}s_{0}^{2}s_{1}^{2} + c_{1}e_{2}e_{3}^{2}e_{4}\tv \tw^{2}s_{1} + b_{1}\tu \tv \tw s_{0} s_{1} $}\\
&\multicolumn{3}{c|}{$+ d_{1}e_{1}e_{2}e_{3}\tu^{2}\tw s_{0}s_{1}^{2} + c_{2,1}e_{0}e_{4}\tv^{2} \tw s_{0} +b_{2}e_{1}e_{2}^{2}e_{3}^{3}e_{4} \tu \tw^{2} s_{1}^{2}$ }\\
\hline
$P$ & \multicolumn{3}{c|}{$\frac{1}{16}b_{1}^{4}c_{1}c_{2,1}(b_{1}b_{2} - c_{1}d_{1})d_{2,1}(b_{0,3}b_{1}^{2} + c_{2,1}(c_{2,1}d_{2,1} - b_{1}d_{0,2}))$} \\
\hline
Intersection numbers & $S_{0}: \delta_{j0}$ & $S_{1}:\delta_{j3}$ & $U: \delta_{j1}$ \\
\hline
\multirow{2}{*}{Shioda-map} & \multicolumn{3}{c|}{$\text{w}_{1} = 5(S_{1} - S_{0} - \mathcal{\bar K}) + \sum_{i}m_{i}E_{i}, \quad m_{i}=(2,4,6,3)$}\\
\hhline{~---}
& \multicolumn{3}{c|}{$\text{w}_{2} = 5(U - S_{0} - \mathcal{\bar K} - [c_{1}] )+ \sum_{i}l_{i}E_{i}, \quad l_{i}=(4,3,2,1)$}\\
\hline
\multirow{4}{*}{Matter curves} & $\{ b_1=0\}$ & $\{c_{1}=0\}$ & $\{ c_{2,1}=0\}$\\
& $\mathbf{10_{-1,-2}}$ + $\mathbf{\overline{10}_{1,2}}$ & $\mathbf{5_{-3,-6}}$ + $\mathbf{\overline{5}_{3,6}}$ & $\mathbf{5_{-3,4}}$ + $\mathbf{\overline{5}_{3,-4}}$ \\
\hhline{~---}
& $\{ d_{2,1}=0\}$ & $\{b_{1}b_{2} - c_{1}d_{1}=0\}$ &\parbox[c][10mm][c]{40mm}{\center  \vspace*{-4mm} $\{b_{0,3}b_{1}^{2} + c_{2,1}(c_{2,1}d_{2,1}$\\ $ - b_{1}d_{0,2})=0\}$} \\
& $\mathbf{5_{-3,-1}}$ + $\mathbf{\overline{5}_{3,1}}$ & $\mathbf{5_{2,4}}$ + $\mathbf{\overline{5}_{-2,-4}}$& $\mathbf{5_{2,-1}}$ + $\mathbf{\overline{5}_{-2,1}}$\\
\hline
\multirow{4}{*}{Yukawa points}  & $\{b_{1}=c_{1}=0\}$ & $\{b_{1}=d_{1}=0\}$ & $\{b_{1}=d_{2,1}=0\}$\\
& $\mathbf{\overline{10}_{1,2}} \mathbf{5_{2,4}} \mathbf{5_{-3,-6}}$ & $\mathbf{10_{1,2}}\mathbf{10_{1,2}}\mathbf{5_{-2,-4}}$ & $\mathbf{\overline{10}_{1,2}}\mathbf{5_{-3,-1}}\mathbf{5_{2,-1}}$ \\
\hhline{~---}
& $\{b_{1}=c_{2,1}=0\}$ &  & \\
& non-flat fibre &  & \\
\hline
\end{tabular}
\caption{Top 1 and 2 for polygon 5. \label{top1and2}}
\end{table}

\begin{table}[h]
\centering
\begin{tabular}{|>{\centering}c@{$\mspace{6mu}$}|c@{$\mspace{6mu}$}|@{$\mspace{6mu}$}c@{$\mspace{6mu}$}|c@{$\mspace{6mu}$}|}
\hline
$(z_{3},z_{4},z_{5},z_{6},z_{7})$ & \multicolumn{3}{c|}{$(-1,1,1,0,0)$} \\
\hline
\multirow{3}{*}{$P_{T}$} & \multicolumn{3}{c|}{$0 = b_{0,2} e_{0}^{2}e_{1}e_{4}\tu \tv^{2} s_{0}^{2} + d_{0,1}e_{0}e_{1}e_{2}\tu^{2}\tv s_{0}^{2}s_{1}$} \\
&\multicolumn{3}{c|}{$+ d_{2,1}e_{0}e_{1}^{2}e_{2}^{3}e_{3}\tu^{3}s_{0}^{2}s_{1}^{2} + c_{2,2}e_{0}^{2}e_{1}e_{3}e_{4}^{2}\tv^{2} \tw s_{0} + b_{1}\tu \tv \tw s_{0} s_{1} $}\\
&\multicolumn{3}{c|}{$+ d_{1}e_{1}e_{2}^{2}e_{3}\tu^{2}\tw s_{0}s_{1}^{2} + c_{1}e_{3}e_{4}\tv \tw^{2}s_{1} +b_{2}e_{1}e_{2}^{2}e_{3}^{2}e_{4} \tu \tw^{2} s_{1}^{2}$ }\\
\hline
$P$ & \multicolumn{3}{c|}{$\frac{1}{16}b_{0,2}b_{1}^{4}c_{1}(b_{0,2}c_{1} - b_{1}c_{2,2})(c_{1}d_{1} - b_{1}b_{2})(b_{1}d_{2,1} - d_{0,1}d_{1})$} \\
\hline
Intersection numbers & $S_{0}: \delta_{j0}$ & $S_{1}:\delta_{j2}$ & $U: \delta_{j2}$ \\
\hline
\multirow{2}{*}{Shioda-map} & \multicolumn{3}{c|}{$\text{w}_{1} = 5(S_{1} - S_{0} - \mathcal{\bar K}) + \sum_{i}m_{i}E_{i}, \quad m_{i}=(3,6,4,2)$}\\
\hhline{~---}
& \multicolumn{3}{c|}{$\text{w}_{2} = 5(U - S_{0} - \mathcal{\bar K} - [c_{1}]) + \sum_{i}l_{i}E_{i}, \quad l_{i}=(3,6,4,2)$}\\
\hline
\multirow{4}{*}{Matter curves} & $\{ b_1=0\}$ & $\{b_{0,2}=0\}$  & $\{ c_1=0\}$\\
& $\mathbf{10_{1,1}}$ + $\mathbf{\overline{10}_{-1,-1}}$ & $\mathbf{5_{3,-2}}$ + $\mathbf{\overline{5}_{-3,2}}$ & $\mathbf{5_{-2,-7}}$ + $\mathbf{\overline{5}_{2,7}}$ \\
\hhline{~---}
& $\{b_{0,2}c_{1} - b_{1}c_{2,2}=0\}$ & $\{c_{1}d_{1} - b_{1}b_{2}=0\}$ & $\{b_{1}d_{2,1} - d_{0,1}d_{1}=0\}$ \\
& $\mathbf{5_{-2,3}}$ + $\mathbf{\overline{5}_{2,-3}}$ & $\mathbf{5_{3,3}}$ + $\mathbf{\overline{5}_{-3,-3}}$& $\mathbf{5_{-2,-2}}$ + $\mathbf{\overline{5}_{2,2}}$\\
\hline
\multirow{4}{*}{Yukawa points}  & $\{b_{1}=b_{0,2}=0\}$ & $\{b_{1}=d_{0,1}=0\}$ & $\{b_{1}=d_{1}=0\}$\\
& $\mathbf{\overline{10}_{-1,-1}} \mathbf{5_{3,-2}} \mathbf{5_{-2,3}}$ & $\mathbf{\overline{10}_{-1,-1}}\mathbf{\overline{10}_{-1,-1}}\mathbf{\overline{5}_{2,2}}$ & $\mathbf{\overline{10}_{-1,-1}} \mathbf{5_{3,3}} \mathbf{5_{-2,-2}}$ \\
\hhline{~---}
& $\{b_{1}=c_{1}=0\}$ &  & \\
& non-flat fibre &  & \\
\hline
\hline
$(z_{3},z_{4},z_{5},z_{6},z_{7})$ & \multicolumn{3}{c|}{$(0,1,0,0,0)$} \\
\hline
\multirow{3}{*}{$P_{T}$} & \multicolumn{3}{c|}{$0 = b_{0,1} e_{0}e_{4}\tu \tv^{2} s_{0}^{2} + d_{0,1}e_{0}e_{2}e_{3}^{2}e_{4}^{2}\tu^{2}\tv s_{0}^{2}s_{1} $} \\
&\multicolumn{3}{c|}{$+ d_{2,1}e_{0}e_{2}^{2}e_{3}^{4}e_{4}^{3}\tu^{3}s_{0}^{2}s_{1}^{2} + c_{2,2}e_{0}^{2}e_{1}^{2}e_{2}e_{4}\tv^{2} \tw s_{0} + b_{1}\tu \tv \tw s_{0} s_{1} $}\\
&\multicolumn{3}{c|}{$+ d_{1}e_{2}e_{3}^{2}e_{4}\tu^{2}\tw s_{0}s_{1}^{2} + c_{1,1}e_{0}e_{1}^{2}e_{2}\tv \tw^{2}s_{1} +b_{2}e_{1}e_{2}e_{3} \tu \tw^{2} s_{1}^{2}$ }\\
\hline
$P$ & \multicolumn{3}{c|}{$\frac{1}{16}b_{0,1}b_{1}^{4}b_{2}c_{1,1}(b_{0,1}c_{1,1} - b_{1}c_{2,2})(b_{0,1}d_{1}^{2} - b_{1}d_{0,1}d_{1} + b_{1}^{2}d_{2,1})$} \\
\hline
Intersection numbers & $S_{0}: \delta_{j4}$ & $S_{1}:\delta_{j3}$ & $U: \delta_{j3}$ \\
\hline
\multirow{2}{*}{Shioda-map} & \multicolumn{3}{c|}{$\text{w}_{1} = 5(S_{1} - S_{0} - \mathcal{\bar K}) + \sum_{i}m_{i}E_{i} - m_{4}W, \quad m_{i}=(1,2,3,-1)$}\\
\hhline{~---}
& \multicolumn{3}{c|}{$\text{w}_{2} = 5(U - S_{0} - \mathcal{\bar K} - [c_{1}]) + \sum_{i}l_{i}E_{i} - m_{4}W, \quad l_{i}=(1,2,3,-1)$}\\
\hline
\multirow{4}{*}{Matter curves} & $\{ b_1=0\}$ & $\{b_{0,1}=0\}$  & $\{b_{2}=0\}$\\
& $\mathbf{10_{2,2}}$ + $\mathbf{\overline{10}_{-2,-2}}$ & $\mathbf{5_{-4,1}}$ + $\mathbf{\overline{5}_{4,-1}}$ & $\mathbf{5_{-4,-4}}$ + $\mathbf{\overline{5}_{4,4}}$ \\
\hhline{~---}
& $\{c_{1,1}=0\}$  & $\{b_{0,1}c_{1,1} - b_{1}c_{2,2}=0\}$ & \parbox[c][10mm][c]{40mm}{\center  \vspace*{-4mm} $\{b_{0,1}d_{1}^{2} - b_{1}d_{0,1}d_{1}$ \\ $ + b_{1}^{2}d_{2,1}=0\}$} \\
& $\mathbf{5_{1,6}}$ + $\mathbf{\overline{5}_{-1,-6}}$ & $\mathbf{5_{1,-4}}$ + $\mathbf{\overline{5}_{-1,4}}$ & $\mathbf{5_{1,1}}$ + $\mathbf{\overline{5}_{-1,-1}}$\\
\hline
\multirow{4}{*}{Yukawa points}  & $\{b_{1}=b_{2}=0\}$ & $\{b_{1}=c_{1,1}=0\}$ & $\{b_{1}=d_{1}=0\}$\\
& $\mathbf{10_{2,2}} \mathbf{10_{2,2}}\mathbf{5_{-4,-4}}$ & $\mathbf{10_{2,2}} \mathbf{\overline{5}_{-1,4}}\mathbf{\overline{5}_{-1,-6}}$ & $\mathbf{10_{2,2}} \mathbf{\overline{5}_{-1,-1}}\mathbf{\overline{5}_{-1,-1}}$ \\
\hhline{~---}
& $\{b_{1}=b_{0,1}=0\}$ &  & \\
& non-flat fibre &  & \\
\hline
\end{tabular}
\caption{Top 3 and 4 for polygon 5}\label{tab:top3_poly5}
\end{table}
\clearpage

\section*{\texorpdfstring{$SU(5)$}{SU(5)} on polygon 3}
\begin{figure}[h!]
    \centering
\def\svgwidth{0.4 \textwidth}
 \executeiffilenewer{polygon3.svg}{polygon3.pdf}%
 {inkscape -z -D --file=polygon3.svg %
  --export-pdf=polygon3.pdf --export-latex}%
   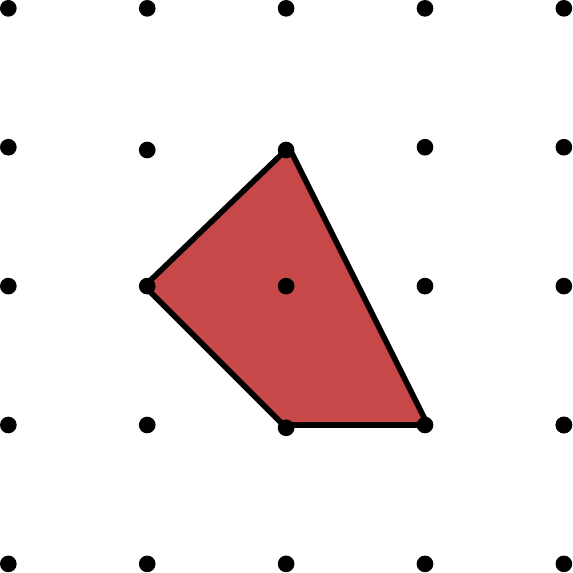%

      \caption{Polygon for $\textmd{Bl}_{1}\mathbb{P}^{2}$ }\label{fig:polygon3}
\end{figure}

\begin{figure}[h!]
    \centering
\def\svgwidth{0.9 \textwidth}
 \executeiffilenewer{su5-tops_polygon3.svg}{su5-tops_polygon3.pdf}%
 {inkscape -z -D --file=su5-tops_polygon3.svg %
  --export-pdf=su5-tops_polygon3.pdf --export-latex}%
   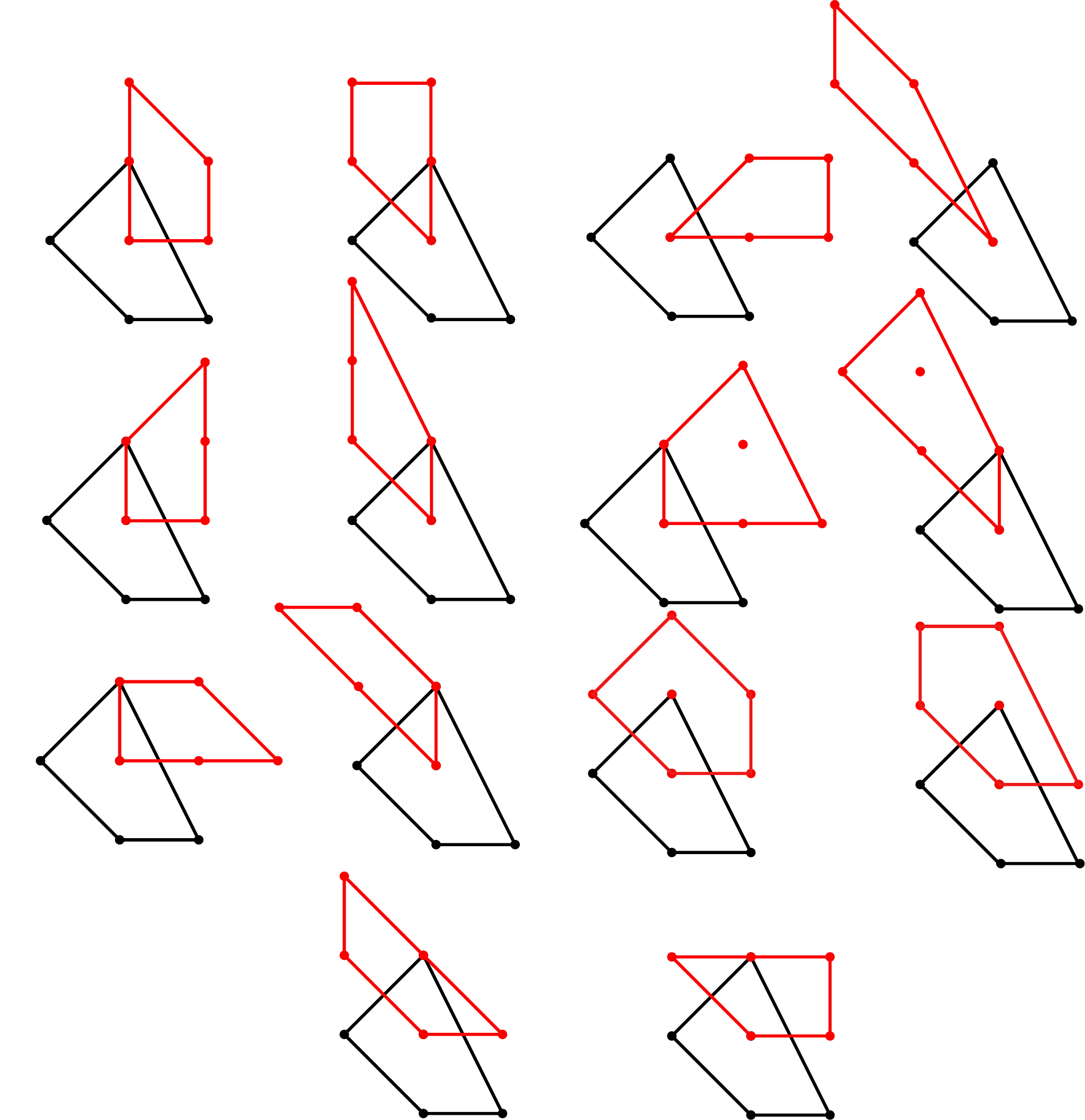%

      \caption{The tops over polygon 3 \cite{Bouchard:2003bu}. We have drawn 14 tops, half of which are related by symmetry.} \label{fig:su5-tops_polygon3}
\end{figure}

\begin{table}[h]
\centering
\begin{tabular}{|>{\centering}c|c|c|c|}
\hline
$(z_{3},z_{4},z_{5},z_{6},z_{7},z_{8})$ & \multicolumn{3}{c|}{$(-1,1,1,1,2,0)$} \\
\hline
\multirow{3}{*}{$P_{T}$} & \multicolumn{3}{c|}{$0 =   d_{2,2}s^{2}x^{3}e_{0}^{2}e_{1}^{3}e_{2}^{2}e_{4} + d_{0}sx^{2}ye_{1}e_{2} $} \\
&\multicolumn{3}{c|}{$ + b_{0}xy^{2}e_{1}e_{2}^{2}e_{3}^{2}e_{4} + d_{1,2}s^{2}x^{2}ze_{0}^{2}e_{1}^{2}e_{2}e_{4} + b_{1}sxyz + c_{0}y^{2}ze_{2}e_{3}^{2}e_{4}$}\\
&\multicolumn{3}{c|}{$ + b_{2,2}s^{2}xz^{2}e_{0}^{2}e_{1}e_{4} + c_{1,1}syz^{2}e_{0}e_{3}e_{4} + c_{2,3}s^{2}z^{3}e_{0}^{3}e_{1}e_{3}e_{4}^{2}$}\\
\hline
$P$ & \multicolumn{3}{c|}{$\frac{1}{16} b_{1}^4 c_{0} (b_{2,2}c_{1,1} - b_{1}c_{2,3}) (b_{0}b_{1} - c_{0}d_{0}) \left(b_{2}d_{0}^{2} + b_{1} (b_{1}d_{2,2} - d_{0}d_{1,2})\right)$} \\
\hline
Intersection numbers & $Z: \delta_{j0} + \delta_{j3}$ & $S:\delta_{j0}$ &  \\
\hline
\multirow{4}{*}{Matter curves} & $\{ b_1=0\}$ & $\{c_{0}=0\}$  & $\{b_{2,2}c_{1,1} - b_{1}c_{2,3}=0\}$\\
& $\mathbf{10}$ + $\mathbf{\overline{10}}$ & $\mathbf{5}$ + $\mathbf{\overline{5}}$ & $\mathbf{5}$ + $\mathbf{\overline{5}}$ \\
\hhline{~---}
& $\{b_{0}b_{1} - c_{0}d_{0}=0\}$ & \parbox[c][10mm][c]{40mm}{\center  \vspace*{-4mm}  $\{b_{2}d_{0}^{2}$\\$ + b_{1} (b_{1}d_{2,2} - d_{0}d_{1,2})=0\}$} & \\
&$\mathbf{5}$ + $\mathbf{\overline{5}}$ & $\mathbf{5}$ + $\mathbf{\overline{5}}$ & \\
\hline
\multirow{4}{*}{Yukawa points}  &$\{b_{1}=c_{0}=0\}$ &$\{b_{1}=b_{2}=0\}$ &$\{b_{1}=c_{1}=0\}$ \\
& $\mathbf{10} \mathbf{\overline{5}}\mathbf{\overline{5}}$ & $\mathbf{\overline{10}} \mathbf{5}\mathbf{5}$ &   $\mathbf{10}\mathbf{10}\mathbf{5}$\\
\hhline{~---}
 &$\{b_{1}=d_{0}=0\}$ & & \\
 & non-flat fibre & & \\
\hline
\end{tabular}
\caption{Top 1 on polygon 3}
\end{table}

\begin{table}[h]
\centering
\begin{tabular}{|>{\centering}c|c|c|c|}
\hline
$(z_{3},z_{4},z_{5},z_{6},z_{7},z_{8})$ & \multicolumn{3}{c|}{$(-1,0,1,2,3,0)$} \\
\hline
\multirow{3}{*}{$P_{T}$} & \multicolumn{3}{c|}{$0 =   d_{2,1}s^{2}x^{3}e_{0}e_{1}^{2}e_{2} + d_{0}sx^{2}ye_{1}e_{2}e_{3} $} \\
&\multicolumn{3}{c|}{$ + b_{0}xy^{2}e_{1}e_{2}^{2}e_{3}^{3}e_{4} + d_{1,2}s^{2}x^{2}ze_{0}^{2}e_{1}^{2}e_{2}e_{4} + b_{1}sxyz + c_{0}y^{2}ze_{2}e_{3}^{2}e_{4}$}\\
&\multicolumn{3}{c|}{$ + b_{2,3}s^{2}xz^{2}e_{0}^{3}e_{1}^{2}e_{2}e_{4}^{2} + c_{1,1}syz^{2}e_{0}e_{4} + c_{2,4}s^{2}z^{3}e_{0}^{4}e_{1}^{2}e_{2}e_{4}^{3}$}\\
\hline
$P$ & \multicolumn{3}{c|}{$\frac{1}{16} b_{1}^4 c_{0}d_{2} (b_{0}b_{1} - c_{0}d_{0}) (b_{1}^{3}c_{2} - b_{1}^{2}b_{2}c_{1} + b_{1}c_{1}^{2}d_{1} - c_{1}^{3}d_{2})$} \\
\hline
Intersection numbers & $Z: \delta_{j0} + \delta_{j4}$ & $S:\delta_{j0}$ &  \\
\hline
\multirow{4}{*}{Matter curves} & $\{ b_1=0\}$ & $\{c_{0}=0\}$  & $\{d_{2}=0\}$\\
& $\mathbf{10}$ + $\mathbf{\overline{10}}$ & $\mathbf{5}$ + $\mathbf{\overline{5}}$ & $\mathbf{5}$ + $\mathbf{\overline{5}}$ \\
\hhline{~---}
& $\{b_{0}b_{1} - c_{0}d_{0}=0\}$ & \parbox[c][10mm][c]{40mm}{\center  \vspace*{-4mm}  $\{b_{1}^{3}c_{2} - b_{1}^{2}b_{2}c_{1}$\\$ + b_{1}c_{1}^{2}d_{1} - c_{1}^{3}d_{2}=0\}$} & \\
&$\mathbf{5}$ + $\mathbf{\overline{5}}$ & $\mathbf{5}$ + $\mathbf{\overline{5}}$ & \\
\hline
\multirow{2}{*}{Yukawa points}  &$\{b_{1}=c_{0}=0\}$ &$\{b_{1}=d_{0}=0\}$ &$\{b_{1}=c_{1}=0\}$ \\
& $\mathbf{10} \mathbf{\overline{5}}\mathbf{\overline{5}}$ & $\mathbf{10} \mathbf{10}\mathbf{5}$ &   non-flat fibre\\
\hline
\hline
$(z_{3},z_{4},z_{5},z_{6},z_{7},z_{8})$ & \multicolumn{3}{c|}{$(-1,0,0,1,3,1)$} \\
\hline
\multirow{3}{*}{$P_{T}$} & \multicolumn{3}{c|}{$0 =   d_{2,1}s^{2}x^{3}e_{0}e_{1}^{2}e_{2}^{2}e_{3} + d_{0}sx^{2}ye_{1}e_{2}^{2}e_{3} $} \\
&\multicolumn{3}{c|}{$ + b_{0}xy^{2}e_{1}e_{2}^{2}e_{3}^{2}e_{4} + d_{1,1}s^{2}x^{2}ze_{0}e_{1}e_{2} + b_{1}sxyz + c_{0}y^{2}ze_{3}e_{4}$}\\
&\multicolumn{3}{c|}{$ + b_{2,2}s^{2}xz^{2}e_{0}^{2}e_{1}e_{4} + c_{1,2}syz^{2}e_{0}^{2}e_{1}e_{3}e_{4}^{2} + c_{2,4}s^{2}z^{3}e_{0}^{4}e_{1}^{2}e_{3}e_{4}^{3}$}\\
\hline
$P$ & \multicolumn{3}{c|}{$\frac{1}{16} b_{1}^4 c_{0} (-b_{0}b_{1} + c_{0}d_{0})(b_{1}d_{2} - d_{0}d_{1}) (b_{2}^{2}c_{0} - b_{1}b_{2}c_{1} + b_{1}^{2}c_{2})$} \\
\hline
Intersection numbers & $Z: \delta_{j0} + \delta_{j4}$ & $S:\delta_{j0}$ &  \\
\hline
\multirow{4}{*}{Matter curves} & $\{ b_1=0\}$ & $\{c_{0}=0\}$  & $\{b_{1}d_{2} - d_{0}d_{1}=0\}$\\
& $\mathbf{10}$ + $\mathbf{\overline{10}}$ & $\mathbf{5}$ + $\mathbf{\overline{5}}$ & $\mathbf{5}$ + $\mathbf{\overline{5}}$ \\
\hhline{~---}
& $\{-b_{0}b_{1} + c_{0}d_{0}=0\}$ & \parbox[c][10mm][c]{40mm}{\center  \vspace*{-4mm}  $\{b_{2}^{2}c_{0}$\\$ - b_{1}b_{2}c_{1} + b_{1}^{2}c_{2}=0\}$} & \\
&$\mathbf{5}$ + $\mathbf{\overline{5}}$ & $\mathbf{5}$ + $\mathbf{\overline{5}}$ & \\
\hline
\multirow{4}{*}{Yukawa points}  &$\{b_{1}=b_{2}=0\}$ &$\{b_{1}=d_{0}=0\}$ &$\{b_{1}=d_{1}=0\}$ \\
& $\mathbf{10} \mathbf{\overline{5}}\mathbf{\overline{5}}$ & $\mathbf{\overline{10}} \mathbf{5}\mathbf{5}$ &   $\mathbf{10}\mathbf{10}\mathbf{5}$\\
\hhline{~---}
 &$\{b_{1}=c_{0}=0\}$ & & \\
 & non-flat fibre & & \\
\hline
\end{tabular}
\caption{Top 2 and 3 on polygon 3}
\end{table}

\begin{table}[h]
\centering
\begin{tabular}{|>{\centering}c|c|c|c|}
\hline
$(z_{3},z_{4},z_{5},z_{6},z_{7},z_{8})$ & \multicolumn{3}{c|}{$(-1,-1,0,2,4,1)$} \\
\hline
\multirow{3}{*}{$P_{T}$} & \multicolumn{3}{c|}{$0 =   d_{2}s^{2}x^{3}e_{1}e_{2}^{2}e_{3} + d_{0}sx^{2}ye_{1}e_{2}^{2}e_{3}^{2}e_{4} $} \\
&\multicolumn{3}{c|}{$ + b_{0}xy^{2}e_{1}e_{2}^{2}e_{3}^{3}e_{4}^{2} + d_{1,1}s^{2}x^{2}ze_{0}e_{1}e_{2} + b_{1}sxyz + c_{0}y^{2}ze_{3}e_{4}$}\\
&\multicolumn{3}{c|}{$ + b_{2,3}s^{2}xz^{2}e_{0}^{3}e_{1}^{2}e_{2}e_{4} + c_{1,2}syz^{2}e_{0}^{2}e_{1}e_{4} + c_{2,5}s^{2}z^{3}e_{0}^{5}e_{1}^{3}e_{2}e_{4}^{2}$}\\
\hline
$P$ & \multicolumn{3}{c|}{$\frac{1}{16} b_{1}^4 c_{0}d_{2}(b_{1}^{2}c_{2} - b_{1}b_{2}c_{1} + c_{1}^{2}d_{1}) (b_{1}^{2}b_{0} + c_{0}(c_{0}d_{2} - b_{1}d_{0}))$} \\
\hline
Intersection numbers & $Z: 2\delta_{j0}$ & $S:\delta_{j0}$ &  \\
\hline
\multirow{4}{*}{Matter curves} & $\{ b_1=0\}$ & $\{c_{0}=0\}$  & $\{d_{2}=0\}$\\
& $\mathbf{10}$ + $\mathbf{\overline{10}}$ & $\mathbf{5}$ + $\mathbf{\overline{5}}$ & $\mathbf{5}$ + $\mathbf{\overline{5}}$ \\
\hhline{~---}
&  \parbox[c][10mm][c]{40mm}{\center  \vspace*{-4mm}  $\{b_{1}^{2}c_{2}$\\$ - b_{1}b_{2}c_{1} + c_{1}^{2}d_{1}=0\}$} & \parbox[c][10mm][c]{40mm}{\center  \vspace*{-4mm}  $\{b_{1}^{2}b_{0}$\\$ + c_{0}(c_{0}d_{2} - b_{1}d_{0})=0\}$} & \\
&$\mathbf{5}$ + $\mathbf{\overline{5}}$ & $\mathbf{5}$ + $\mathbf{\overline{5}}$ & \\
\hline
\multirow{2}{*}{Yukawa points}  &$\{b_{1}=d_{1}=0\}$ &$\{b_{1}=d_{2}=0\}$ &$\{b_{1}=c_{0}=0\}$ \\
& $\mathbf{10} \mathbf{10}\mathbf{5}$ & $\mathbf{10} \mathbf{\overline{5}}\mathbf{\overline{5}}$ &   non-flat fibre\\
\hline
\hline
$(z_{3},z_{4},z_{5},z_{6},z_{7},z_{8})$ & \multicolumn{3}{c|}{$(0,2,1,0,1,0)$} \\
\hline
\multirow{3}{*}{$P_{T}$} & \multicolumn{3}{c|}{$0 =   d_{2,3}s^{2}x^{3}e_{0}^{3}e_{1}^{4}e_{2}^{2}e_{4} + d_{0,1}sx^{2}ye_{0}e_{1}^{2}e_{2} $} \\
&\multicolumn{3}{c|}{$ + b_{0}xy^{2}e_{1}e_{2}e_{3} + d_{1,2}s^{2}x^{2}ze_{0}^{2}e_{1}^{2}e_{2}e_{4} + b_{1}sxyz + c_{0}y^{2}ze_{2}e_{3}^{2}e_{4}$}\\
&\multicolumn{3}{c|}{$ + b_{2,1}s^{2}xz^{2}e_{0}e_{4} + c_{1,1}syz^{2}e_{0}e_{2}e_{3}^{2}e_{4}^{2} + c_{2,2}s^{2}z^{3}e_{0}^{2}e_{2}e_{3}^{2}e_{4}^{3}$}\\
\hline
$P$ & \multicolumn{3}{c|}{$\frac{1}{16} b_{1}^4 b_{0}c_{0} (b_{2}^{2}c_{0} - b_{1}b_{2}c_{1} + b_{1}^{2}c_{2})(b_{2}d_{0}^{2} + b_{1}(b_{1}d_{2} - d_{0}d_{1}))$} \\
\hline
Intersection numbers & $Z: \delta_{j3} + \delta_{j4}$ & $S:\delta_{j0}$ &  \\
\hline
\multirow{4}{*}{Matter curves} & $\{ b_1=0\}$ & $\{b_{0}=0\}$  & $\{c_{0}=0\}$\\
& $\mathbf{10}$ + $\mathbf{\overline{10}}$ & $\mathbf{5}$ + $\mathbf{\overline{5}}$ & $\mathbf{5}$ + $\mathbf{\overline{5}}$ \\
\hhline{~---}
& \parbox[c][10mm][c]{40mm}{\center  \vspace*{-4mm}  $\{b_{2}^{2}c_{0}$\\$  - b_{1}b_{2}c_{1} + b_{1}^{2}c_{2}=0\}$} & \parbox[c][10mm][c]{40mm}{\center  \vspace*{-4mm}  $\{b_{2}d_{0}^{2}$\\$ + b_{1}(b_{1}d_{2} - d_{0}d_{1})=0\}$} & \\
&$\mathbf{5}$ + $\mathbf{\overline{5}}$ & $\mathbf{5}$ + $\mathbf{\overline{5}}$ & \\
\hline
\multirow{2}{*}{Yukawa points}  &$\{b_{1}=b_{0}=0\}$ &$\{b_{1}=c_{0}=0\}$ &$\{b_{1}=b_{2}=0\}$ \\
& $\mathbf{10}\mathbf{10}\mathbf{5}$ & $\mathbf{\overline{10}} \mathbf{5}\mathbf{5}$ &   non-flat fibre\\
\hline
\end{tabular}
\caption{Top 4 and 7 on polygon 3}
\end{table}
\clearpage

\subsection*{\texorpdfstring{$SU(5)$}{SU(5)} on Polygon 6}
\begin{figure}[h!]
    \centering
\def\svgwidth{0.7 \textwidth}
  \hspace*{2cm}%
 \executeiffilenewer{polygon6.svg}{polygon6.pdf}%
 {inkscape -z -D --file=polygon6.svg %
  --export-pdf=polygon6.pdf --export-latex}%
   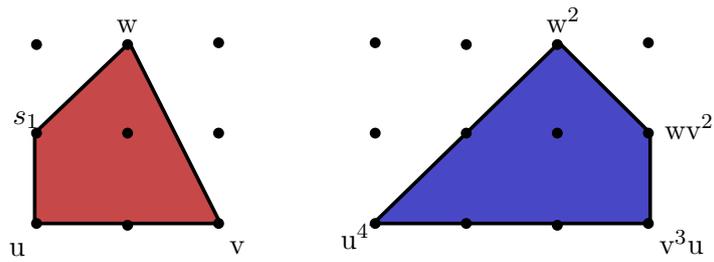%

      \caption{Polygon for $\textmd{Bl}_{1}\mathbb{P}_{[1,1,2]}$}\label{fig:polygon6}
\end{figure}

\begin{figure}[h!]
    \centering
\def\svgwidth{0.6 \textwidth}
 \executeiffilenewer{su5-tops_polygon6.svg}{su5-tops_polygon6.pdf}%
 {inkscape -z -D --file=su5-tops_polygon6.svg %
  --export-pdf=su5-tops_polygon6.pdf --export-latex}%
   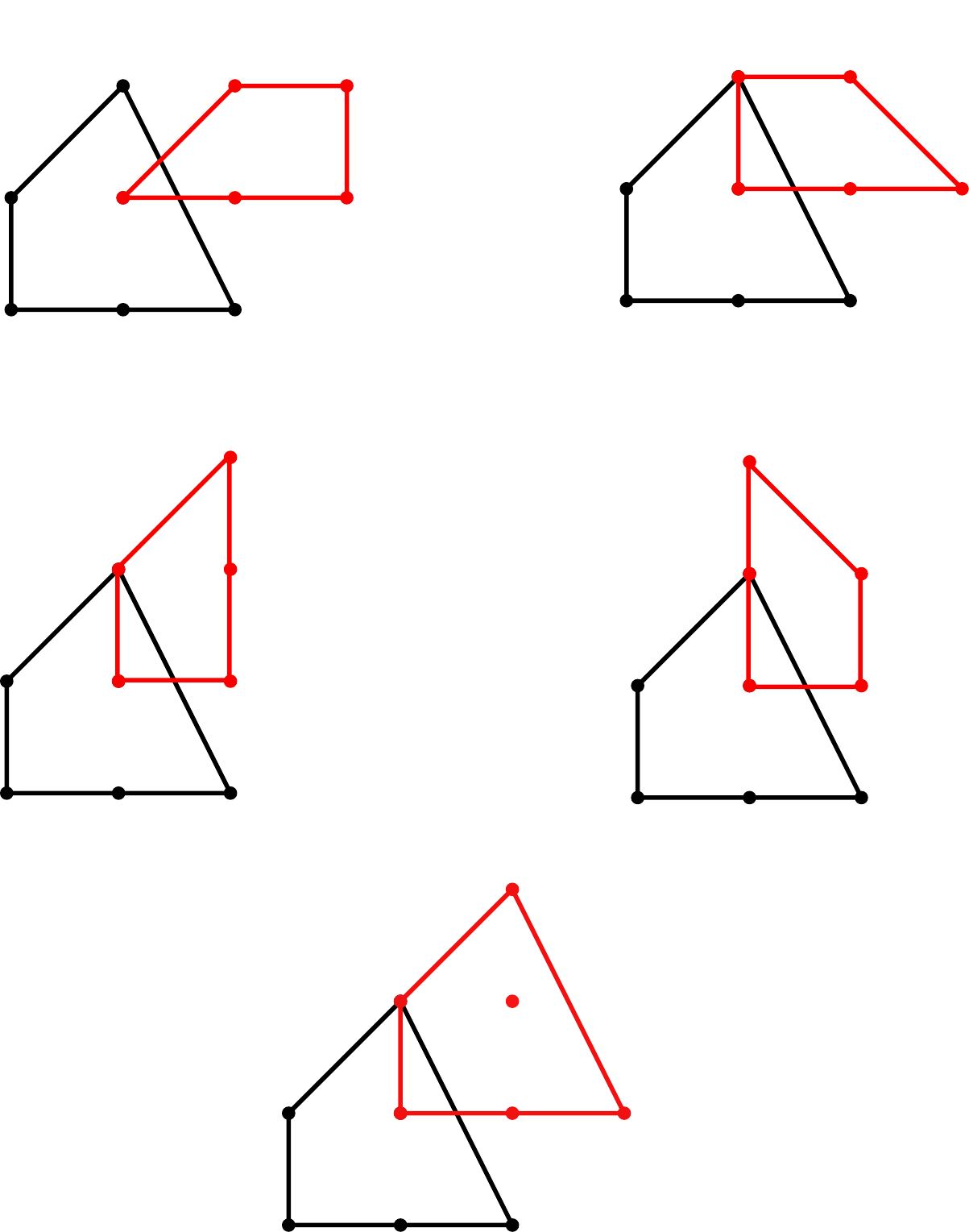%

      \caption{The tops over polygon 6 \cite{Bouchard:2003bu}}\label{fig:su5-tops_polygon6}
\end{figure}

\begin{table}[h]
\centering
\begin{tabular}{|>{\centering}c|c|c|c|}
\hline
$(z_{3},z_{4},z_{5},z_{6},z_{7})$ & \multicolumn{3}{c|}{$(-1,0,2,4,1)$} \\
\hline
\multirow{3}{*}{$P_{T}$} & \multicolumn{3}{c|}{$0 = \tw^{2}s_{1}e_{3}e_{4} + b_{0,2} \tw \tu^{2} s_{1}^{2} e_{0}^{2}e_{1}e_{4} $} \\
&\multicolumn{3}{c|}{$+ b_{1}\tu \tv \tw s_{1}+ b_{2}\tv^{2} \tw  e_{1}e_{2}^{2}e_{3}^{2}e_{4} - c_{0,5}\tu^{4}s_{1}^{3}e_{0}^{5}e_{1}^{3}e_{2}e_{4}^{2}$}\\
&\multicolumn{3}{c|}{$- c_{1,3} \tu^{3} \tv s_{1}^{2} e_{0}^{3} e_{1}^{2}e_{2}e_{4} - c_{2,1}\tu^{2} \tv^{2}s_{1}e_{0}e_{1}e_{2} - c_{3}\tu \tv^{3}e_{1}e_{2}^{2}e_{3}$ }\\
\hline
$P$ & \multicolumn{3}{c|}{$\frac{1}{16}$} \\
\hline
Intersection numbers & $U: \delta_{j0}$ & $S_{1}:\delta_{j0}$ &  \\
\hline
Shioda-map & \multicolumn{3}{c|}{$\text{w}_{1} = 5(S_{1} - U - \mathcal{\bar K} - [b_{2}])$}\\
\hline
\multirow{4}{*}{Matter curves} & $\{ b_1=0\}$ & $\{c_{3}=0\}$  & $\{b_{1}b_{2} + c_{3}=0\}$\\
& $\mathbf{10_{0}}$ + $\mathbf{\overline{10}_{0}}$ & $\mathbf{5_{-1}}$ + $\mathbf{\overline{5}_{1}}$ & $\mathbf{5_{1}}$ + $\mathbf{\overline{5}_{-1}}$ \\
\hhline{~---}
& \parbox[c][10mm][c]{35mm}{\center  \vspace*{-4mm} $\{b_{1}^{2}c_{0,5} - b_{0,2}b_{1}c_{1,3}$ \\ $ + b_{0,2}^{2}c_{2,1}=0\}$}  &  & \\
& $\mathbf{5_{0}}$ + $\mathbf{\overline{5}_{0}}$ &  & \\
\hline
\multirow{2}{*}{Yukawa points}  &$\{b_{1}=c_{2}=0\}$ & $\{b_{1}=c_{3}=0\}$ & \\
& $\mathbf{\overline{10}_{0}} \mathbf{\overline{10}_{0}}\mathbf{\overline{5}_{0}}$ & $\mathbf{10_{0}} \mathbf{\overline{5}_{1}} \mathbf{\overline{5}_{-1}}$ &   \\
\hline
\hline
$(z_{3},z_{4},z_{5},z_{6},z_{7})$ & \multicolumn{3}{c|}{$(0,0,1,3,1)$} \\
\hline
\multirow{3}{*}{$P_{T}$} & \multicolumn{3}{c|}{$0 = \tw^{2}s_{1}e_{3}e_{4} + b_{0,2} \tw \tu^{2} s_{1}^{2} e_{0}^{2}e_{1}e_{3}e_{4}^{2}$} \\
&\multicolumn{3}{c|}{$+ b_{1}\tu \tv \tw s_{1}  + b_{2}\tv^{2} \tw  e_{1}e_{2}^{2}e_{3} - c_{0,4}\tu^{4}s_{1}^{3}e_{0}^{4}e_{1}^{2}e_{3}e_{4}^{3}$}\\
&\multicolumn{3}{c|}{$- c_{1,2} \tu^{3} \tv s_{1}^{2} e_{0}^{2}e_{1}e_{4} - c_{2,1}\tu^{2} \tv^{2}s_{1}e_{0}e_{1}e_{2} - c_{3,1}\tu \tv^{3}e_{0}e_{1}^{2}e_{2}^{3}e_{3}$ }\\
\hline
$P$ & \multicolumn{3}{c|}{$\frac{1}{16}$} \\
\hline
Intersection numbers & $U: \delta_{j0}$ & $S_{1}:\delta_{j4}$ &  \\
\hline
Shioda-map & \multicolumn{3}{c|}{$\text{w}_{1} = 5(S_{1} - U - \mathcal{\bar K} - [b_{2}]) + \sum_{i}m_{i}E_{i}, \quad m_{i}=(1,2,3,4)$}\\
\hline
\multirow{4}{*}{Matter curves} & $\{ b_1=0\}$ & $\{b_{2}=0\}$  & $\{b_{1}c_{3,1} + b_{2}c_{2,1}=0\}$\\
& $\mathbf{10_{2}}$ + $\mathbf{\overline{10}_{-2}}$ & $\mathbf{5_{6}}$ + $\mathbf{\overline{5}_{-6}}$ & $\mathbf{5_{-4}}$ + $\mathbf{\overline{5}_{4}}$ \\
\hhline{~---}
&\parbox[c][10mm][c]{35mm}{\center  \vspace*{-4mm}  $\{b_{1}^{2}c_{0,4} - b_{0,2}b_{1}c_{1,2}$ \\ $ - c_{1,2}^{2}=0\}$}  &  & \\
& $\mathbf{5_{1}}$ + $\mathbf{\overline{5}_{-1}}$ &  & \\
\hline
\multirow{2}{*}{Yukawa points}  &$\{b_{1}=b_{2}=0\}$ & $\{b_{1}=c_{1,2}=0\}$ & $\{b_{1}=c_{2,1}=0\}$\\
& $\mathbf{\overline{10}_{-2}} \mathbf{5_{6}}\mathbf{5_{-4}}$ & $\mathbf{\overline{10}_{-2}} \mathbf{5_{1}} \mathbf{5_{1}}$ & $\mathbf{\overline{10}_{-2}}\mathbf{\overline{10}_{-2}}  \mathbf{\overline{5}_{4}}$  \\
\hline
\end{tabular}
\caption{Top 1 and 2 for polygon 6}
\end{table}

\begin{table}[h]
\centering
\begin{tabular}{|>{\centering}c|c|c|c|}
\hline
$(z_{3},z_{4},z_{5},z_{6},z_{7})$ & \multicolumn{3}{c|}{$(0,1,2,3,0)$} \\
\hline
\multirow{3}{*}{$P_{T}$} & \multicolumn{3}{c|}{$0 =  \tw^{2}s_{1}e_{3}^{2}e_{4} + b_{0,1} \tw \tu^{2} s_{1}^{2}e_{0}e_{4}$} \\
&\multicolumn{3}{c|}{$+ b_{1}\tu \tv \tw s_{1} + b_{2}\tv^{2} \tw e_{1}e_{2}e_{3} - c_{0,4}\tu^{4}s_{1}^{3}e_{0}^{4}e_{1}^{2}e_{2}e_{4}^{3}$}\\
&\multicolumn{3}{c|}{$ - c_{1,3} \tu^{3} \tv s_{1}^{2} e_{0}^{3}e_{1}^{2}e_{2}e_{4}^{2} - c_{2,2}\tu^{2} \tv^{2}s_{1}e_{0}^{2}e_{1}^{2}e_{2}e_{4} - c_{3,1}\tu \tv^{3}e_{0}e_{1}^{2}e_{2}$ }\\
\hline
$P$ & \multicolumn{3}{c|}{$\frac{1}{16}$} \\
\hline
Intersection numbers & $U: \delta_{j0}$ & $S_{1}:\delta_{j4}$ &  \\
\hline
Shioda-map & \multicolumn{3}{c|}{$\text{w}_{1} = 5(S_{1} - U - \mathcal{\bar K} - [b_{2}]) + \sum_{i}m_{i}E_{i}, \quad m_{i}=(1,2,3,4)$}\\
\hline
\multirow{4}{*}{Matter curves} & $\{ b_1=0\}$ & $\{b_{2}=0\}$  & $\{c_{3,1}=0\}$\\
& $\mathbf{10_{-3}}$ + $\mathbf{\overline{10}_{3}}$ & $\mathbf{5_{6}}$ + $\mathbf{\overline{5}_{-6}}$ & $\mathbf{5_{-4}}$ + $\mathbf{\overline{5}_{4}}$ \\
\hhline{~---}
& \parbox[c][10mm][c]{40mm}{\center  \vspace*{-4mm}$\{b_{1}^{3}c_{0,4} - b_{0,1}b_{1}^{2}c_{1,3} $\\ $ + b_{0,1}^{2}b_{1}c_{2,2} - b_{0,1}^{3}c_{3,1}=0\}$}  &  & \\
& $\mathbf{5_{1}}$ + $\mathbf{\overline{5}_{-1}}$ &  & \\
\hline
\multirow{2}{*}{Yukawa points}  &$\{b_{1}=b_{2}=0\}$ & $\{b_{1}=c_{3,1}=0\}$ & $\{b_{1}=b_{0,1}=0\}$\\
& $\mathbf{10_{-3}} \mathbf{10_{-3}}\mathbf{5_{6}}$ & $\mathbf{10_{-3}} \mathbf{\overline{5}_{-1}} \mathbf{\overline{5}_{4}}$ & non-flat fibre  \\
\hline
\hline
$(z_{3},z_{4},z_{5},z_{6},z_{7})$ & \multicolumn{3}{c|}{$(1,1,1,2,0)$} \\
\hline
\multirow{3}{*}{$P_{T}$} & \multicolumn{3}{c|}{$0 =   \tw^{2}s_{1}e_{2}e_{3}^{2}e_{4} + b_{0,1} \tw \tu^{2} s_{1}^{2}e_{0}e_{3}e_{4}$} \\
&\multicolumn{3}{c|}{$+ b_{1}\tu \tv \tw s_{1} + b_{2}\tv^{2} \tw e_{1}e_{2} - c_{0,3}\tu^{4}s_{1}^{3}e_{0}^{3}e_{1}e_{3}e_{4}^{2}$}\\
&\multicolumn{3}{c|}{$ - c_{1,2} \tu^{3} \tv s_{1}^{2} e_{0}^{2}e_{1}e_{4} - c_{2,2}\tu^{2} \tv^{2}s_{1}e_{0}^{2}e_{1}^{2}e_{2}e_{4} - c_{3,2}\tu \tv^{3}e_{0}^{2}e_{1}^{3}e_{2}^{2}e_{4}$ }\\
\hline
$P$ & \multicolumn{3}{c|}{$\frac{1}{16}$} \\
\hline
Intersection numbers & $U: \delta_{j0}$ & $S_{1}:\delta_{j3}$ &  \\
\hline
Shioda-map & \multicolumn{3}{c|}{$\text{w}_{1} = 5(S_{1} - U - \mathcal{\bar K} - [b_{2}]) + \sum_{i}m_{i}E_{i}, \quad m_{i}=(2,4,6,3)$}\\
\hline
\multirow{4}{*}{Matter curves} & $\{ b_1=0\}$ & $\{b_{2}=0\}$  & $\{b_{1}c_{0,3} - b_{0,1}c_{1,2}=0\}$\\
& $\mathbf{10_{-1}}$ + $\mathbf{\overline{10}_{1}}$ & $\mathbf{5_{7}}$ + $\mathbf{\overline{5}_{-7}}$ & $\mathbf{5_{2}}$ + $\mathbf{\overline{5}_{-2}}$ \\
\hhline{~---}
& \parbox[c][10mm][c]{35mm}{\center  \vspace*{-4mm} $\{b_{2}^{2}c_{1,2} - b_{1}b_{2}c_{2,2}$ \\ $ + b_{1}^{2}c_{3,2}=0\}$}  &  & \\
& $\mathbf{5_{-3}}$ + $\mathbf{\overline{5}_{3}}$ &  & \\
\hline
\multirow{2}{*}{Yukawa points}  &$\{b_{1}=b_{0,1}=0\}$ & $\{b_{1}=c_{1,2}=0\}$ & $\{b_{1}=b_{2}=0\}$\\
& $\mathbf{\overline{10}_{1}} \mathbf{\overline{10}_{1}}\mathbf{\overline{5}_{-2}}$  & $\mathbf{\overline{10}_{1}} \mathbf{5_{-3}} \mathbf{5_{2}}$ & non-flat fibre  \\
\hline
\end{tabular}
\caption{Top 3 and 4 for polygon 6}
\end{table}
\clearpage

\subsection*{\texorpdfstring{$SU(5)$}{SU(5)} on polygon 8}
\begin{figure}[h!]
    \centering
\def\svgwidth{0.4 \textwidth}
 \executeiffilenewer{polygon8.svg}{polygon8.pdf}%
 {inkscape -z -D --file=polygon8.svg %
  --export-pdf=polygon8.pdf --export-latex}%
   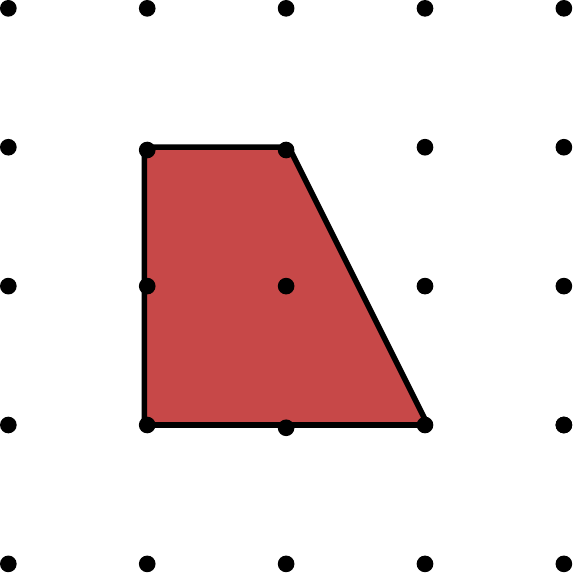%

      \caption{Polygon for $\textmd{Bl}_{2}\mathbb{P}_{[1,1,2]}$}\label{fig:polygon8}
\end{figure}

\begin{figure}[h!]
    \centering
\def\svgwidth{0.6 \textwidth}
 \executeiffilenewer{su5-tops_polygon8.svg}{su5-tops_polygon8.pdf}%
 {inkscape -z -D --file=su5-tops_polygon8.svg %
  --export-pdf=su5-tops_polygon8.pdf --export-latex}%
   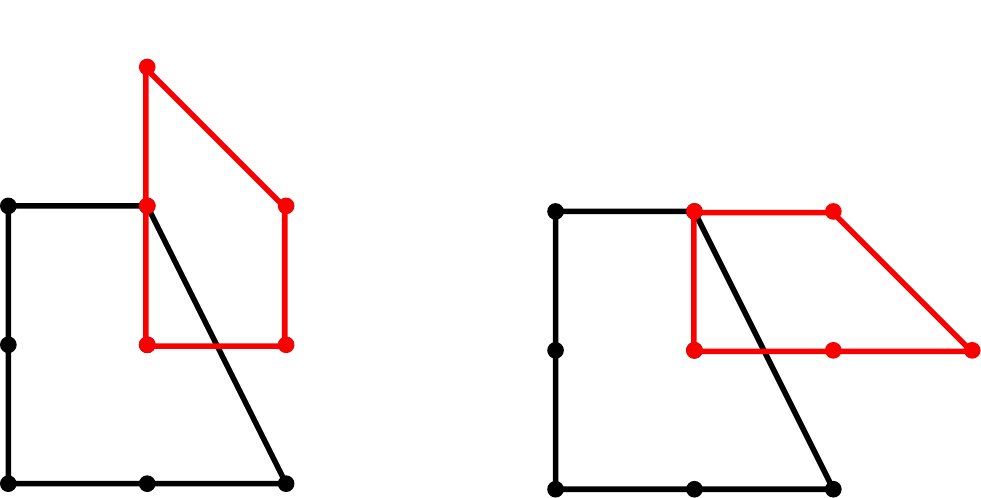%

      \caption{The tops over polygon 8 \cite{Bouchard:2003bu}}\label{fig:su5-tops_polygon8}
\end{figure}
\begin{table}[h]
\centering
\begin{tabular}{|>{\centering}c|c|c|c|}
\hline
$(z_{3},z_{4},z_{5},z_{6})$ & \multicolumn{3}{c|}{$(0,2,1,1)$} \\
\hline
\multirow{2}{*}{$P_{T}$} & \multicolumn{3}{c|}{$0 = c_{0,3}\tu^{4} s^{2}e_{0}^{3}e_{1}e_{3}e_{4}^{2} + c_{1,2}\tu^{3}\tv se_{0}^{2}e_{1}e_{4} + c_{2,2}\tu^{2}\tv^{2}e_{0}^{2}e_{1}^{2}e_{2}e_{4}$} \\
&\multicolumn{3}{c|}{$ +b_{0,1}\tu^{2}\tw s^{2}e_{0}e_{3}e_{4} + b_{1}\tu \tv \tw s + b_{2}\tv^{2}\tw e_{1}e_{2} + \tw^{2}s^{2}e_{3}^{2}e_{4}$}\\
\hline
$P$ & \multicolumn{3}{c|}{$\frac{1}{16}b_{1}^{4}b_{2}^{2}(b_{1}c_{0} + b_{0}c_{1})(b_{2}c_{1} + b_{1}c_{2})$} \\
\hline
Intersection numbers & $U: \delta_{j0}$ & $S:\delta_{j2}$ &  \\
\hline
Shioda-map & \multicolumn{3}{c|}{$\text{w} = 5(S - U - \mathcal{\bar K}) + \sum_{i}m_{i}E_{i}, \quad m_{i}=(2,4,6,3)$}\\
\hline
\multirow{4}{*}{Matter curves} & $\{ b_1=0\}$ & $\{b_{2}=0\}$  & $\{b_{1}c_{0} + b_{0}c_{1}=0\}$\\
& $\mathbf{10_{-1}}$ + $\mathbf{\overline{10}_{1}}$ & $\mathbf{5_{2}}$ + $\mathbf{\overline{5}_{-2}}$ & $\mathbf{5_{2}}$ + $\mathbf{\overline{5}_{-2}}$ \\
\hhline{~---}
& $\{b_{2}c_{1} + b_{1}c_{2}=0\}$ &  & \\
&$\mathbf{5_{-3}}$ + $\mathbf{\overline{5}_{3}}$ &  & \\
\hline
\multirow{2}{*}{Yukawa points}  &$\{b_{1}=c_{1}=0\}$ & $\{b_{1}=b_{0}=0\}$ & \\
& $\mathbf{10_{-1}} \mathbf{\overline{5}_{-2}}\mathbf{\overline{5}_{3}}$ & $\mathbf{10_{-1}} \mathbf{10_{-1}}\mathbf{5_{2}}$ &   \\
\hline
\hline
$(z_{3},z_{4},z_{5},z_{6})$ & \multicolumn{3}{c|}{$(1,3,1,0)$} \\
\hline
\multirow{2}{*}{$P_{T}$} & \multicolumn{3}{c|}{$0 = c_{0,3}\tu^{4} s^{2}e_{0}^{3}e_{1}e_{3}e_{4}^{2} + c_{1,2}\tu^{3}\tv se_{0}^{2}e_{1}e_{4} + c_{2,2}\tu^{2}\tv^{2}e_{0}^{2}e_{1}^{2}e_{2}e_{4}$} \\
&\multicolumn{3}{c|}{$ +b_{0,1}\tu^{2}\tw s^{2}e_{0}e_{3}e_{4} + b_{1}\tu \tv \tw s + b_{2}\tv^{2}\tw e_{1}e_{2} + \tw^{2}s^{2}e_{3}^{2}e_{4}$}\\
\hline
$P$ & \multicolumn{3}{c|}{$-\frac{1}{16}b_{1}^{4}b_{2}^{2}c_{2}(b_{1}^{2}c_{0} + b_{0}b_{1}c_{1} - c_{1}^{2})$} \\
\hline
Intersection numbers & $U: \delta_{j0}$ & $S:\delta_{j2}$ &  \\
\hline
Shioda-map & \multicolumn{3}{c|}{$\text{w} = 5(S - U - \mathcal{\bar K}) + \sum_{i}m_{i}E_{i}, \quad m_{i}=(2,4,6,3)$}\\
\hline
\multirow{4}{*}{Matter curves} & $\{ b_1=0\}$ & $\{b_{2}=0\}$  & $\{c_{2}=0\}$\\
& $\mathbf{10_{-2}}$ + $\mathbf{\overline{10}_{2}}$ & $\mathbf{5_{-1}}$ + $\mathbf{\overline{5}_{1}}$ & $\mathbf{5_{4}}$ + $\mathbf{\overline{5}_{-4}}$ \\
\hhline{~---}
& $\{b_{1}^{2}c_{0} + b_{0}b_{1}c_{1} - c_{1}^{2}=0\}$ &  & \\
&$\mathbf{5_{-1}}$ + $\mathbf{\overline{5}_{1}}$ &  & \\
\hline
\multirow{2}{*}{Yukawa points}  &$\{b_{1}=b_{2}=0\}$ & $\{b_{1}=c_{2}=0\}$ & \\
& $\mathbf{10_{-2}} \mathbf{\overline{5}_{1}}\mathbf{\overline{5}_{1}}$ & $\mathbf{10_{-2}} \mathbf{10_{-2}}\mathbf{\overline{5}_{4}}$ &   \\
\hline
\end{tabular}
\caption{Top 1 and 2 for polygon 8}
\end{table}
\clearpage

\subsection*{\texorpdfstring{$SU(5)$}{SU(5)} on polygon 11}
\begin{figure}[h!]
    \centering
\def\svgwidth{0.4 \textwidth}
 \executeiffilenewer{polygon11.svg}{polygon11.pdf}%
 {inkscape -z -D --file=polygon11.svg %
  --export-pdf=polygon11.pdf --export-latex}%
   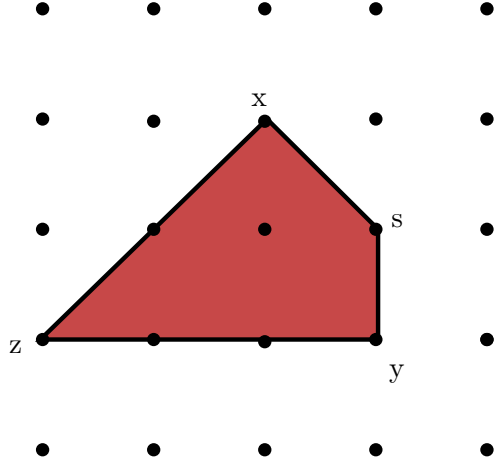%

      \caption{Polygon for $\textmd{Bl}_{1}\mathbb{P}_{[2,3,1]}$}\label{fig:polygon11}
\end{figure}

\begin{figure}[h!]
    \centering
\def\svgwidth{0.4 \textwidth}
 \executeiffilenewer{su5-tops_polygon11.svg}{su5-tops_polygon11.pdf}%
 {inkscape -z -D --file=su5-tops_polygon11.svg %
  --export-pdf=su5-tops_polygon11.pdf --export-latex}%
   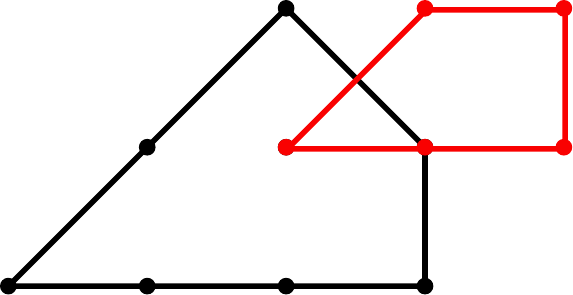%

      \caption{The tops over polygon 11 \cite{Bouchard:2003bu}}\label{fig:su5-tops_polygon11}
\end{figure}
\begin{table}[h!]
\centering
\begin{tabular}{|>{\centering}c|c|c|c|}
\hline
$(z_{3},z_{4},z_{5})$ & \multicolumn{3}{c|}{$(0,2,1)$} \\
\hline
\multirow{2}{*}{$P_{T}$} & \multicolumn{3}{c|}{$0 =   y^{2}se_{3}e_{4} - x^{3}s^{2}e_{1}e_{2}^{2}e_{3} - a_{1}xyzs$} \\
&\multicolumn{3}{c|}{$ + a_{3,2}yz^{3}e_{0}^{2}e_{1}e_{4} - a_{2,1}x^{2}z^{2}se_{0}e_{1}e_{2} - a_{4,3}xz^{4}e_{0}^{3}e_{1}^{2}e_{2}e_{4}$}\\
\hline
$P$ & \multicolumn{3}{c|}{$\frac{1}{16}a_{1}^{4}a_{3}(a_{2}a_{3} - a_{1}a_{4})$} \\
\hline
Intersection numbers & $Z: \delta_{j0}$ & $S:\delta_{j2}$ &  \\
\hline
Shioda-map & \multicolumn{3}{c|}{$\text{w} = 5(S - Z - \mathcal{\bar K}) + \sum_{i}m_{i}E_{i}, \quad m_{i}=(2,4,6,3)$}\\
\hline
\multirow{2}{*}{Matter curves} & $\{ a_1=0\}$ & $\{a_{3,2}=0\}$  & $\{a_{2,1}a_{3,2} - a_{1}a_{4,3}=0\}$\\
& $\mathbf{10_{-1}}$ + $\mathbf{\overline{10}_{1}}$ & $\mathbf{5_{-3}}$ + $\mathbf{\overline{5}_{3}}$ & $\mathbf{5_{2}}$ + $\mathbf{\overline{5}_{-2}}$ \\
\hline
\multirow{2}{*}{Yukawa points}  &$\{a_{1}=a_{2,1}=0\}$ & $\{a_{1}=a_{3,2}=0\}$ & \\
& $\mathbf{10_{-1}} \mathbf{10_{-1}}\mathbf{5_{2}}$ & $\mathbf{10_{-1}} \mathbf{\overline{5}_{3}}\mathbf{\overline{5}_{-2}}$ &   \\
\hline
\end{tabular}
\caption{Top 1 on polygon 11}
\end{table}
\clearpage

\section{\texorpdfstring{$SU(4)$}{SU(4)}-symmetry} \label{app-SU4}

In the previous appendix we constructed tops that lead to an $SU(5)$ singularity. Tops constructions are such that the associated singularity is present for generic coefficients. However the class of models with non-generic coefficients can often be very interesting, and one such class was constructed in \cite{Mayrhofer:2012zy}. This class was such that one starts with a top that leads to an $SU(4)$ singularity with generic coefficients, rather than $SU(5)$, but then restricts the coefficients so as to induce an additional singularity thereby enhancing the $SU(4) \rightarrow SU(5)$. It was shown there that this leads to $SU(5)$ models which have more than a single $\te$-matter curves, in contrast to all the $SU(5)$ models constructed so far. The final resolution step performed in \cite{Mayrhofer:2012zy} gives rise to complete intersection, as opposed to a hypersurface. This is interesting both from a phenomenological perspective, since multiple $\te$-curves can be used for various model building purposes, 
for example to understand flavour physics \cite{Dudas:2009hu}, and from a formal perspective, especially since such multiple $\te$-curves are not possible in IIB.

In this appendix we fill in some details regarding the smoothness of the $SU(4)\rightarrow SU(5)$ top construction of \cite{Mayrhofer:2012zy}, and present further $SU(4)$ top constructions that can be used as a basis to perform a similar enhancement to $SU(5)$ through non-generic coefficients, and we expect, multiple $\te$-curves.

The particular top studied in \cite{Mayrhofer:2012zy} is the top 2 of polygon 6 in this appendix. The particular map to non-generic forms of the coefficients is, mapping to the notation of \cite{Mayrhofer:2012zy} for the 3-2 split case,
\begin{equation}
b_{0,1} = - d_3 \alpha \;,\;\; b_1=-c_2 d_3\;,\;\; b_2 = \delta \;,\;\;c_{0,3}= \alpha \gamma \;,\;\; c_{1,2} = d_2 \alpha + c_2 \gamma\;,\;\;c_{2,1} = c_2 d_2 \;,\;\; c_{3,1} = \beta \;.
\end{equation}
This was shown to lead to two different $\te$-matter curves localised on $c_2=0$ and $d_3=0$. 

An interesting feature of this fibration is that it is still not smooth over the locus $\alpha=\gamma=0$. The solution proposed in \cite{Mayrhofer:2012zy} is that we avoid such points, one way being through a similar mechanism we used to avoid the non-flat points in Section~\ref{sec:Flatness} which is setting the homology class of $\alpha$ to be trivial. Indeed this constraint was already applied to the study of this model, albeit in the Tate form side of the rational map, in Section~\ref{sec:recombe8}. To emphasise that avoiding this point does not turn off one of the $\te$-matter curves we perform a similar calculation of the possible homology classes of the curves in the base and show that setting $\alpha$ trivial does not set the classes of either of the $\te$-curves trivial. As mentioned in the main text, because the coefficient of the $\tw^2$ term is set to a constant this fibration allows for only one free parameter in terms of the base classes, 
which we call $B$. However, the further restriction of the coefficients introduces one more freedom, so that one can consider the class of $c_2$ as a free parameter. In terms of these two classes, the GUT class $w$, and the anti-canonical class of the base $\bar {\cal K}$, the homology classes of the sections are shown in Table~\ref{coeffsu4fact}.
\begin{table}
\centering
\begin{tabular}{c|c|c|c|c|c}
 $\delta$ & $\beta$ & $\alpha$ & $\gamma$ & $d_{2} $ & $d_{3}$\\
\hline
 $B$ & $B+\bar{\cal K}-w$ & $\bar {\cal K} - w - B + c_2$ & $3\bar {\cal K} - 2 w - B - c_2$ & $2\bar {\cal K}-w-c_2$ & $\bar {\cal K}-c_2$ \\
\end{tabular}
\caption{Classes of the coefficients with $B$ and $c_2$ arbitrary classes. Here $\bar {\cal K}$ is the anti-canonical class on $B_3$ and $w$ is the class of $S_{}$.}
\label{coeffsu4fact}
\end{table}
With these results it is evident that setting the class of $\alpha$ trivial does not necessarily turn off a $\te$-curves. A simple solution is for example $\left[c_2\right]= \left[w\right]$ and $\left[B\right]= \left[\bar {\cal K}\right]$ which leaves only a mild constraint on $\left[\bar {\cal K}\right]$ and $\left[w\right]$ such that all the sections are positive: $2\left[\bar {\cal K}\right] > 3\left[w\right]$ (for example the embedding into ${\mathbb P}^3$ of $\left[\bar {\cal K}\right]=4H$ and $\left[w\right]=H$ used in Section~\ref{sec:Flatness} would satisfy these).

We now go on to present other $SU(4)$ tops which can form a base for exploring enhancements to $SU(5)$ with multiple $\te$-curves as in the example above.

\subsection*{\texorpdfstring{$SU(4)$}{SU(4)} on polygon 5}

\begin{table}[h]
\centering
\begin{tabular}{|>{\centering}c|c|c|c|}
\hline
$(z_{3},z_{4},z_{5},z_{6},z_{7})$ & \multicolumn{3}{c|}{$(-1,0,1,0,0)$} \\
\hline
\multirow{3}{*}{$P_{T}$} & \multicolumn{3}{c|}{$0 = b_{0,2} e_{0}^{2}e_{1}e_{3}\tu \tv^{2} s_{0}^{2} + d_{0,1}e_{0}e_{1}\tu^{2}\tv s_{0}^{2}s_{1}$} \\
&\multicolumn{3}{c|}{$+ d_{2,1}e_{0}e_{1}^{2}e_{2}\tu^{3}s_{0}^{2}s_{1}^{2} + c_{2,1}e_{0}e_{3}\tv^{2} \tw s_{0} + b_{1}\tu \tv \tw s_{0} s_{1}$}\\
&\multicolumn{3}{c|}{$+ d_{1}e_{1}e_{2}\tu^{2}\tw s_{0}s_{1}^{2} + c_{1}e_{3}\tv \tw^{2}s_{1} +b_{2}e_{1}e_{2}^{2}e_{3} \tu \tw^{2} s_{1}^{2}$ }\\
\hline
$P$ & \multicolumn{3}{c|}{$\frac{1}{16}b_{1}^{4}c_{1}c_{2,1}(b_{0,2}b_{1} - c_{2,1}d_{0,1})(b_{1}b_{2} - c_{1}d_{1})(b_{1}d_{2,1} - d_{0,1}d_{1})$} \\
\hline
Intersection numbers & $S_{0}: \delta_{j0}$ & $S_{1}:\delta_{j2}$ & $U: \delta_{j1}$ \\
\hline
\multirow{2}{*}{Shioda-map} & \multicolumn{3}{c|}{$\text{w}_{1} = 4(S_{1} - S_{0} - \mathcal{\bar K}) + \sum_{i}m_{i}E_{i}, \quad m_{i}=(2,4,2)$}\\
\hhline{~---}
& \multicolumn{3}{c|}{$\text{w}_{2} = 4(U - S_{0} - \mathcal{\bar K} - [c_{1}] ) + \sum_{i}l_{i}E_{i}, \quad l_{i}=(3,2,1)$}\\
\hline
\multirow{4}{*}{Matter curves} & $\{ b_1=0\}$ & $\{c_{1}=0\}$  & $\{c_{2,1}=0\}$\\
& $\mathbf{6_{0,2}}$ + $\mathbf{6_{0,-2}}$ & $\mathbf{4_{-2,-5}}$ + $\mathbf{\overline{4}_{2,5}}$ & $\mathbf{4_{-2,3}}$ + $\mathbf{\overline{4}_{2,-3}}$ \\
\hhline{~---}
& $\{b_{0,2}b_{1} - c_{2,1}d_{0,1}=0\}$  & $\{b_{1}d_{2,1} - d_{0,1}d_{1}=0\}$ & $\{b_{1}b_{2} - c_{1}d_{1}=0\}$ \\
& $\mathbf{4_{2,-1}}$ + $\mathbf{\overline{4}_{-2,1}}$ & $\mathbf{4_{-2,-1}}$ + $\mathbf{\overline{4}_{2,1}}$ & $\mathbf{4_{2,3}}$ + $\mathbf{\overline{4}_{-2,-3}}$\\
\hline
\multirow{4}{*}{Yukawa points}  & $\{b_{1}=c_{1}=0\}$ & $\{b_{1}=c_{2}=0\}$ & $\{b_{1}=d_{0}=0\}$\\
& $\mathbf{6_{0,2}} \mathbf{4_{2,3}}\mathbf{4_{-2,-5}}$ & $\mathbf{\overline{6}_{0,2}} \mathbf{\overline{4}_{-2,1}}\mathbf{\overline{4}_{2,-3}}$ & $\mathbf{\overline{6}_{0,-2}} \mathbf{\overline{4}_{-2,1}}\mathbf{\overline{4}_{2,1}}$  \\
\hhline{~---}
& $\{b_{1}=d_{1}=0\}$ &  & \\
& $\mathbf{6_{0,-2}} \mathbf{4_{-2,-1}}\mathbf{4_{2,3}}$ &  & \\
\hline
\hline
$(z_{3},z_{4},z_{5},z_{6},z_{7})$ & \multicolumn{3}{c|}{$(-1,1,1,0,-1)$} \\
\hline
\multirow{3}{*}{$P_{T}$} & \multicolumn{3}{c|}{$0 = b_{0,2} e_{0}^{2}e_{1}\tu \tv^{2} s_{0}^{2} + d_{0,1}e_{0}e_{1}e_{2}\tu^{2}\tv s_{0}^{2}s_{1}$} \\
&\multicolumn{3}{c|}{$+ d_{2}e_{1}e_{2}\tu^{3}s_{0}^{2}s_{1}^{2} + c_{2,2}e_{0}^{2}e_{1}e_{3}\tv^{2} \tw s_{0} + b_{1}\tu \tv \tw s_{0} s_{1}$}\\
&\multicolumn{3}{c|}{$+ d_{1}e_{1}e_{2}^{2}e_{3}\tu^{2}\tw s_{0}s_{1}^{2} + c_{1}e_{3}\tv \tw^{2}s_{1} +b_{2}e_{1}e_{2}^{2}e_{3}^{2} \tu \tw^{2} s_{1}^{2}$ }\\
\hline
$P$ & \multicolumn{3}{c|}{$\frac{1}{16}b_{0,2}b_{1}^{4}c_{1}(b_{0,2}c_{1} - b_{1}c_{2,2})d_{2}(b_{1}^{2}b_{2} - b_{1}c_{1}d_{1} + c_{1}^{2}d_{2})$} \\
\hline
Intersection numbers & $S_{0}: \delta_{j0}$ & $S_{1}:\delta_{j2}$ & $U: \delta_{j2}$ \\
\hline
\multirow{2}{*}{Shioda-map} & \multicolumn{3}{c|}{$\text{w}_{1} = 4(S_{1} - S_{0} - \mathcal{\bar K}) + \sum_{i}m_{i}E_{i}, \quad m_{i}=(2,4,2)$}\\
\hhline{~---}
& \multicolumn{3}{c|}{$\text{w}_{2} = 4(U - S_{0} - \mathcal{\bar K} - [c_{1}] ) + \sum_{i}l_{i}E_{i}, \quad l_{i}=(2,4,2)$}\\
\hline
\multirow{4}{*}{Matter curves} & $\{ b_1=0\}$ & $\{b_{0,2}=0\}$  & $\{c_{1}=0\}$\\
& $\mathbf{6_{0,0}}$ + $\mathbf{6_{0,0}}$ & $\mathbf{4_{2,-2}}$ + $\mathbf{\overline{4}_{-2,2}}$ & $\mathbf{4_{-2,-6}}$ + $\mathbf{\overline{4}_{2,6}}$ \\
\hhline{~---}
& $\{d_{2}=0\}$  & $\{b_{0,2}c_{1} - b_{1}c_{2,2}=0\}$ & \parbox[c][10mm][c]{30mm}{\center  \vspace*{-4mm}  $\{b_{1}^{2}b_{2} - b_{1}c_{1}d_{1}$ \\ $ + c_{1}^{2}d_{2}=0\}$} \\
& $\mathbf{4_{-2,-2}}$ + $\mathbf{\overline{4}_{2,2}}$ & $\mathbf{4_{-2,2}}$ + $\mathbf{\overline{4}_{2,-2}}$ & $\mathbf{4_{2,2}}$ + $\mathbf{\overline{4}_{-2,-2}}$\\
\hline
\multirow{2}{*}{Yukawa points}  & $\{b_{1}=b_{0,2}=0\}$ & $\{b_{1}=d_{2}=0\}$ & $\{b_{1}=c_{1}=0\}$\\
& $\mathbf{\overline{6}_{0,0}} \mathbf{\overline{4}_{2,-2}}\mathbf{\overline{4}_{-2,2}}$ & $\mathbf{6_{0,0}} \mathbf{4_{2,2}}\mathbf{4_{-2,-2}}$ & non-flat fibre  \\
\hline
\end{tabular}
\caption{Top 1 and 2 for polygon 5}
\end{table}

\begin{table}[h]
\centering
\begin{tabular}{|>{\centering}c|c|c|c|}
\hline
$(z_{3},z_{4},z_{5},z_{6},z_{7})$ & \multicolumn{3}{c|}{$(0,0,0,0,0)$} \\
\hline
\multirow{3}{*}{$P_{T}$} & \multicolumn{3}{c|}{$0 = b_{0,1} e_{0}e_{3}\tu \tv^{2} s_{0}^{2} + d_{0,1}e_{0}e_{2}e_{3}^{2}\tu^{2}\tv s_{0}^{2}s_{1}$} \\
&\multicolumn{3}{c|}{$+ d_{2,1}e_{0}e_{2}^{2}e_{3}^{3}\tu^{3}s_{0}^{2}s_{1}^{2} + c_{2,1}e_{0}e_{1}\tv^{2} \tw s_{0} + b_{1}\tu \tv \tw s_{0} s_{1}$}\\
&\multicolumn{3}{c|}{$+ d_{1}e_{2}e_{3}\tu^{2}\tw s_{0}s_{1}^{2} + c_{1,1}e_{0}e_{1}^{2}e_{2}\tv \tw^{2}s_{1} +b_{2}e_{1}e_{2} \tu \tw^{2} s_{1}^{2}$ }\\
\hline
$P$ & \multicolumn{3}{c|}{$\frac{1}{16}b_{0,1}b_{1}^{4}b_{2}c_{2,1}(b_{1}c_{1,1} - b_{2}c_{2,1})(b_{0,1}d_{1}^{2} - b_{1}d_{0,1}d_{1} + b_{1}^{2}d_{2,1})$} \\
\hline
Intersection numbers & $S_{0}: \delta_{j3}$ & $S_{1}:\delta_{j2}$ & $U: \delta_{j3}$ \\
\hline
\multirow{2}{*}{Shioda-map} & \multicolumn{3}{c|}{$\text{w}_{1} = 4(S_{1} - S_{0} - \mathcal{\bar K}) + \sum_{i}m_{i}E_{i}-m_{3}W, \quad m_{i}=(1,2,-1)$}\\
\hhline{~---}
& \multicolumn{3}{c|}{$\text{w}_{2} = 4(U - S_{0} - \mathcal{\bar K} - [c_{1}] )$}\\
\hline
\multirow{4}{*}{Matter curves} & $\{ b_1=0\}$ & $\{b_{0,1}=0\}$  & $\{b_{2}=0\}$\\
& $\mathbf{6_{2,0}}$ + $\mathbf{6_{-2,0}}$ & $\mathbf{4_{-3,0}}$ + $\mathbf{\overline{4}_{3,0}}$ & $\mathbf{4_{-3,-4}}$ + $\mathbf{\overline{4}_{3,4}}$ \\
\hhline{~---}
& $\{c_{2,1}=0\}$  & $\{b_{1}c_{1,1} - b_{2}c_{2,1}=0\}$ &\parbox[c][9mm][c]{35mm}{\center  \vspace*{-4mm} $\{b_{0,1}d_{1}^{2} - b_{1}d_{0,1}d_{1}$ \\ $ + b_{1}^{2}d_{2,1}=0\}$} \\
& $\mathbf{4_{1,-4}}$ + $\mathbf{\overline{4}_{-1,4}}$ & $\mathbf{4_{1,4}}$ + $\mathbf{\overline{4}_{-1,-4}}$ & $\mathbf{4_{1,0}}$ + $\mathbf{\overline{4}_{-1,0}}$\\
\hline
\multirow{2}{*}{Yukawa points}  &$\{b_{1}=b_{2}=0\}$ & $\{b_{1}=b_{0,1}=0\}$ & $\{b_{1}=c_{2,1}=0\}$\\
& $\mathbf{6_{2,0}} \mathbf{4_{-3,-4}}\mathbf{4_{1,4}}$ & $\mathbf{6_{2,0}} \mathbf{4_{-3,0}}\mathbf{4_{1,0}}$ & $\mathbf{6_{-2,0}} \mathbf{4_{1,4}}\mathbf{4_{1,-4}}$  \\
\hline
\hline
$(z_{3},z_{4},z_{5},z_{6},z_{7})$ & \multicolumn{3}{c|}{$(0,0,1,0,-1)$} \\
\hline
\multirow{3}{*}{$P_{T}$} & \multicolumn{3}{c|}{$0 = b_{0,2} e_{0}^{2}e_{1}^{2}e_{2}\tu \tv^{2} s_{0}^{2} + d_{0,1}e_{0}e_{1}^{2}e_{2}\tu^{2}\tv s_{0}^{2}s_{1}$} \\
&\multicolumn{3}{c|}{$+ d_{2}e_{1}^{2}e_{2}\tu^{3}s_{0}^{2}s_{1}^{2} + c_{2,1}e_{0}\tv^{2} \tw s_{0} + b_{1}\tu \tv \tw s_{0} s_{1}$}\\
&\multicolumn{3}{c|}{$+ d_{1}e_{1}e_{2}e_{3}\tu^{2}\tw s_{0}s_{1}^{2} + c_{1,1}e_{0}e_{2}e_{3}^{2}\tv \tw^{2}s_{1} +b_{2}e_{2}e_{3}^{2} \tu \tw^{2} s_{1}^{2}$ }\\
\hline
$P$ & \multicolumn{3}{c|}{$\frac{1}{16}b_{1}^{4}b_{2}c_{2,1}(b_{1}c_{1,1} - b_{2}c_{2,1})d_{2}(b_{0,2}b_{1}^{2} - b_{1}c_{2,1}d_{0,1} + c_{2,1}^{2}d_{2})$} \\
\hline
Intersection numbers & $S_{0}: \delta_{j1}$ & $S_{1}:\delta_{j3}$ & $U: \delta_{j1}$ \\
\hline
\multirow{2}{*}{Shioda-map} & \multicolumn{3}{c|}{$\text{w}_{1} = 4(S_{1} - S_{0} - \mathcal{\bar K}) + \sum_{i}m_{i}E_{i}-m_{1}W, \quad m_{i}=(-2,0,2)$}\\
\hhline{~---}
& \multicolumn{3}{c|}{$\text{w}_{2} = 4(U - S_{0} - \mathcal{\bar K} - [c_{1}] )$}\\
\hline
\multirow{4}{*}{Matter curves} & $\{ b_1=0\}$ & $\{b_{2}=0\}$  & $\{c_{2,1}=0\}$\\
& $\mathbf{6_{0,0}}$ + $\mathbf{6_{0,0}}$ & $\mathbf{4_{2,1}}$ + $\mathbf{\overline{4}_{-2,-1}}$ & $\mathbf{4_{-2,1}}$ + $\mathbf{\overline{4}_{2,-1}}$ \\
\hhline{~---}
& $\{d_{2}=0\}$  & $\{b_{1}c_{1,1} - b_{2}c_{2,1}=0\}$ & \parbox[c][10mm][c]{35mm}{\center  \vspace*{-4mm} $\{b_{0,2}b_{1}^{2} - b_{1}c_{2,1}d_{0,1}$ \\ $ + c_{2,1}^{2}d_{2}=0\}$} \\
& $\mathbf{4_{-2,0}}$ + $\mathbf{\overline{4}_{2,0}}$ & $\mathbf{4_{1,-1}}$ + $\mathbf{\overline{4}_{-1,1}}$ & $\mathbf{4_{2,0}}$ + $\mathbf{\overline{4}_{-2,0}}$\\
\hline
\multirow{2}{*}{Yukawa points}  &$\{b_{1}=b_{2}=0\}$ & $\{b_{1}=d_{2}=0\}$ & $\{b_{1}=c_{2,1}=0\}$\\
& $\mathbf{6_{0,0}} \mathbf{4_{-2,-1}}\mathbf{4_{2,1}}$ & $\mathbf{6_{0,0}} \mathbf{4_{2,0}}\mathbf{4_{-2,0}}$ & non-flat fibre  \\
\hline
\end{tabular}
\caption{Top 3 and 4 for polygon 5}
\end{table}

\begin{table}[h]
\centering
\begin{tabular}{|>{\centering}c|c|c|c|}
\hline
$(z_{3},z_{4},z_{5},z_{6},z_{7})$ & \multicolumn{3}{c|}{$(0,1,0,-1,-1)$} \\
\hline
\multirow{3}{*}{$P_{T}$} & \multicolumn{3}{c|}{$0 = b_{0,1} e_{0}e_{3}\tu \tv^{2} s_{0}^{2} + d_{0}e_{2}e_{3}\tu^{2}\tv s_{0}^{2}s_{1} $} \\
&\multicolumn{3}{c|}{$+ d_{2}e_{1}e_{2}^{2}e_{3}^{2}\tu^{3}s_{0}^{2}s_{1}^{2} + c_{2,2}e_{0}^{2}e_{1}e_{3}\tv^{2} \tw s_{0} + b_{1}\tu \tv \tw s_{0} s_{1}$}\\
&\multicolumn{3}{c|}{$+ d_{1}e_{1}e_{2}^{2}e_{3}\tu^{2}\tw s_{0}s_{1}^{2} + c_{1,1}e_{0}\tv \tw^{2}s_{1} +b_{2}e_{1}e_{2} \tu \tw^{2} s_{1}^{2}$ }\\
\hline
$P$ & \multicolumn{3}{c|}{$\frac{1}{16}b_{0,1}b_{1}^{4}b_{2}c_{1,1}(b_{0,1}c_{1,1} - b_{1}c_{2,2})(b_{2}d_{0}^{2} - b_{1}d_{0}d_{1} + b_{1}^{2}d_{2})$} \\
\hline
Intersection numbers & $S_{0}: \delta_{j3}$ & $S_{1}:\delta_{j2}$ & $U: \delta_{j2}$ \\
\hline
\multirow{2}{*}{Shioda-map} & \multicolumn{3}{c|}{$\text{w}_{1} = 4(S_{1} - S_{0} - \mathcal{\bar K}) + \sum_{i}m_{i}E_{i}-m_{3}W, \quad m_{i}=(1,2,-1)$}\\
\hhline{~---}
& \multicolumn{3}{c|}{$\text{w}_{2} = 4(U - S_{0} - \mathcal{\bar K} - [c_{1}] ) + \sum_{i}l_{i}E_{i}-l_{3}W, \quad l_{i}=(1,2,-1)$}\\
\hline
\multirow{4}{*}{Matter curves} & $\{ b_1=0\}$ & $\{b_{0,1}=0\}$  & $\{b_{2}=0\}$\\
& $\mathbf{6_{2,2}}$ + $\mathbf{6_{-2,-2}}$ & $\mathbf{4_{-3,1}}$ + $\mathbf{\overline{4}_{3,-1}}$ & $\mathbf{4_{-3,-3}}$ + $\mathbf{\overline{4}_{3,3}}$ \\
\hhline{~---}
& $\{c_{1,1}=0\}$  & $\{b_{0,1}c_{1,1} - b_{1}c_{2,2}=0\}$ & \parbox[c][9mm][c]{35mm}{\center  \vspace*{-4mm} $\{b_{2}d_{0}^{2} - b_{1}d_{0}d_{1}$\\$ + b_{1}^{2}d_{2}=0\}$} \\
& $\mathbf{4_{1,5}}$ + $\mathbf{\overline{4}_{-1,-5}}$ & $\mathbf{4_{1,-3}}$ + $\mathbf{\overline{4}_{-1,3}}$ & $\mathbf{4_{1,1}}$ + $\mathbf{\overline{4}_{-1,-1}}$\\
\hline
\multirow{2}{*}{Yukawa points}  &$\{b_{1}=b_{2}=0\}$ & $\{b_{1}=b_{0,1}=0\}$ & $\{b_{1}=c_{1,1}=0\}$\\
& $\mathbf{6_{2,2}} \mathbf{4_{1,1}}\mathbf{4_{-3,-3}}$ & $\mathbf{6_{2,2}} \mathbf{4_{1,-3}}\mathbf{4_{-3,1}}$ & $\mathbf{6_{-2,-2}} \mathbf{4_{1,5}}\mathbf{4_{1,-3}}$  \\
\hline
\end{tabular}
\caption{Top 5 for polygon 5}
\end{table}
\clearpage

\subsection*{\texorpdfstring{$SU(4)$}{SU(4)} on Polygon 6}

\begin{table}[h]
\centering
\begin{tabular}{|>{\centering}c|c|c|c|}
\hline
$(z_{3},z_{4},z_{5},z_{6},z_{7})$ & \multicolumn{3}{c|}{$(-1,0,1,3,1)$} \\
\hline
\multirow{3}{*}{$P_{T}$} & \multicolumn{3}{c|}{$0 =   \tw^{2}s_{1}e_{3} + b_{0,2} s_{1}^{2}\tu^{2}\tw e_{0}^{2}e_{1}e_{3}$} \\
&\multicolumn{3}{c|}{$+ b_{1}\tu \tv \tw s_{1} + b_{2}\tv^{2} \tw e_{1}e_{2}^{2}e_{3} - c_{0,4}\tu^{4}s_{1}^{3}e_{0}^{4}e_{1}^{2}e_{3}$}\\
&\multicolumn{3}{c|}{$ - c_{1,2} \tu^{3} \tv s_{1}^{2} e_{0}^{2}e_{1}  - c_{2,1}\tu^{2} \tv^{2}s_{1}e_{0} e_{1}e_{2}- c_{3}\tu \tv^{3}e_{1}e_{2}^{2}$ }\\
\hline
$P$ & \multicolumn{3}{c|}{$\frac{1}{16}b_{1}^{4}(b_{1}^{2}c_{0,4} - b_{0,2}b_{1}c_{1,2} - c_{1,2}^{2})c_{3}(b_{1}b_{2} + c_{3})$} \\
\hline
Intersection numbers & $U: \delta_{j0}$ & $S_{1}:\delta_{j0}$ &  \\
\hline
Shioda-map & \multicolumn{3}{c|}{$\text{w}_{1} = 4(S_{1} - U - \mathcal{\bar K} - [b_{2}])$}\\
\hline
\multirow{4}{*}{Matter curves} & $\{ b_1=0\}$ & $\{c_{3}=0\}$  & $\{b_{1}^{2}c_{0,4} - b_{0,2}b_{1}c_{1,2} - c_{1,2}^{2}=0\}$\\
& $\mathbf{6_{0}}$ + $\mathbf{\overline{6}_{0}}$ & $\mathbf{4_{-1}}$ + $\mathbf{\overline{4}_{1}}$ & $\mathbf{4_{0}}$ + $\mathbf{\overline{4}_{0}}$ \\
\hhline{~---}
& $\{b_{1}b_{2} + c_{3}=0\}$  &  & \\
& $\mathbf{4_{1}}$ + $\mathbf{\overline{4}_{-1}}$ &  & \\
\hline
\multirow{2}{*}{Yukawa points}  &$\{b_{1}=c_{3}=0\}$ &  & \\
& $\mathbf{6_{0}} \mathbf{4_{-1}}\mathbf{4_{1}}$  &  &   \\
\hline
\hline
$(z_{3},z_{4},z_{5},z_{6},z_{7})$ & \multicolumn{3}{c|}{$(0,0,1,2,0)$} \\
\hline
\multirow{3}{*}{$P_{T}$} & \multicolumn{3}{c|}{$0 =   \tw^{2}s_{1}e_{1}e_{2} + b_{0,1} s_{1}^{2}\tu^{2}\tw e_{0}e_{1}$} \\
&\multicolumn{3}{c|}{$+ b_{1}\tu \tv \tw s_{1} + b_{2}\tv^{2} \tw e_{2}e_{3} - c_{0,3}\tu^{4}s_{1}^{3}e_{0}^{3}e_{1}^{2}e_{3}$}\\
&\multicolumn{3}{c|}{$ - c_{1,2} \tu^{3} \tv s_{1}^{2} e_{0}^{2}e_{1}e_{3}  - c_{2,1}\tu^{2} \tv^{2}s_{1}e_{0} e_{3} - c_{3,1}\tu \tv^{3}e_{0}e_{2}e_{3}^{2}$ }\\
\hline
$P$ & \multicolumn{3}{c|}{$\frac{1}{16}b_{1}^{4}b_{2}(b_{1}^{2}c_{0,3} - b_{0,1}b_{1}c_{1,2} + b_{0,1}^{2}c_{2})(b_{1}c_{3,1} - b_{2}c_{2,1})$} \\
\hline
Intersection numbers & $U: \delta_{j0}$ & $S_{1}:\delta_{j1}$ &  \\
\hline
Shioda-map & \multicolumn{3}{c|}{$\text{w}_{1} = 4(S_{1} - U - \mathcal{\bar K} - [b_{2}]) + \sum_{i}m_{i}E_{i}, \quad m_{i}=(3,2,1)$}\\
\hline
\multirow{4}{*}{Matter curves} & $\{ b_1=0\}$ & $\{b_{2}=0\}$  & $\{b_{1}^{2}c_{0} - b_{0}b_{1}c_{1} + b_{0}^{2}c_{2}=0\}$\\
& $\mathbf{6_{2}}$ + $\mathbf{\overline{6}_{-2}}$ & $\mathbf{4_{-5}}$ + $\mathbf{\overline{4}_{5}}$ & $\mathbf{4_{-1}}$ + $\mathbf{\overline{4}_{1}}$ \\
\hhline{~---}
& $\{b_{1}c_{3} - b_{2}c_{2}=0\}$  &  & \\
& $\mathbf{4_{3}}$ + $\mathbf{\overline{4}_{-3}}$ &  & \\
\hline
\multirow{2}{*}{Yukawa points}  &$\{b_{1}=b_{2}=0\}$ & $\{b_{1}=c_{2}=0\}$ & \\
& $\mathbf{\overline{6}_{-2}} \mathbf{\overline{4}_{-3}}\mathbf{\overline{4}_{5}}$ & $\mathbf{6_{2}} \mathbf{4_{-1}}\mathbf{4_{3}}$ &   \\
\hline
\end{tabular}
\caption{Top 1 and 2 for polygon 6}
\end{table}

\begin{table}[h]
\centering
\begin{tabular}{|>{\centering}c|c|c|c|}
\hline
$(z_{3},z_{4},z_{5},z_{6},z_{7})$ & \multicolumn{3}{c|}{$(1,1,1,1,0)$} \\
\hline
\multirow{3}{*}{$P_{T}$} & \multicolumn{3}{c|}{$0 =   \tw^{2}s_{1}e_{1}e_{2}^{2} + b_{0,1} s_{1}^{2}\tu^{2}\tw e_{0}e_{1}e_{2}$} \\
&\multicolumn{3}{c|}{$+ b_{1}\tu \tv \tw s_{1} + b_{2}\tv^{2} \tw e_{3} - c_{0,2}\tu^{4}s_{1}^{3}e_{0}^{2}e_{1}$}\\
&\multicolumn{3}{c|}{$ - c_{1,2} \tu^{3} \tv s_{1}^{2} e_{0}^{2}e_{1}e_{3}  - c_{2,2}\tu^{2} \tv^{2}s_{1}e_{0}^{2} e_{1}e_{3}^{2} - c_{3,2}\tu \tv^{3}e_{0}^{2}e_{1}e_{3}^{3}$ }\\
\hline
$P$ & \multicolumn{3}{c|}{$\frac{1}{16}b_{1}^{4}b_{2}c_{0,2}(b_{1}b_{2}^{2}c_{1,2} - b_{2}^{3}c_{0,2} - b_{1}^{2}b_{2}c_{2,2} + b_{1}^{3}c_{3,2})$} \\
\hline
Intersection numbers & $U: \delta_{j0}$ & $S_{1}:\delta_{j2}$ &  \\
\hline
Shioda-map & \multicolumn{3}{c|}{$\text{w}_{1} = 4(S_{1} - U - \mathcal{\bar K} - [b_{2}]) + \sum_{i}m_{i}E_{i}, \quad m_{i}=(2,4,2)$}\\
\hline
\multirow{4}{*}{Matter curves} & $\{ b_1=0\}$ & $\{b_{2}=0\}$  &\parbox[c][10mm][c]{40mm}{\center  \vspace*{-4mm}  $\{b_{1}b_{2}^{2}c_{1} - b_{2}^{3}c_{0}$ \\ $ - b_{1}^{2}b_{2}c_{2} + b_{1}^{3}c_{3}=0\}$}\\
& $\mathbf{6_{0}}$ + $\mathbf{\overline{6}_{0}}$ & $\mathbf{4_{-6}}$ + $\mathbf{\overline{4}_{6}}$ & $\mathbf{4_{2}}$ + $\mathbf{\overline{4}_{-2}}$  \\
\hhline{~---}
& $\{c_{0}=0\}$ &  & \\
& $\mathbf{4_{-2}}$ + $\mathbf{\overline{4}_{2}}$&  & \\
\hline
\multirow{2}{*}{Yukawa points}  &$\{b_{1}=c_{0,2}=0\}$ & $\{b_{1}=b_{2}=0\}$ & \\
& $\mathbf{\overline{6}_{0}} \mathbf{\overline{4}_{-2}}\mathbf{\overline{4}_{2}}$ & non-flat fibre &   \\
\hline
\end{tabular}
\caption{Top 3 for polygon 6}
\end{table}
\clearpage

\newpage
\bibliography{papers}  
\bibliographystyle{custom1}

\end{document}